%% file: fin.tex
\documentclass[11pt,leqno]{article}
\usepackage{latexsym}
\usepackage{epsfig}

\usepackage{color}%%%%%%%%%
\usepackage{subfigure}
\usepackage{framed}

\usepackage{amssymb,amsmath}

\numberwithin{equation}{section}

\newtheorem{theorem}{Theorem}[section]
\newtheorem{lemma}[theorem]{Lemma}
\newtheorem{proposition}[theorem]{Proposition}

\newtheorem{definition}{Definition}[section]
\newtheorem{corollary}[theorem]{Corollary}
\newtheorem{example}{Example}[section]

\def\remark #1{\noindent{\bf Remark:} #1\\}
\def\example #1{\noindent{\bf Example:} #1\\}
\long\def\remarks #1{\noindent{\bf Remarks:} #1\\}
\long\def\claim #1 #2{\bigskip\noindent{\bf Claim {#1}} {\it #2}\bigskip}
\def\xclaim #1 #2{\noindent{\bf Claim {#1}} {\it #2}\bigskip}
\newenvironment{proof}{\noindent{\bf Proof:}}{\hfill $\Box $\\}

\newcommand{\eproof}{\hfill $\Box $\\}

\renewcommand{\thetheorem}{\arabic{section}.\arabic{theorem}}
\def\sqr#1#2{{\vcenter{\vbox{\hrule height .#2pt
              \hbox{\vrule width .#2pt height#1pt \kern#1pt
              \vrule width .#2pt} \hrule height .#2pt}}}}
\def\ncas #1 {\noindent {\bf Case #1.}\ }

\def\bipart #1 #2{\bigskip \noindent {\bf #1} {\it #2}}
\def\xbipart #1 #2{\noindent {\bf #1} {\it #2}}
\def\iipart #1 #2{\bigskip \noindent {\it #1} {\it #2}}
\def\xiipart #1 #2{\noindent {\it #1} {\it #2}}
\def\brpart #1 #2{\bigskip \noindent {\bf #1} {#2}}
\def\xbrpart #1 #2{\noindent {\bf #1} {#2}}
\def\irpart #1 #2{\bigskip \noindent {\it #1} {#2}}
\def\xirpart #1 #2{\noindent {\it #1} {#2}}

\def\o {\overline}

\def\case #1{\bigskip\noindent{{\bf Case} {\em #1}:}}

\def\numcase #1 #2{\bigskip\noindent{{\bf Case #1} {\em #2}:}}

\def\bclaim{\bigskip \noindent{\bf Claim: }}

\def\nclaim #1 {\noindent{\bf Claim #1: }}

\def\obs #1 {\bigskip\noindent{\bf Observation #1: }} 

\def\mathy #1{\ifmmode {#1}\else{$#1$}\fi}
\def\o{\overline}

\newlength{\figwidth}
\newlength{\figindent}
\newlength{\vmargin}
\newlength{\Efigwidth}

\begin{document}
%\tracinggroups=1

\title{Blocking Trails for $f$-factors of Multigraphs} 

\author{%
Harold N.~Gabow%
\thanks{Department of Computer Science, University of Colorado at Boulder,
Boulder, Colorado 80309-0430, USA. 
E-mail: {\tt hal@cs.colorado.edu} 
}%thanks
}%author

\maketitle

\input prelude

\begin{abstract} 
  Blocking flows 
were introduced by Dinic \cite{D} to speed up the computation of 
maximum network flows. They have been used in algorithms for
problems such as 
maximum cardinality matching of bipartite 
graphs 
\cite[Hopcroft and Karp]{HK} and 
general graphs \cite[Micali and Vazirani]{MV},
maximum weight 
%$f$-matching of bipartite graphs and ordinary 
matching of general graphs
\cite[Gabow and Tarjan]{GT}, %{GT89, GT}, 
and many others.
The blocking algorithm of \cite{GT} for matching is based on 
depth-first search.
We extend the depth-first search approach to 
find $f$-factors of general multigraphs. 
Here $f$ is an arbitrary integral-valued
function
on vertices, an {\em $f$-matching}
is a subgraph where every vertex $x$ has degree $\le f(x)$,
an {\em $f$-factor} has equality in every  degree bound.
A set of {\em blocking trails} for an $f$-matching $M$ is a
maximal  
collection \A. of edge-disjoint augmenting trails such that
$M\bigoplus_{A\in \A.} A $ is a valid
$f$-matching.

Blocking trails are needed in efficient algorithms for
maximum cardinality $f$-matching
\cite{HP},
maximum weight $f$-factors/matchings by scaling
\cite{DHZ, G20},
and approximate maximum weight $f$-factors and $f$-edge covers \cite{HP}.
Since these algorithms find many sets of blocking trails, the time to
find blocking trails  is a dominant factor in 
the running time.
Our blocking trail algorithm runs in linear time $O(m)$.
In independent work and using a different approach,
Huang and Pettie \cite{HP} present
a blocking trails algorithm using time $O(m\alpha(m,n))$.
As examples of the time bounds for the above applications,
an approximate maximum weight $f$-factor 
is found in time 
$O(m\,\alpha(m,n))$ using \cite{HP}, and our algorithm eliminates
the factor $\alpha(m,n)$. Similarly a maximum weight $f$-factor
is found in time
$O(\sqrt {\Phi \log \Phi}\, m\,\alpha(m,n)\, \log (\Phi W))\,$ using \cite{HP},
($\Phi =\sum _{v\in V} f(v)$, $W$ the maximum edge weight) and our algorithm
eliminates the $\alpha(m,n)$ factor, making the time within a factor
$\sqrt {\log {\Phi}}$
of the bound for bipartite multigraphs.

  The technical difficulty for this work stems from the fact 
that previous algorithms for both matching and $f$-matching
use vertex contractions to form blossoms. The dfs approach
necessitates using edge contractions.
As an example difficulty, edge contractions require  extending the definition of blossom
to "skew blossoms" -- configurations that must be
reorganized in order to become valid blossoms.
(The algorithm of \cite{HP} uses vertex contractions.)

\end{abstract}

\iffalse
Keywords
matching, $f$-factor, graph, scaling, optimization, algorithm

ORCID ID
 0000-0002-9775-3492

The submission id is: ALGO-D-20-00259

HGabow-938
\fi

\iffalse 
We present an algorithm to find a maximum weight $f$-factor of a multigraph
in time
%$O(\sqrt {f(V)\log n}\; m\log nW)$ simple graph.
$O(\sqrt {\Phi\log \Phi}\; m\log \Phi W)$, where
$\Phi$ is the size of the $f$-factor, i.e., 
$\Phi=\sum{v\in V}f(v)$, $m\ge \Phi/2$ is the number of edges,
and $W$ is an upper bound on the given integral edge weights.

%The best known time bound for the simple
This is within a factor $\sqrt \Phi$ of the best-known bound
for the simple special case of bipartite multigraphs.
\fi

\def\switch{0}

\ifcase \switch
\input intro

\input blossoms
\input alg

\input blocking

\input app       %DONE
\or
\input maximal
\fi

\ifcase \switch %%%%%%%%%%%%%%%%%%%%%%%%%%%%%%
%CASE 0 IS THE BIBLIOGRAPHY

%\section*{Acknowledgments}

\or
\fi
\end{document}

%% file: prelude.tex
\input pmacros
\def\o{\overline} 
\def\u{\underline}
\def\opn{\hangindent=40pt\hangafter=1}
\def\h{{\hskip 20pt}}
\def\v{\vfill}
\def\hi{\advance \parindent by 20pt}
\def\d{\cdot}
\def \il #1{\log^{(#1)} }
\def\al.{{\it add\_leaf}}
\def\alm{{\it add\_leaf}$\,$}
\def\O{o\hbox{-}smallest}
\def\os.{\ifmmode{ \o{\cal S} }\else{$\o {\cal S}$}\fi}
\def\oP.{\ifmmode{ \o{\cal P} }\else{$\o {\cal P}$}\fi}
\def\ot.{\mathy{ \o{\cal T} }}
\def\oG{\o G}
\def\oB{\o B}
\def\oE.{\mathy{\overline E}}
\def\p(#1,#2){\ifmmode p(#1,#2) \else{$p(#1,#2)$}\fi}
\def\op(#1,#2){\ifmmode \o{p}(#1,#2) \else{$\o{p}(#1,#2)$}\fi}
\def\lb{\ifmmode \,{ \rm log}_\beta \else{\it log XX }\fi}
\def\wh{\widehat}
\def\wx.{\ifmmode \wh x \else$\wh x$\fi}
\def\wy.{\ifmmode \wh y \else$\wh y$\fi}
\def\wz.{\ifmmode \wh z \else$\wh z$\fi}
\def\wv.{\ifmmode \wh v \else$\wh v$\fi}
\def\Px.{\ifmmode \wh x \else$\wh x$\fi}
\def\Py.{\ifmmode \wh y \else$\wh y$\fi}
\def\Pz.{\ifmmode \wh z \else$\wh z$\fi}
\def\Pv.{\ifmmode \wh v \else$\wh v$\fi}
\def\Pr.{\ifmmode \wh r \else$\wh r$\fi}
\def\Pr.{\ifmmode \wh r \else$\wh r$\fi}
\def\A.{\mathy{{\cal A}}}
\def\B.{\mathy{{\cal B}}}
\def\C.{\mathy{{\cal C}}}
\def\D.{\mathy{{\cal D}}}
\def\E.{\ifmmode {{\cal E}}\else{{$\cal E$}}\fi}
\def\F.{\mathy{\cal F}}
\def\G.{\mathy{\cal G}}
\def\H#1{\widehat{#1}} 

\def\L.{\mathy{\cal L}}
\def\M.{\mathy{\cal M}}
\def\P.{\mathy{\cal P}}
\def\Q.{\mathy{\cal Q}}
\def\R.{\mathy{\cal R}}
\def\S.{\ifmmode {{\cal S}}\else{{$\cal S$}}\fi}
\def\T.{\mathy{\cal T}}
\def\U.{\mathy{\cal U}}
\def\W.{\mathy{\cal W}}
\def\mathy #1{\ifmmode {#1}\else{$#1$}\fi}
\iffalse
\def\rt{\mathy{\rho}}
\def\pt{\pi}
\long\def\example #1. #2{\bigskip \noindent{\bf Example #1.} 
{#2}\bigskip} 
%{#2}\hfill $\circ$ \bigskip} %\bigcirc$\bigskip}
\long\def\xexample #1 {\bigskip \noindent{\bf Example.} 
{#1}\bigskip}
\fi
\def\goin{\hspace{17pt}}
\def \algname #1.{{\sc #1}}

\def \shell (#1,#2){\mathy {(#1\backslash #2)}}
\def\mydef #1{\bigskip{\narrower#1}\bigskip}
\def\mmpr(#1){\mathy{\widehat{\,#1\,}}}
\def\mpr(#1){\mathy{\overline{#1}}}

%% file: pmacros.tex
\def\today{\ifcase\month\or
January\or February\or March\or April\or May\or June\or
July\or August\or September\or October\or November\or December\fi
\ \number\day, \number\year}
\def\date#1.#2.{\ifcase#1\or
January\or February\or March\or April\or May\or June\or
July\or August\or September\or October\or November\or December\fi
\ #2, \number\year}
\def\ydate#1.#2.#3.{\ifcase#1\or
January\or February\or March\or April\or May\or June\or
July\or August\or September\or October\or November\or December\fi
\ #2, 199#3}
\def\nydate#1.#2.{\ifcase#1\or
January\or February\or March\or April\or May\or June\or
July\or August\or September\or October\or November\or December\fi
\ #2}
\def\doublespace{\multiply\baselineskip by3\divide\baselineskip by2%
                 \def\doublespace{}}%don't doublespace again!
\def\bigdoublespace{\multiply\baselineskip by2%
                 \def\bigdoublespace{}}%don't doublespace again!
\def\imp{\ifmmode {\ \Longrightarrow \ }\else{$\ \Longrightarrow \ $}\fi}
\def\rimp{\ifmmode {\ \Longleftarrow \ }\else{$\ \Longleftarrow \ $}\fi}
\def\ximp{\ifmmode {\Longrightarrow\ }\else{$\Longrightarrow\ $}\fi}
\def\xrimp{\ifmmode {\Longleftarrow\ }\else{$\Longleftarrow\ $}\fi}
\def\iff{\ifmmode {\ \Longleftrightarrow \ }\else{$\ \Longleftrightarrow \ $}\fi}
\def\xiff{\ifmmode {\Longleftrightarrow\ }\else{$\Longleftrightarrow\ $}\fi}
\def\tru{\ {\bf true}\ }
\def\fal{\ {\bf false}\ }
\def\wrt{\ {\it wrt}\ }
\def\endskip{\medskip}%\smallskip
% Nov 19 2013: i redefined \qed to a hollow box
%\def\qed{\  \vrule width4pt depth-1pt height7pt\endskip} %\bigskip }
\def\qed{$\Box$}
\def\qedn{\ \vrule width4pt depth-1pt height7pt }
\def\rqed{\hfill\hbox to 24 pt{\vrule width4pt depth-1pt
height7pt\hfil}\bigskip}
\def\rqedn{\hfill\hbox to 24 pt{\vrule width4pt depth-1pt height7pt\hfil}}
%\def\rsqed{\hfill\vrule width4pt depth-0.5pt height5.5pt\endskip}
% old version:\def\log{\ifmmode \,\hbox{log}\,\else{\it log }\fi}
% this puts 10point roman into superscritps
% new version
\def\log{\ifmmode \,{ \rm log}\,\else{\it log }\fi}
\def\con {\subseteq}
\def\pcon{\subset}
% the old step macro, \numstp (formerly \stp) gives a numbered step
\def\firstnumstp#1 {\bigskip \noindent{\it Step} #1.\newquad}
\def\numstp#1 {\endskip\noindent{\it Step} #1.\newquad}
% the new macro, \stp gives a named step
\def\newquad{\hskip1ex}
\def\stp#1.{\endskip
%\penalty-1000
\noindent{\it #1 Step.}\newquad}
\def\firststp#1.{\bigskip
\penalty-1000
\noindent{\it #1 Step.}\newquad}
\def\cas#1 {\smallskip\noindent{\bf Case} #1.\ } %\quad}
%
%the 1st sec macro allowed breaks at the end of a page
%\def\sec#1{\bigskip\bigskip\noindent{\bf #1}\bigskip}
%
%the 2nd encouraged it to go to the next page
%\def\sec#1{\bigskip
%\penalty-2000%
%\noindent{\bf #1\hfill\break}%\hskip\parindent
%\hbox to \parindent{\hfill}\ignorespaces}
%%\bigskip}
%
%the 3rd set section heading in 12 point type
\long\def\sec#1{\bigskip
\penalty-2000%
\noindent{\twelvebf #1}\par\ignorespaces\noindent\ignorespaces}
%with this sec macro for acknowledgments and bibliography
\def\aorbsec#1{\noindent{\twelvebf #1}}
\def\nsec#1{\penalty-2000%
\noindent{\bf #1\hfill\break}%\hskip\parindent
\hbox to \parindent{\hfill}\ignorespaces}
\long\def\res #1. #2{\bigskip
\penalty-1000
\noindent {\bf #1.}\newquad%
#2 \bigskip}
\long\def\nres #1. #2{\bigskip
%\penalty-1000
\noindent {\bf #1.}\newquad%
#2}
\def\pf{\noindent {\bf Proof.}\newquad}
\def\cont{\ifmmode\star\else$\star$\fi}
% {\ \ \hbox{---}\kern-9.8pt\mid\kern-6pt/\kern-5pt\backslash\ }
% \else{$\ \ \hbox{---}\kern-9.8pt\mid\kern-6pt/\kern-5pt\backslash\ $}\fi}
\def\+{\tabalign} %ordinarily \+ is \outer
\def\nskp{\def\bigskip{}}
\def\i{($i$) } \def\xi{($i$)}
\def\ii{($ii$) } \def\xii{($ii$)}
\def\iii{($iii$) } \def\xiii{($iii$)}
\def\iv{($iv$) } \def\xiv{($iv$)}
\def\pv{($v$) } \def\xpv{($v$)}
\def\pa{({\it a}) } \def\xpa{({\it a})} 
\def\pb{({\it b}) } \def\xpb{({\it b})}
\def\pc{({\it c}) } \def\xpc{({\it c})}
\def\hi{\hskip20pt\i} \def\hii{\hskip20pt\ii} \def\hiii{\hskip20pt\iii}
\def\ha{\hskip20pt\pa} \def\hb{\hskip20pt\pb} \def\hc{\hskip20pt\pc}
\def\tran{{\buildrel*\over\to}}
\def\n{\rlap{$\>/$}}
\def\({{\rm(}} \def\){{\rm)}}
\def\c#1{\lceil {#1} \rceil}
\def\f#1{\lfloor {#1} \rfloor}
\long\def\boxit#1{\vtop{\hrule
\hbox{\vrule\quad\vtop{\vskip5pt\hbox{#1}\vskip5pt}\quad\vrule}
\hrule}} %usually #1 is a \vtop
\def\iboxit#1{\vtop{\hrule
\hbox{\vrule\quad\vtop{\vskip5pt\hbox{{\it #1}}\vskip5pt}\quad\vrule}
\hrule}} %usually #1 is a \vtop
\def\x{\iffalse}
\def\b{\bigskip}
\def\set #1#2{\{ #1:#2 \}}
\def\pset #1#2{( #1:#2 )}
\def\h{\hskip20pt}
\def\hi{\advance\parindent by 20pt}

%% file: intro.tex
\section {Introduction} 
\label{IntroSec}
\def\ed #1#2{\mathy{\{#1,#2\}}}

%%%%%%%%%%%%%%%%%%%%%%%%%%%%%%%%%%%%%%%%

Blocking flows (aka blocking paths) are a fundamental
tool for speeding up augmenting-type algorithms.
They were introduced by Dinic \cite{D} to obtain the first strongly polynomial
algorithm for maximum network flow. Hopcroft and Karp \cite{HK}
(and independently Karzanov \cite{K})
applied
the idea to maximum cardinality bipartite matching. They also proving its
applicability to general graph matching, and  Micali and Vazirani \cite{MV}
gave the first blocking paths algorithm for general graph cardinality matching.
Gabow and Tarjan applied blocking flows in cost-scaling algorithms
for minimum cost network flow (aka degree-constrained subgraphs or $f$-matchings)
\cite{GT89} and minimum cost general graph matching
\cite{GT}.
Here we have only cited some first applications of the idea,
see reference \cite[Sec. 9.6,10.8b,16.7a,17.5a,24.4a]{S}
for a comprehensive bibliography for the many uses of blocking
paths.

\iffalse
ET 
DHZ
DPS

HP card gen alpha
G17 cardinalit

\fi

We present an algorithm for blocking trails for $f$-factors of general multigraphs.
Here ``general'' is to be contrasted with ``bipartite'', an important but much simpler special
case. For the definitions of $f$-factors, blocking trails, etc., please see the above abstract or the last paragraph of this section.

We turn to discussing
recent work that applies blocking trails to multigraph $f$-factors.
Huang and Pettie \cite{HP} give algorithms for a host of  problems
on general multigraphs:
They achieve time 
$O(\sqrt {\Phi}\, m\, \alpha(m,n))$ for maximum cardinality $f$-matching.
Here $\Phi=\sum_{v\in V}f(v)$, i.e.,  twice the size of an $f$-factor.
The same bound holds for a minimum cardinality $f$-edge cover
(i.e., $f$ is a lower bound for degrees in the desired subgraph).
These algorithms use a scaling algorithm that finds an approximation
to the weighted version of the problem.
The scaling algorithm uses
time
$O(m\, \alpha(m,n)\, \epsilon^{-1}\log \epsilon^{-1})$ to find a
$(1-\epsilon)$-approximate maximum weight $f$-matching, or a
$(1+\epsilon)$-approximate minimum weight $f$-edge cover.

The next applications are scaling algorithms for maximum weight $f$-factors.
To put the results in perspective first recall the classic time bounds for (unweighted)
bipartite
$f$-factors \cite{K,ET}:
\begin{equation}
  \label{fTimeBoundsEqn}
  \begin{cases}
    O(n^{2/3}\; m)&G \text{ a simple graph}\\
    O(\sqrt {\Phi} \; m)&G \text{ a multigraph.}
\end{cases}
\end{equation}
\iffalse

\label{sBoundEqn}  O(n^{2/3}\; m)&G \text{ a simple graph}\\
\label{mBoundEqn}  O(\sqrt {f(V)} \; m)&G \text{ a multigraph.}
\end{eqnarray}
\fi
For maximum weight bipartite $f$-factors, Gabow and Tarjan \cite{GT89}  presented  a scaling algorithm
whose time bound
is just a scaling factor above \eqref{fTimeBoundsEqn}, i.e., 
 time  $O(n^{2/3}\; m \;\log (nW))$ for simple graphs and 
$O(\sqrt {\Phi} \; m\; \log (\Phi W))$ for multigraphs. Here $W$ is the maximum
(integral) edge weight.

Now consider general graphs. Duan, He and Zhang \cite{DHZ} give a scaling
algorithm for maximum weight $f$-factors of simple general graphs.
The time is only logarithmic factors above \eqref{fTimeBoundsEqn}, i.e.,
$\widetilde{O} (n^{2/3}\, m\, \log W)$.
In fact when all weights are 1 this is the first algorithm
to achieve the 
bound of  \eqref{fTimeBoundsEqn}, to within logarithmic factors,
for unweighted $f$-factors of simple general graphs.
Returning to weighted $f$-factors, Gabow \cite{G20} gives a scaling algorithm
that achieves the multigraph bound of \eqref{fTimeBoundsEqn}, to within logarithmic factors,
specifically
$O(\sqrt {\Phi\log \Phi} \,m \,\alpha(m,n) \, \log  (\Phi W) )$.

All of the above algorithms
use a subroutine that finds a set of blocking trails.
The time bounds cited above assume the blocking set is found by the subroutine
presented by Huang and Pettie \cite {HP}. It finds a blocking set (more specifically,
a set of blocking trails and cycles) in time $O(m\alpha(m,n))$. Our blocking trail algorithm
runs in time $O(m)$. Using our algorithm the above time bounds all decrease by a factor
$\alpha(m,n)$. (The decrease occurs in \cite {DHZ} but is hidden by the use of
$\widetilde{O}$.) 

\iffalse

\begin{equation*}
\begin{cases}
  O(n^{2/3}\; m \log nW)&G \text{ a simple graph}\\
  O(\sqrt {\Phi} \; m \log \Phi W)&G \text{ a multigraph.}
  \end{cases}
\end{equation*}
, in time 
$\widetilde{O} (n^{2/3}\, m\, \log W)$. Here $W$ is the maximum edge weight.
Ignoring logarithmic factors hidden by $\widetilde{O}$ their bound matches
the bound of $O (n^{2/3}\, m\, \log W)$ in \cite{GT89} for simple bipartite graphs.
\fi

Our algorithm is based on
the depth-first-search approach to blocking paths for matching, introduced in
\cite{GT}; see also  \cite{G17}.
Dfs is a natural approach for finding blocking paths,
since backing up prematurely may necessitate reexploring edges later on.
However it introduces complications for managing blossoms -- they are
usually processed immediately on discovery, but they are necessarily
postponed in a dfs
regime \cite{GT,G17}. Extending
the dfs approach 
to
$f$-factors of general multigraphs introduces new
complications, which we now summarize.

Fundamentally,
in previous settings blossoms are ``shrunk'' using
vertex contraction. This holds even in
previous $f$-factor algorithms, e.g., see the definition of blossom in \cite{G18}.
However dfs requires shrinking by
edge contraction.
A first consequence is new way of discovering blossoms, 
which we call  ``skew blossoms''.
Skew blossoms have the same properties as ordinary blossoms.
However they require a
reorganization in order to achieve valid blossom structure
(Lemma \ref{SkewBlossomLemma}). 

A second consequence of dfs is
incomplete blossoms, i.e., blossoms that do not get completely
processed because of the discovery of an augmenting trail.
Incomplete blossoms occur in ordinary matching, but they present additional
complexity in $f$-factors. Specifically because of edge contraction,
an alternating trail from a later search can reenter an incomplete
blossom.  (See Sec.\ref{BlossomSec}.) The analysis of Sec.\ref{BlockingSec}
shows these reentries present no problem.
\iffalse
NO
??????????????????????
Furthermore incomplete blossoms
introduce new structures in the residual graph (e.g., Fig.\ref{BigIncompleteFig}
\fi

Finally edge-contracted blossoms are not immediately
handled by the standard
min-max formula for maximum cardinality $f$-matching (which we require to prove
the blocking property). We show the essential structure of vertex-contracted
blossoms is preserved by our algorithm (e.g., 
Lemma \ref{TBarResidualLemma}).

These complications also reveal the high-level difference between our approach and
that of 
Huang and Pettie \cite{HP}. Their algorithm uses vertex contraction
rather than edge contraction. Its fundamental idea is cycle cancellation, which does
not occur in our algorithm.

\iffalse
Regarding intuition, multiple occurrences of the same $G$-vertex in the search
structure, in conjunction
the edge-contracted blossoms, complicates visualization and figures. ions
tree view becomes ambiguous since destination of nontree edges
can originate and heads and tails are unclear.

Figures are dificuilt to draw
because of the multigraph

 Fig.\ref{badBlossomFig}. (We remark that these impossible pictures are unfortunately easy to
 draw and believe!)
 \fi
 
\bigskip

The paper is organized as follows.  
Section \ref{BlossomSec} discusses blossoms, first reviewing
the definition of $f$-factor blossoms, and discussing the new types
of blossoms in our algorithm. Appendix \ref{GTApp}
gives the depth-first search algorithm
for blocking ordinary matchings \cite{GT, G17}, for possible
help in the reading of Section \ref{BlossomSec}.
Section
\ref{AlgSec} presents the blocking algorithm.
Section \ref{WellDefinedSec} proves the 
algorithm constructs a valid search structure, most importantly,
the blossoms are valid.
Section \ref{BlockingSec} proves
the augmenting trails found by the algorithm form a blocking set.
This completes the analysis of the algorithm, which is summarized in 
Theorem \ref{MainTheorem}.

\paragraph*{Terminology and conventions}
We often omit set braces from singleton sets, denoting $\{v\}$ as
$v$. So $S-v$ denotes $S-\{v\}$. We abbreviate expressions $\{v\}\cup S$ to $v+S$.
We use a common summing notation: If $f$ is a function on elements
and $S$ is a set of elements then $f(S)$ denotes $\sum_{s\in S}f(s)$.
%$\log n$ denotes logarithm to the base two.

The trees in this paper are out-trees. Writing $xy$ for an arc of a tree
implies the arc joins parent $x$ to child $y$. We extend parent and child relations to
tree arcs, e.g., arc $xy$ is the parent of $yz$. A node $x$ is an {\em ancestor}
of a node $y$ allows the possibility $x=y$, unless $x$ is a {\em proper} ancestor
of $y$. Similarly for arcs. A {\em pendant} edge has no children.

Graphs in this paper are undirected multigraphs. An edge joining
vertices $x$ and $y$ is denoted \ed xy.
Note this notation ignores distinctions between multiple copies of an edge.
Such distinctions are irrelevant to our algorithm.
Also note that 
a loop at $x$ is denoted \ed xx.
Finally note that we use the usual shorthand notation $xy$ to denote edge
\ed xy if context makes it clear that \ed xy is a graph edge, not a tree arc.
(This point is reiterated at the start of Section \ref{AlgSec}.)

In  graph $G=(V,E)$ for $S\con V$ and $M\con E$,
$\delta(S,M)$
($\gamma(S,M)$) denotes the set of edges of $M$ with exactly one
(respectively two) endpoints in $S$.
\iffalse
$\delta_M(S)$
($\gamma_M(S)$)
It is sometimes more convenent
to make $M$ an argument rather than a subscript, e.g., $\delta(S,M)$.
\fi
We omit $M$ (writing $\delta(S)$ or $\gamma(S)$) when
$M=E$. A loop at $v\in S$ belongs to
$\gamma(S)-\delta(S)$.
For multigraphs $G$ all of these sets are
multisets.

Figures in this paper use the following conventions.
Matched edges are drawn heavy, unmatched edges light.
Free vertices are drawn as rectangles.
Trails are indicated by arrowheads on their edges.
A figure illustrating the algorithm can be drawn in the given graph $G$
or in an auxiliary graph \T. (which is essentially
the search tree; 
Fig.\ref{dExamplesFig}(d.1) and (d.2) shows the two views).
The view in $G$ is intuitive and informative but ambiguous:
A given vertex $v$ can occur many times in the search, and an edge of the search
leading to $v$ can  potentially be used at any of
these occurrences (again see Fig.\ref{dExamplesFig}(d)).
The same edge drawn in \T. goes to a new occurrence
of $v$, clearly less informative. Of course there is no such difficulty
in ordinary matching.

\iffalse
%
\footnote{We have attempted to follow Vazirani's dictum on figures for ordinary matching
-- ``It is difficult to overemphasize the importance of well-chosen examples for understanding this result''\cite[p.4 of the May 13, 2014 version; see also the many examples in
the 2020 version]{V}.}
\fi

For an undirected multigraph $G=(V,E)$ with function $f:V\to \mathbb Z_+$,
an {\em $f$-factor} is a subgraph where each vertex $v\in V$ has
degree exactly $f(v)$.
In a {\em partial factor} each  $v$ has
degree $\le f(v)$. $v$ is {\em free} if strict inequality holds.

We often call the edges of a partial $f$-factor a {\em matching} or $f$-{\em matching}.%
\footnote{An $f$-matching is not to be confused with a $b$-matching \cite{S}. This paper
never discusses the latter.}
So we refer to an ordinary matching as a 
{\em 1-matching} (i.e., $f$ is 
identically 1 for all vertices).
$def(x)$ is the deficiency of vertex $x$ in the current matching $M$,
$def(x)= f(x)- |\delta(x,M)|-2|\gamma(x,M)|$.

When discussing a matching $M$, 
the {\em M-type} of an edge $e$ is $M$ or $\o M$ depending on whether
$e$ is matched or unmatched, respectively.
We usually denote an arbitrary M-type as $\mu$,
and 
$\mu(e)$ denotes the M-type of edge $e$.

Consider a graph $G$ with an $f$-matching $M$.
An {\em augmenting trail} is an alternating trail $A$ that
begins and ends at a free vertex,
such that $M\oplus A$ is
a valid matching, i.e., the two ends of the trail still satisfy
their degree bound $f$. (The trail may be closed,
i.e., $A$ begins and ends at the same vertex. {\em Alternating} means
consecutive edges of $A$ have opposite M-types.)
An {\em augmenting set} is a  collection of edge-disjoint augmenting trails
\A. such that $M\bigoplus_{A\in \A.} A $ is a valid $f$-matching
(i.e.,
rematching all the trails keeps every free vertex of $M$
within its degree bound $f$.)
A {\em blocking trail set} is a maximal
augmenting set. It is the analog of a blocking flow. 

\iffalse
A collection of edge-disjoint augmenting trails \A. is {\em blocking (for $M$)} if it is maximal, i.e., there is no augmenting trail 
it cannot be enlarged, i.e., the matching $M-\A.$ on 
$G-\A.$ has no augmenting trail.
Equivalently
define the {\em residual degree constraint function} by
\begin{equation*}
  \label{fPrimeEqn}
f'(x)=f(x) - deg(x,M\oplus \A.)
\end{equation*}
for the degree function $deg$. Blocking means
$M-\A.$ is a maximum cardinality $f'$-matching on $G-\A.$.
\fi

\iffalse
multigraph preprocessing

rematch eligible pairs
edge $e$ has matched copy of weighte $w$, unmatched copy $w-2$
rematching this pair makes them ineligible and preserves
$|M|$ and every degree $deg(v,M)$

multigraph notation
$e^i$ designates the $i$th parallel copy of edge $e$

$xy^i$

\fi

\iffalse
For a set of vertices $S\subseteq V$ and a subgraph $H$ of $G$,
$\delta(S,H)$ ($\gamma(S,H)$) denotes the set of edges with exactly
one (respectively two) endpoint in $S$.
$d(v,H)$ denotes the degree of vertex $v$ in $H$.  
When
referring to the given graph $G$ we often omit the last argument and write, e.g.,
$\delta(S)$. 
(For example a vertex $v$ has $d(v)=|\delta(v)|+2|\gamma(v)|$.)

For a graph $G$, $V(G)$ denotes its vertices and $E(G)$ its edges.
For vertices $x,y$ an {\em $xy$-path} has ends $x$ and $y$.

\fi

%% file: blossoms.tex
\section{Blossoms in {\boldmath $d$}}
\label{BlossomSec}
\def\nb{\bigskip\noindent $\bullet$\ }
\def\edx#1#2{\mathy{#1#2}}
Blossoms are the main issue, and the main stumbling block, in any
algorithm for matching general graphs. So it is appropriate to start with
the new features of blossoms in our algorithm. This section starts
by reviewing the
natural definition of $f$-factor blossoms presented in \cite{G18}.
Then it introduces our new blossom variant, skew blossoms,
and overviews the new potential difficulty for our algorithm,
incomplete blossoms. Finally
it presents a blossom substitute structure that allows our algorithm to
treat weighted blossoms as ordinary
vertices. Appendix \ref{GTApp} gives the predecessor of our algorithm
and may be useful for Fig.\ref{IncompleteFig} showing incomplete blossoms.

\subsection*{Definition of Blossom, in {\boldmath $G$}}

Our algorithm constructs blossoms that satisfy the
definition presented in \cite{G18}.
We call these {\em ordinary blossoms}. We briefly restate
their definition below.
Section \ref{AlgSec} gives a similar definition for ordinary blossoms,
but as they occur in our auxiliary graph \T. rather than $G$.

Let $G$ be a multigraph with an $f$-matching.
A blossom $B$ is a subgraph of $G$ that has a distinguished vertex,
called the {\em base vertex $\beta$}, and a distinguished incident edge,
called the {\em base edge $\eta$}, whose end in $B$ is $\beta$.
(If $\beta$ is free then $\eta$ is an artificial unmatched edge.)
The detailed definition is recursive.
We begin with a graph $\o G$, the original graph $G$ with zero or more
recursively defined blossoms contracted.
A vertex in $\o G$ is either a contracted blossom or
a vertex of $G$ called an  {\em atom}.
The new blossom $B$ is defined as a closed trail $C$.
$C$ begins and ends at a vertex of $\o G$  called the {\em starter}.
Removing the starter from $C$ gives the {\em blossom trail}.
The blossom trail must be  alternating.
{\em Alternating} means two consecutive edges meeting at an atom
have opposite M-types;
there is no restriction on edges meeting at a
contracted blossom.
Any atom may occur arbitrarily many times in $C$.
However a  contracted blossom $A$ 
occurs at most  once (a starter blossom occurs as both ends of $C$). Further if $A$ is in the blossom trail
its base edge must be one of its two incident $C$-edges.

In the base case of the recursion the starter 
is an atom, which  is taken to be $\beta$.
The first and last edges of $C$ must have the same $M$-type, which is called the M-type
of $B$.
$\eta$ is an edge incident to $C$, whose M-type is opposite that of $B$.
%If $\beta$ is free then $\eta$ does not exist and $B$ must have M-type $\o M$.
(So a blossom whose base vertex is free has M-type  $\o M$.)

In the recursive case the starter is a blossom (that is being enlarged).
There is no restriction on the M-type of the
starter's two incident edges. The base vertex of $B$
is the base vertex of the starter. The same holds for the
blossom's M-type and its
base
edge, which must be incident to the trail $C$.
\hfill\qed

\bigskip

Our definition of blossoms differs from \cite{G18} in an important way.
The algorithms of \cite{G18} contract the vertex sets of blossoms.
This is not compatible with the depth-first search strategy of our algorithm.
Instead our definition of blossoms contracts edge sets.
%In this paper we contract edge sets to form blossoms.
(The first example is Fig.\ref{dExamplesFig}(d).)
This means that  a vertex of $G$ occuring in a contracted blossom of $C$
may also have atomic occurrences in $C$.
(Fig.\ref{PDsecendFig} illustrates the common case of pendant edges,
introduced in Lemma \ref{e1Lemma}\xi.) In fact a vertex may also occur in
many different blossoms of $C$.

Our edge-contracted blossoms have three properties allowing them to function like
vertex-contracted blossoms and make our algorithm correct.
First is the most fundamental property of blossoms:
An augmenting trail in $\o G$ (graph $G$ with
all blossoms contracted) gives an augmenting trail in $G$, if we add
appropriate trails through contracted blossoms.
The appropriate trails are called $P_i(v,\beta)$ in \cite{G18}. Here
$v$ is a vertex in a blossom $B$ with base vertex $\beta$, and the edges of
this trail are in the blossom subgraph of $B$. The trails are constructed in Lemma
4.4 of \cite{G18}. This construction works for edge-contracted blossoms if we make
one simple change: 
In every closed trail $C$ of a blossom, replace each occurrence of an atomic vertex $x$
by a new vertex $x_i$. Clearly the edge contractions used in our definition correspond
to vertex contractions in the modified graph.
So the $P_i$ trails exist for edge-contracted blossoms.

The second property is a relaxed version of vertex-contracted blossoms:
At any point in our algorithm a given vertex $x$ of $G$ occurs in at most one blossom.
This is property $(\dagger)$, stated after Proposition \ref{Noe1ExistsProp}.
$(\dagger)$ insures a given blossom occurs at most once in $C$.
It is also crucial in establishing the blocking property (starting with the definition of labels in
\eqref{LabelDefEqn}).

The third property is
that in the search structure, a blossom has one entering edge (its base) and every other edge
is leaving. This property fails for edge-contracted blossoms. However
it holds when our algorithm
halts (Corollary \ref{CompleteBlossomCor}).
This allows us to use the 
min-max formula for maximum cardinality $f$-matching in
Section \ref{BlockingSec} to establish the blocking property of our algorithm.

\iffalse
comes from our proof of the blocking property, i.e., deleting the augmenting trails
found by our algorithm gives a
maximum cardinality $f$-matching. Our proof
uses the

The formula 
refers to connected components in a certain decomposition of the given graph $G$.
Since a vertex of $G$ can occur  in only one such connected component, it is intuitively
clear that 
blossoms must be vertex-contracted, not edge-contracted.
This property holds when it is required, i.e., when
our algorithm halts. In fact it holds at the end of every search for an augmenting trail,
as proved in Lemma \ref.
\fi

We continue with several more comments on the definition of blossom.
First note that the aforementioned trails  $P_i(v,\beta)$ explain
the interpretation of ``alternating'' in the definition of blossom
(the alternating trails through the contracted blossom exist regardless of the edges incident
to it in $C$).
Second, a loop $xx$ qualifies as a closed trail. So a blossom may have
base vertex $x$, closed trail $xx$, whose first and last edges are identical.
Finally,
a blossom is called  {\em heavy} ({\em light}) when its M-type is
$M$ ($\o M$), respectively.

\iffalse
%
\footnote{In this paper these atomic occurrences only exist
for starter blossoms.  Such blossoms $B$
can also be formed using vertex contractions: Simply break
$C$ up into subtrails, each beginning and ending in the starter.
For examples of these atomic occurrences see Lemma \ref{DefinitionOfBlossomSatisfiedLemma}.}
\fi

\subsection*{Skew Blossoms}
\begin{figure}[h]
\centering
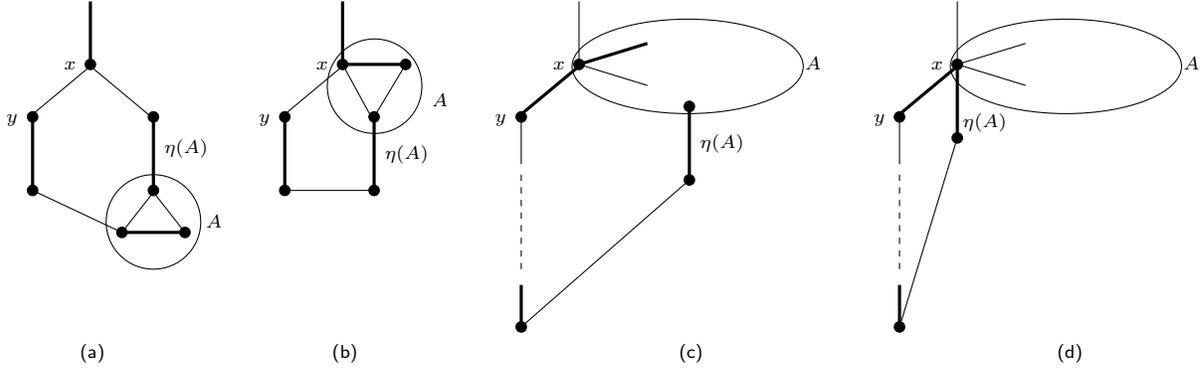
\caption{Example blossoms: (a) A minimal (ordinary) blossom. 
(b) A skew blossom.
Structure of a (heavy) skew blossom: (c) $x\ne \beta(A)$.
(d) $x=\beta(A)$.}
\label{SkewBlossomFig}
\end{figure}

Skew blossoms extend the algorithmic definition of  blossom.
They have the same structure as ordinary blossoms after a reorganization.
The analysis of skew blossoms is similar to the construction of
$P_i$ trails in \cite{G18}. 
Skew blossom are  illustrated in 
Fig.\ref{SkewBlossomFig} and defined as follows.

Consider a  graph with various blossoms contracted, including
a blossom $A$ that contains a vertex $x\in V(G)$.
Let $T$ be an alternating trail,
with first
vertex $x$ and last vertex (the contracted)  blossom $A$,
first edge \edx xy and last edge $\eta(A)$.
$T$ is a {\em skew blossom}.

Note the first vertex $x$ is an atom in $G$ and does not belong to $A$.
Also it is possible that
$\edx xy =\eta(A)$. We may also have \edx xy a loop at $x$. 

\begin{lemma}
  \label{SkewBlossomLemma}
  $T$ is a valid blossom, with base vertex $x$ and M-type $\mu(\edx xy)$.
\end{lemma}

\iffalse
  Define a slight variant of an ordinary blossom: An {\em open
    blossom} has the same recursive definition as an ordinary blossom
  but with a modified base case: The starter $C$ is an alternating
  trail that has no constraint at its two ends -- they can be
  different vertices, and their two incident edges can have arbitrary
  M-type. As an extreme example we allow the starter to be a single
  edge.

\bclaim {Let $B$ be an arbitrary ordinary blossom with base edge
  $\eta=a\beta$. Let $x$ be a vertex in $V(B)$ and $\mu$
  a given M-type. There is an open blossom $B'$ whose vertices
  and edges are precisely those of $B\cup \eta$. The initial open blossom
  has $x$ at the start, with an incident edge of M-type $\mu$.
  The other end is at vertex $a$, with incident edge $eta$.}

The claim proves the lemma, by showing
a skew blossom can be converted to an ordinary
blossom, as follows: Convert $A$ to an open blossom
whose starter is a trail with first vertex $x$, first edge of M-type
$\mu(xy)$,  last edge $\eta(A)$. Form blossom The skew blossom
is a starter blossom, with first vertex the atom $x$, first edge $xy$,
whose trail starts with atom $x$ and edge $xy$, follows $T$ to
$\eta(A)$, and then follows the open blossom starter from $\eta(A)$
to $x$.
\fi

\begin{proof}
  We wish to construct a blossom decomposition $B$ for the given
  skew blossom $S$.
  This decomposition will be a sequence of closed trails $C_i$, $i\ge 0$,
  satisfying the definition of a valid blossom, and having the same set
  of edges as $S$.
The initial blossom trail $C_0$ will
start with the given trail $T$ from the atom $x$ to $\eta(A)$,
and follow a trail in $A$ to an occurrence of $x$ on an edge of
M-type $\mu(\edx xy)$.

  To achieve this goal
we will construct a sequence of alternating trails $T_i$, $i=1,\ldots, k$, where each
$T_{i}$ is a prefix of $T_{i+1}$. The last trail
$T_k$ will be the above initial blossom trail $C_0$.
Along the way we also construct the desired sequence of blossom trails
$C_i, i>0$.
The construction maintains the invariant that starting with $C_0$,
adding the trails $C_i$ in order
gives a valid blossom. Furthermore the construction ends with the $C_i$
consisting of the same edges as $S$.

To begin observe that
the vertex $x$ occurs as an atom in a unique closed trail $C$ of $A$.
Let $x$ refer to some fixed occurrence of $x$ in $C$ (chosen arbitrarily
if there are more than one). Let $\beta$ and $\eta$ be the base vertex
and base edge of the
blossom corresponding to 
$C$.

Suppose $x \ne \beta$.  Let $T_1$ be the subtrail of $C \cup \eta$
that starts with the edge of $(\delta(x)\cup \gamma(x)) \cap \mu(\edx xy)$, follows $C$ to
$\beta$ and then traverses $\eta$.  $T_1$ exists since $C$ alternates
at $x$.  Let $C_1$ be the trail $C-T_1$.  Clearly adding $C_1$ to
$T_k$ gives a valid blossom, which contains all of $C$.

Now suppose $x = \beta$.  We proceed exactly as before.  If the M-type
of $C$ is $\mu(\edx xy)$ then $T_1$ is the entire trail $C\cup \eta$ and
$C_1$ is empty.  If the M-type of $C$ is $\mu(\edx xy)$ then $T_1$ is the
single edge $\eta$ and $C_1$ is $C$.

Now inductively assume $T_{i-1}$ ends with the edge $\eta(A_{i-1})$,
where $A_{i-1}$ is a blossom in the closed trail $D$ of $A$.
Proceed
similar to the base case: Let $\beta$ and $\eta$ be the base vertex
and base edge of $D$.
Let $T_i$ be the subtrail of $D \cup \eta$
that starts with edge $\eta(A_{i-1})$,
follows $D$ to $\beta$ and then traverses $\eta$.
Let $C_i$ be the trail $D-T_i$.
Adding $C_i$ to $T_k \cup \bigcup_1^{i-1} C_j$
gives a valid blossom, which contains all of
$D$. As a special case it is possible that
$A_{i-1}$
is the starter blossom for $D$. In that case
$T_{i}=T_{i-1}$ and $C_i=D$.

Eventually we have $A_{i-1}=A$. In that case $i=k$ and $T_k$ is as specified above. 
\end{proof}

\subsection*{Incomplete Blossoms}

\begin{figure}[h]
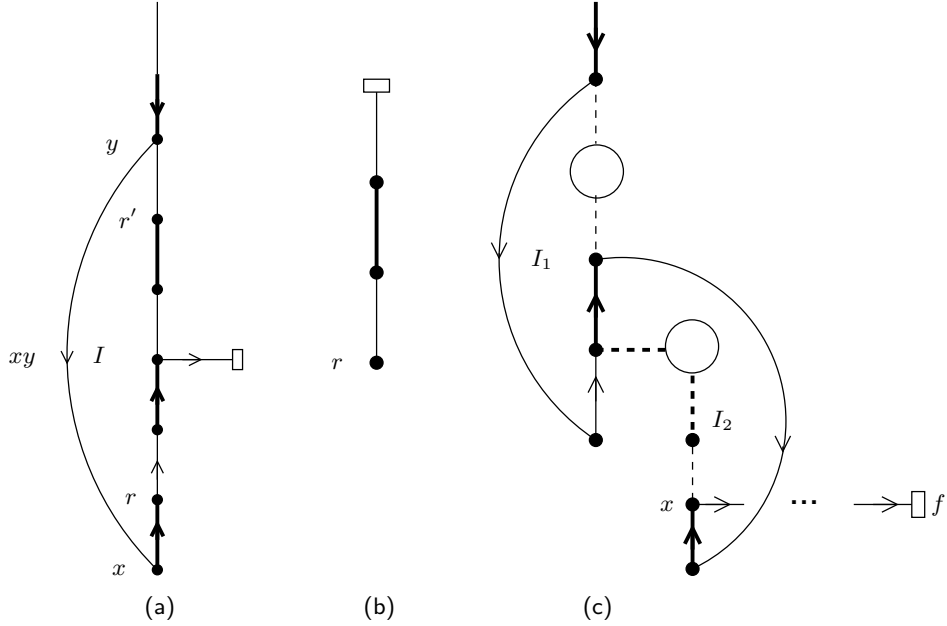

\centering
\input Incomplete.pstex_t
\caption{(a) An incomplete blossom $I$ formed in successful search $S$,
with the trail from $I$
to a free vertex. Arrows show the augmenting trail. 
$I$ is formed by edge $\edx xy$.
Vertex $r'$ is completely scanned in search $S$, but unmatched edges from
$r$ are not scanned.
(b) A subsequent search $S'$ enters $I$ at vertex $r$.
$S'$ cannot enter $I$ at vertex $r'$.
(c)
An incomplete  blossom may contain smaller  incomplete blossoms:
The search forms incomplete  blossom $I_1$ and then incomplete blossom $I_2$.
The trail from $x$ to free vertex $f$ may contain other incomplete blossoms.
Residual edges are drawn dashed.}
\label{IncompleteFig}
\end{figure}

A {\em successful search} finds an augmenting trail.
This leads to the possibility of incomplete blossoms.
A blossom $B$ is {\em incomplete }  if the blossom step has
some $d(u_i,f_i)$ invocation leading to a free vertex, so not all of the vertices in $B$ get scanned. Fig.\ref{IncompleteFig} gives examples of incomplete blossoms.

In 1-matching incomplete blossoms $I$ present little problem.
In detail, the augment step removes all edges in the augmenting path $P$.
In 1-matching this removes all vertices on $P$.
The remaining vertices
in $I$ have all been completely scanned (by the dfs order). So $I$
has the same properites as an ordinary blossom.
This is not the case for multigraphs, since as illustrated in (a),
a vertex $r$ on the augmenting trail remains in the graph with
unscanned edges.

\subsection*{Eliminating Weighted Blossoms}
Blocking trails are required in two types of applications of our algorithm:
algorithms for maximum cardinality $f$-factors
and scaling algorithms for maximum weight $f$-factors.
Cardinality algorithms are handled directly by the algorithm of Section
\ref{AlgSec}.
But weighted algorithms require a modification of the graph, which
we now describe.

\begin{figure}[h]
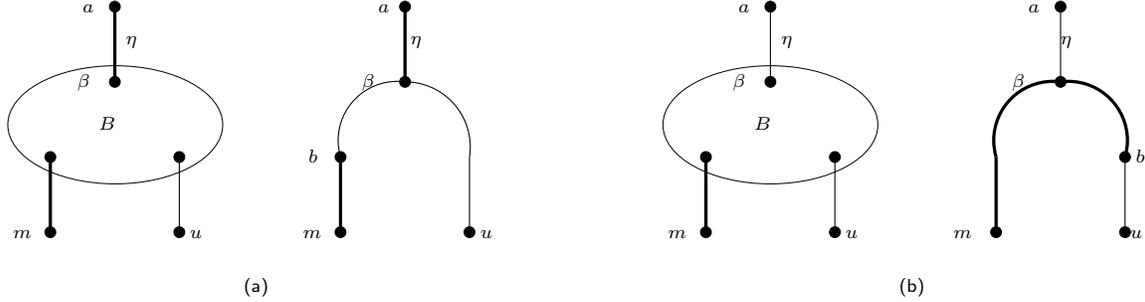

\centering
\input WBsubstitute.pstex_t
\caption{Weighted blossom substitute:
(a) Light blossom. (b) Heavy blossom.
}
\label{WBsubstituteFig}
\end{figure}

In scaling algorithms,
a blocking trail is found and rematched
after each dual adjustment. The graph for the blocking trail is formed
from the input graph by contracting every 
weighted blossom (i.e., blossom with positive dual variable $z(B)$).
A blocking set trail can pass through such a blossom only once.
In contrast an ordinary vertex that is not in a weighted blossom
can appear in
many different blocking set trails, and many different times in
each trail.
In order to have all vertices alike we replace each weighted blossom
by a blossom substitute, illustrated in Fig.\ref{WBsubstituteFig}.

In detail consider a contracted blossom $B$ with base vertex $\beta$,
and base edge $\eta=\edx a \beta$. ($\eta$ does not exist if  $\beta$ is a free vertex.)
Let $Bm$ ($Bu$) denote a typical matched (unmatched)
edge incident to $B$ other than $\eta$. The blossom substitute for $B$ discards the vertices of
$B-\beta$ and replaces them by a
new vertex denoted $b$ and a new  edge $\edx \beta b$. It defines $f(\beta)=f(b)=1$.
If $B$ is a light blossom,  $\edx \beta b$ is unmatched, 
each  $Bm$ edge is replaced by matched $\edx bm$, and
each $Bu$ edge is replaced by unmatched $\edx \beta u$.
(Note that aside from $\eta$, edges incident to $\beta$ in the original graph
are treated as $Bm$ or $Bu$ edges in the substitute.)
If $B$ is heavy then $\edx \beta b$ is matched, 
each  $Bm$ edge is replaced by matched $\edx \beta m$, and
each $Bu$ edge is replaced by unmatched $\edx b u$.

It is easy to see that blocking trails in the original graph $G$
correspond to blocking trails in the graph with substitutes, $G'$. In
detail consider an alternating trail $T$ in $G$ and a weighted blossom
$B$.  $T$ either contains $\beta$ or it does not.  If $T$ contains
$\beta$, it contains $\eta$ (if it exists) and exactly one of the
$Bm$, $Bu$ edges.  If $T$ does not contain $\beta$ it does not contain
$\eta$ or $Bm$ or $Bu$ edge.  Let $T'$ be an alternating trail in
$G'$.  We give the details for a light blossom, heavy blossoms are
symmetric.  If $T'$ contains $\beta$, it contains $\eta$ (if it
exists), and it either contains $\edx \beta b$, exactly one $\edx bm$
edge and no $\edx bu$ edge, or it contains exactly one edge $\edx \beta
u$ edge and no $\edx bm$ edge.  If $T'$ does not contain $\beta$ it
does not contain $\eta$, $\edx \beta b$ or any $\edx \beta u$ or $\edx
bm$ edge.

%% file: SkewBlossom.pstex_t
\begin{picture}(0,0)%
\includegraphics{SkewBlossom.pstex}%
\end{picture}%
\setlength{\unitlength}{1737sp}%
\begingroup\makeatletter\ifx\SetFigFont\undefined%
\gdef\SetFigFont#1#2#3#4#5{%
  \reset@font\fontsize{#1}{#2pt}%
  \fontfamily{#3}\fontseries{#4}\fontshape{#5}%
  \selectfont}%
\fi\endgroup%
\begin{picture}(16830,5235)(-14,-8102)
\put(11401,-3886){\makebox(0,0)[lb]{\smash{{\SetFigFont{7}{8.4}{\rmdefault}{\mddefault}{\updefault}{\color[rgb]{0,0,0}$A$}%
}}}}
\put(5401,-5161){\makebox(0,0)[lb]{\smash{{\SetFigFont{7}{8.4}{\rmdefault}{\mddefault}{\updefault}{\color[rgb]{0,0,0}$\eta(A)$}%
}}}}
\put(6076,-4411){\makebox(0,0)[lb]{\smash{{\SetFigFont{7}{8.4}{\rmdefault}{\mddefault}{\updefault}{\color[rgb]{0,0,0}$A$}%
}}}}
\put(4426,-3886){\makebox(0,0)[lb]{\smash{{\SetFigFont{7}{8.4}{\rmdefault}{\mddefault}{\updefault}{\color[rgb]{0,0,0}$x$}%
}}}}
\put(3601,-4636){\makebox(0,0)[lb]{\smash{{\SetFigFont{7}{8.4}{\rmdefault}{\mddefault}{\updefault}{\color[rgb]{0,0,0}$y$}%
}}}}
\put(4651,-8011){\makebox(0,0)[lb]{\smash{{\SetFigFont{7}{8.4}{\sfdefault}{\mddefault}{\updefault}{\color[rgb]{0,0,0}(b)}%
}}}}
\put(9901,-5011){\makebox(0,0)[lb]{\smash{{\SetFigFont{7}{8.4}{\rmdefault}{\mddefault}{\updefault}{\color[rgb]{0,0,0}$\eta(A)$}%
}}}}
\put(2251,-5086){\makebox(0,0)[lb]{\smash{{\SetFigFont{7}{8.4}{\rmdefault}{\mddefault}{\updefault}{\color[rgb]{0,0,0}$\eta(A)$}%
}}}}
\put(2851,-6136){\makebox(0,0)[lb]{\smash{{\SetFigFont{7}{8.4}{\rmdefault}{\mddefault}{\updefault}{\color[rgb]{0,0,0}$A$}%
}}}}
\put(826,-3886){\makebox(0,0)[lb]{\smash{{\SetFigFont{7}{8.4}{\rmdefault}{\mddefault}{\updefault}{\color[rgb]{0,0,0}$x$}%
}}}}
\put(  1,-4636){\makebox(0,0)[lb]{\smash{{\SetFigFont{7}{8.4}{\rmdefault}{\mddefault}{\updefault}{\color[rgb]{0,0,0}$y$}%
}}}}
\put(1051,-8011){\makebox(0,0)[lb]{\smash{{\SetFigFont{7}{8.4}{\sfdefault}{\mddefault}{\updefault}{\color[rgb]{0,0,0}(a)}%
}}}}
\put(13201,-3886){\makebox(0,0)[lb]{\smash{{\SetFigFont{7}{8.4}{\rmdefault}{\mddefault}{\updefault}{\color[rgb]{0,0,0}$x$}%
}}}}
\put(12376,-4636){\makebox(0,0)[lb]{\smash{{\SetFigFont{7}{8.4}{\rmdefault}{\mddefault}{\updefault}{\color[rgb]{0,0,0}$y$}%
}}}}
\put(15001,-8011){\makebox(0,0)[lb]{\smash{{\SetFigFont{7}{8.4}{\sfdefault}{\mddefault}{\updefault}{\color[rgb]{0,0,0}(d)}%
}}}}
\put(13651,-4711){\makebox(0,0)[lb]{\smash{{\SetFigFont{7}{8.4}{\rmdefault}{\mddefault}{\updefault}{\color[rgb]{0,0,0}$\eta(A)$}%
}}}}
\put(7801,-3886){\makebox(0,0)[lb]{\smash{{\SetFigFont{7}{8.4}{\rmdefault}{\mddefault}{\updefault}{\color[rgb]{0,0,0}$x$}%
}}}}
\put(6976,-4636){\makebox(0,0)[lb]{\smash{{\SetFigFont{7}{8.4}{\rmdefault}{\mddefault}{\updefault}{\color[rgb]{0,0,0}$y$}%
}}}}
\put(9601,-8011){\makebox(0,0)[lb]{\smash{{\SetFigFont{7}{8.4}{\sfdefault}{\mddefault}{\updefault}{\color[rgb]{0,0,0}(c)}%
}}}}
\put(16801,-3886){\makebox(0,0)[lb]{\smash{{\SetFigFont{7}{8.4}{\rmdefault}{\mddefault}{\updefault}{\color[rgb]{0,0,0}$A$}%
}}}}
\end{picture}%

%% file: Incomplete.pstex_t
\begin{picture}(0,0)%
\includegraphics{Incomplete.pstex}%
\end{picture}%
\setlength{\unitlength}{2131sp}%
\begingroup\makeatletter\ifx\SetFigFont\undefined%
\gdef\SetFigFont#1#2#3#4#5{%
  \reset@font\fontsize{#1}{#2pt}%
  \fontfamily{#3}\fontseries{#4}\fontshape{#5}%
  \selectfont}%
\fi\endgroup%
\begin{picture}(10755,7260)(9061,-7427)
\put(10651,-7336){\makebox(0,0)[lb]{\smash{{\SetFigFont{9}{10.8}{\sfdefault}{\mddefault}{\updefault}{\color[rgb]{0,0,0}(a)}%
}}}}
\put(9076,-4411){\makebox(0,0)[lb]{\smash{{\SetFigFont{9}{10.8}{\rmdefault}{\mddefault}{\updefault}{\color[rgb]{0,0,0}$xy$}%
}}}}
\put(10201,-1936){\makebox(0,0)[lb]{\smash{{\SetFigFont{9}{10.8}{\rmdefault}{\mddefault}{\updefault}{\color[rgb]{0,0,0}$y$}%
}}}}
\put(10276,-6886){\makebox(0,0)[lb]{\smash{{\SetFigFont{9}{10.8}{\rmdefault}{\mddefault}{\updefault}{\color[rgb]{0,0,0}$x$}%
}}}}
\put(10051,-4411){\makebox(0,0)[lb]{\smash{{\SetFigFont{9}{10.8}{\rmdefault}{\mddefault}{\updefault}{\color[rgb]{0,0,0}$I$}%
}}}}
\put(10351,-2836){\makebox(0,0)[lb]{\smash{{\SetFigFont{9}{10.8}{\rmdefault}{\mddefault}{\updefault}{\color[rgb]{0,0,0}$r'$}%
}}}}
\put(10426,-6061){\makebox(0,0)[lb]{\smash{{\SetFigFont{9}{10.8}{\rmdefault}{\mddefault}{\updefault}{\color[rgb]{0,0,0}$r$}%
}}}}
\put(15151,-3286){\makebox(0,0)[lb]{\smash{{\SetFigFont{9}{10.8}{\rmdefault}{\mddefault}{\updefault}{\color[rgb]{0,0,0}$I_1$}%
}}}}
\put(17251,-5161){\makebox(0,0)[lb]{\smash{{\SetFigFont{9}{10.8}{\rmdefault}{\mddefault}{\updefault}{\color[rgb]{0,0,0}$I_2$}%
}}}}
\put(15751,-7336){\makebox(0,0)[lb]{\smash{{\SetFigFont{9}{10.8}{\sfdefault}{\mddefault}{\updefault}{\color[rgb]{0,0,0}(c)}%
}}}}
\put(13201,-7336){\makebox(0,0)[lb]{\smash{{\SetFigFont{9}{10.8}{\sfdefault}{\mddefault}{\updefault}{\color[rgb]{0,0,0}(b)}%
}}}}
\put(12826,-4486){\makebox(0,0)[lb]{\smash{{\SetFigFont{9}{10.8}{\rmdefault}{\mddefault}{\updefault}{\color[rgb]{0,0,0}$r$}%
}}}}
\put(16651,-6136){\makebox(0,0)[lb]{\smash{{\SetFigFont{9}{10.8}{\rmdefault}{\mddefault}{\updefault}{\color[rgb]{0,0,0}$x$}%
}}}}
\put(19801,-6136){\makebox(0,0)[lb]{\smash{{\SetFigFont{9}{10.8}{\rmdefault}{\mddefault}{\updefault}{\color[rgb]{0,0,0}$f$}%
}}}}
\end{picture}%

%% file: WBsubstitute.pstex_t
\begin{picture}(0,0)%
\includegraphics{WBsubstitute.pstex}%
\end{picture}%
\setlength{\unitlength}{1776sp}%
\begingroup\makeatletter\ifx\SetFigFont\undefined%
\gdef\SetFigFont#1#2#3#4#5{%
  \reset@font\fontsize{#1}{#2pt}%
  \fontfamily{#3}\fontseries{#4}\fontshape{#5}%
  \selectfont}%
\fi\endgroup%
\begin{picture}(15767,4186)(-5101,-12827)
\put(9151,-8836){\makebox(0,0)[lb]{\smash{{\SetFigFont{7}{8.4}{\rmdefault}{\mddefault}{\updefault}{\color[rgb]{0,0,0}$a$}%
}}}}
\put(-974,-11986){\makebox(0,0)[lb]{\smash{{\SetFigFont{7}{8.4}{\rmdefault}{\mddefault}{\updefault}{\color[rgb]{0,0,0}$m$}%
}}}}
\put(1501,-11986){\makebox(0,0)[lb]{\smash{{\SetFigFont{7}{8.4}{\rmdefault}{\mddefault}{\updefault}{\color[rgb]{0,0,0}$u$}%
}}}}
\put(-149,-9886){\makebox(0,0)[lb]{\smash{{\SetFigFont{7}{8.4}{\rmdefault}{\mddefault}{\updefault}{\color[rgb]{0,0,0}$\beta$}%
}}}}
\put(526,-9286){\makebox(0,0)[lb]{\smash{{\SetFigFont{7}{8.4}{\rmdefault}{\mddefault}{\updefault}{\color[rgb]{0,0,0}$\eta$}%
}}}}
\put( 76,-8836){\makebox(0,0)[lb]{\smash{{\SetFigFont{7}{8.4}{\rmdefault}{\mddefault}{\updefault}{\color[rgb]{0,0,0}$a$}%
}}}}
\put(-1799,-12736){\makebox(0,0)[lb]{\smash{{\SetFigFont{7}{8.4}{\sfdefault}{\mddefault}{\updefault}{\color[rgb]{0,0,0}(a)}%
}}}}
\put(7351,-12736){\makebox(0,0)[lb]{\smash{{\SetFigFont{7}{8.4}{\sfdefault}{\mddefault}{\updefault}{\color[rgb]{0,0,0}(b)}%
}}}}
\put(10651,-10936){\makebox(0,0)[lb]{\smash{{\SetFigFont{7}{8.4}{\rmdefault}{\mddefault}{\updefault}{\color[rgb]{0,0,0}$b$}%
}}}}
\put(-4049,-8836){\makebox(0,0)[lb]{\smash{{\SetFigFont{7}{8.4}{\rmdefault}{\mddefault}{\updefault}{\color[rgb]{0,0,0}$a$}%
}}}}
\put(-5024,-11986){\makebox(0,0)[lb]{\smash{{\SetFigFont{7}{8.4}{\rmdefault}{\mddefault}{\updefault}{\color[rgb]{0,0,0}$m$}%
}}}}
\put(-3449,-9286){\makebox(0,0)[lb]{\smash{{\SetFigFont{7}{8.4}{\rmdefault}{\mddefault}{\updefault}{\color[rgb]{0,0,0}$\eta$}%
}}}}
\put(-2549,-11986){\makebox(0,0)[lb]{\smash{{\SetFigFont{7}{8.4}{\rmdefault}{\mddefault}{\updefault}{\color[rgb]{0,0,0}$u$}%
}}}}
\put(-3824,-10486){\makebox(0,0)[lb]{\smash{{\SetFigFont{7}{8.4}{\rmdefault}{\mddefault}{\updefault}{\color[rgb]{0,0,0}$B$}%
}}}}
\put(-4124,-9886){\makebox(0,0)[lb]{\smash{{\SetFigFont{7}{8.4}{\rmdefault}{\mddefault}{\updefault}{\color[rgb]{0,0,0}$\beta$}%
}}}}
\put(5101,-8836){\makebox(0,0)[lb]{\smash{{\SetFigFont{7}{8.4}{\rmdefault}{\mddefault}{\updefault}{\color[rgb]{0,0,0}$a$}%
}}}}
\put(4126,-11986){\makebox(0,0)[lb]{\smash{{\SetFigFont{7}{8.4}{\rmdefault}{\mddefault}{\updefault}{\color[rgb]{0,0,0}$m$}%
}}}}
\put(5701,-9286){\makebox(0,0)[lb]{\smash{{\SetFigFont{7}{8.4}{\rmdefault}{\mddefault}{\updefault}{\color[rgb]{0,0,0}$\eta$}%
}}}}
\put(6601,-11986){\makebox(0,0)[lb]{\smash{{\SetFigFont{7}{8.4}{\rmdefault}{\mddefault}{\updefault}{\color[rgb]{0,0,0}$u$}%
}}}}
\put(5326,-10486){\makebox(0,0)[lb]{\smash{{\SetFigFont{7}{8.4}{\rmdefault}{\mddefault}{\updefault}{\color[rgb]{0,0,0}$B$}%
}}}}
\put(5026,-9886){\makebox(0,0)[lb]{\smash{{\SetFigFont{7}{8.4}{\rmdefault}{\mddefault}{\updefault}{\color[rgb]{0,0,0}$\beta$}%
}}}}
\put(8101,-11986){\makebox(0,0)[lb]{\smash{{\SetFigFont{7}{8.4}{\rmdefault}{\mddefault}{\updefault}{\color[rgb]{0,0,0}$m$}%
}}}}
\put(10576,-11986){\makebox(0,0)[lb]{\smash{{\SetFigFont{7}{8.4}{\rmdefault}{\mddefault}{\updefault}{\color[rgb]{0,0,0}$u$}%
}}}}
\put(8926,-9886){\makebox(0,0)[lb]{\smash{{\SetFigFont{7}{8.4}{\rmdefault}{\mddefault}{\updefault}{\color[rgb]{0,0,0}$\beta$}%
}}}}
\put(9601,-9286){\makebox(0,0)[lb]{\smash{{\SetFigFont{7}{8.4}{\rmdefault}{\mddefault}{\updefault}{\color[rgb]{0,0,0}$\eta$}%
}}}}
\put(-899,-10936){\makebox(0,0)[lb]{\smash{{\SetFigFont{7}{8.4}{\rmdefault}{\mddefault}{\updefault}{\color[rgb]{0,0,0}$b$}%
}}}}
\end{picture}%

%% file: alg.tex
\section{The blocking trail algorithm}
\label{AlgSec}
\def\fbs{find\_trails}
\def\vtx #1#2{\mathy{#1\dot#2}}

This section presents the algorithm. It illustrates the algorithm's execution
and elaborates on details of the algorithm statement.

The overall algorithm is called \fbs. It uses a recursive depth-first search
routine called $d$.
We begin by describing the data structures and giving an overview of
$d$.
Each vertex $x$ has two lists: $GL(x)$
contains the edges $e$ incident to $x$ 
that can be used in a grow step from $x$.
$e$ may be matched or unmatched, and may be a loop. A nonloop
$e$ starts out in two GL's, and is removed from both lists
when a grow step is executed from the first end.
Similarly
the  second list $BL(x)$
contains the edges incident to $x$ that can trigger a blossom
step.
When find\_blocking\_set begins the lists are initialized as
 {\begin{equation*}
   \def\cpt{20}
%\begin{array}{lll}
GL(x)=\delta(x)\cup \gamma(x),\ BL(x)=\emptyset.
 % GL(x)&=&\delta(x),BL(x)&=&\emptyset.
%\end{array}
\end{equation*}}
\noindent
Each search for an augmenting path begins with the GL's
and BL's
as they were at the end of the previous search.
(We shall see that a BL entry is relevant only
in the search that created it.)

We manage each $BL(x)$ using a routine {\em pop} that
removes edges from a list. Specifically pop$(L)$ removes and returns
an element $e$ of list $L$, where $e$ can be chosen arbitrarily with one exception:
The first invocation
of pop$(L)$ must return the first element that was added to $L$.
Obviously this is a special case of 
a FIFO queue, but we use pop for clarity and greater generality.

\iffalse
For BL's,
clearly there are no pending blossoms when a search begins,
so the BL's should be empty. But previous searches, if successful,
may end with nonempty BL's.
The reader may assume that successful searches
end by emptying all BL's.
The algorithm statement below omits this,
because in fact BL entries from previous searches
are never examined. This is proved in
Appendix \ref{OldBLEntriesApp}.
\fi

The $d$ routine works in two phases. It starts by using its GL
to do every possible
grow step. Then it uses the BL to do every possible blossom step.
In this second phase $BL(x)$ contains the edges where
the dfs retreated from $x$.

The algorithm constructs a search tree \T. consisting of
the edges added in grow steps. To distinguish \T. from $G$
we use the terms {\em node} and {\em arc} for elements
of $V(\T.)$ and $E(\T.)$, respectively. 
We view \T. as an out-tree, so every arc is directed, from parent to child.
Consider an
edge of $G$ $e=\ed xy$.
An occurrence of $e$  in \T.
is denoted as $xy$ or $yx$, where the arc is directed from the first vertex to the second,
e.g., $xy$ means $x$ is the parent.

Since a vertex $x\in V(G)$ can occur multiple times in \T., we
identify nodes of \T. by an incident \T.-arc.
Specifically let $yx$ be an arc
in \T.. % directed from $y$ to $x$, i.e., arc $yx$.
The notation {\vtx yx} refers to the node of \T.
at the $x$ end of $yx$, and \vtx xy is the node at the $y$ end.
So node $\vtx yx$ has $y$ the parent of $x$ or a child of $x$.

\iffalse
HAL
THIS NOTATION IS UNAMBIGUOUS IN A SIMPLE GRAPH BUT NOT
 MULTIGRAPH WHERE THERE CAN BE 2 WX's
\fi
 
The low-level algorithm 
represents blossoms using a data structure for set merging.
The universe is the set of \T.-nodes. The sets are the vertex
sets of the current blossoms.
(So these blossoms are complete or incomplete.)
In the pseudocode below,
$B_{\vtx xy}$ is maintained as the set of all nodes in the blossom 
currently containing node $\vtx xy$. It is a simple matter to
transform this to
a low-level linear-time set merging algorithm \cite{GT85} (also see
the simplfied incremental-tree set-merging algorithm of
 \cite{Gnca}).%
\footnote{A node $\vtx zx$ not in any blossom has
$B_{\vtx xy}=\{\vtx zx\}$. This condition also holds for a loop $xx$ when
edge \ed xx is a singleton blossom. This double usage will not cause any confusion.}
At any point in time $\o{\T.}$ denotes the graph with the current blossoms
(i.e., the $B_\cdot$ sets) contracted.

An invocation of $d$ either returns normally or gets {\em terminated},
i.e., it stops execution prematurely. The latter occurs when an augmenting
trail is discovered. When this occurs every invocation of $d$ in the
current call chain is terminated. All those terminated invocations correspond
to edges on the augmenting trail. This trail is added to \A.,
the set of all augmenting trails that have been discovered.

\fbs\ does not rematch the trails of \A., leaving that to the calling routine.
For instance in applications of our algorithm for weighted matching,
the rematching must undo the blossom substitutes.
Rematching is further discussed in detail after Theorem \ref{MainTheorem}
of
Section \ref{BlockingSec}. 
The other advantage of postponing the rematching
is that it simplifies wording in the analysis.

In contrast,
the algorithm maintains the 
quantities $def(x)$, $x\in V(G)$, as the deficiency of vertex $x$
when the trails of \A. {\em have been} rematched.
This allows proper identification of the free vertices.

\input code.tex

\begin{figure}[p]
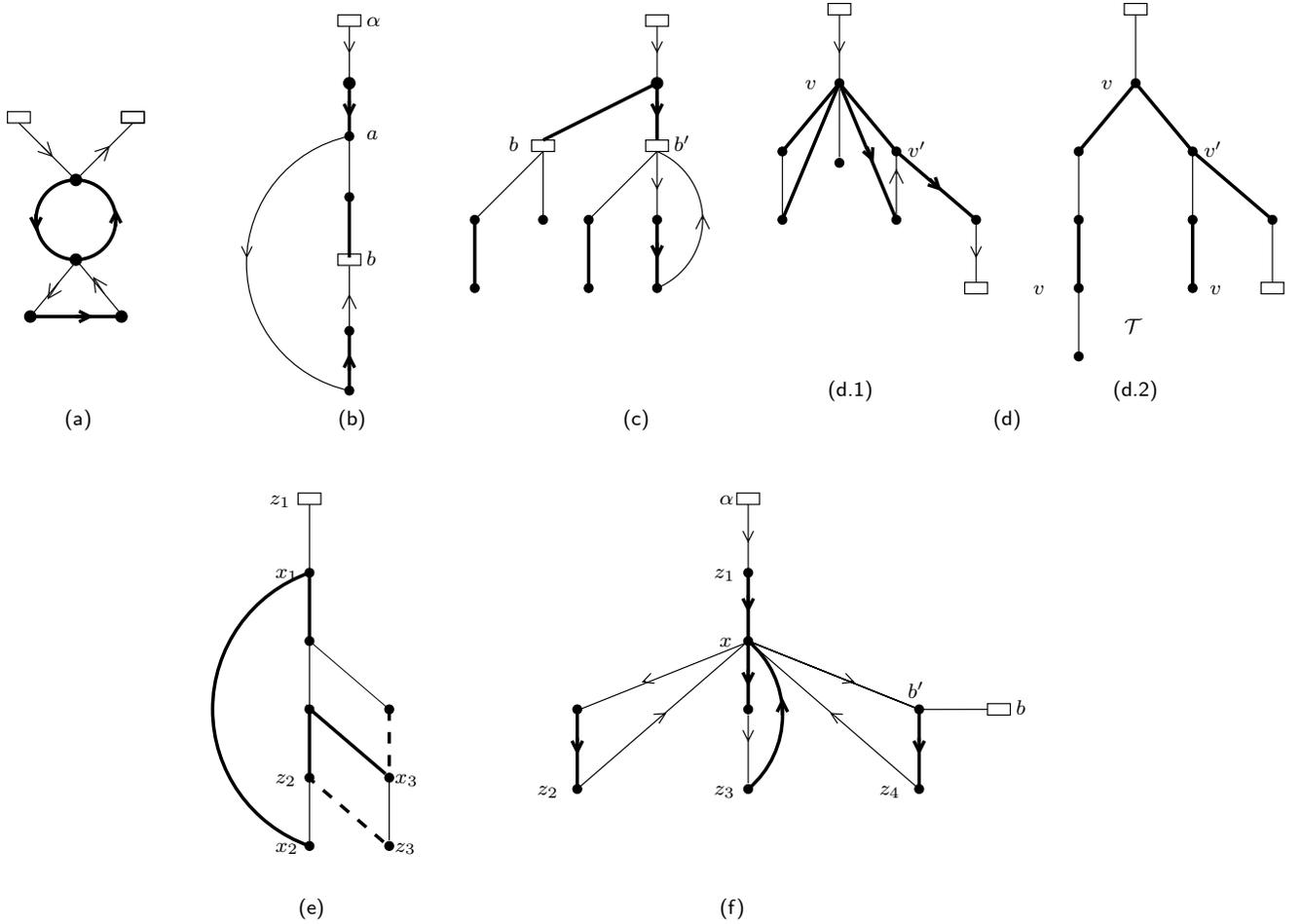

\centering
\input dExamples.pstex_t
\caption{Examples of $d$ search.
  (a) Augmenting trail  in a multigraph.
    Every edge is added in a grow step, no blossom steps.
    (b) A blossom step causes an augment,
    assuming $b\ne \alpha$ or $def(\alpha)\ge 2$. (c) $b'$ is on the augmenting trail.
 The nontrail edges in \T. will not be in future augmenting trails,
 and they are not scanned again.
    (d) (d.1) A search executing 2 blossom steps, the second finding
an augmenting trail. (d.2) Search tree \T. corresponding to (d.1).
The 2 deepest occurrences of $v$ trigger blossom steps.
(e) Cross edges trigger blossom steps:
$d(\vtx {z_1}{x_1})$, executes the first blossom step.
Then 2 cross edges, drawn dashed,
trigger blossom steps in
$d(\vtx {z_i}{x_i})$, $i=2,3$.
%$d(\vtz {z_i}{x_i})$, $i=1,2,3$ 
(f) When \vtx {z_4}x returns \T. contains the alternating path shown 
by arrows. A blossom step for edge $z_4x$  discovers the augmenting trail 
(which contains
every edge shown except $xb'$).
Now suppose edge $b'b$ does not exist.
The blossom for $z_4x$ is discovered in $d(\vtx {z_3}x)$, so its base 
edge is $z_3x$. After that a blossom for  
$z_3x$ is discovered in $d(\vtx {z_2}x)$. Its base 
edge is $z_2x$. The blossom is skew.  Finally a skew blossom for
$z_2x$ is discovered in $d(\vtx {z_1}x)$. Its base 
edge is $z_1x$.}
\label{dExamplesFig}
\end{figure}

Fig.\ref{FBSFig} gives the pseudocode for our algorithm.
Fig.\ref{dExamplesFig} gives examples of searches of $d$.

\begin{figure}[h]
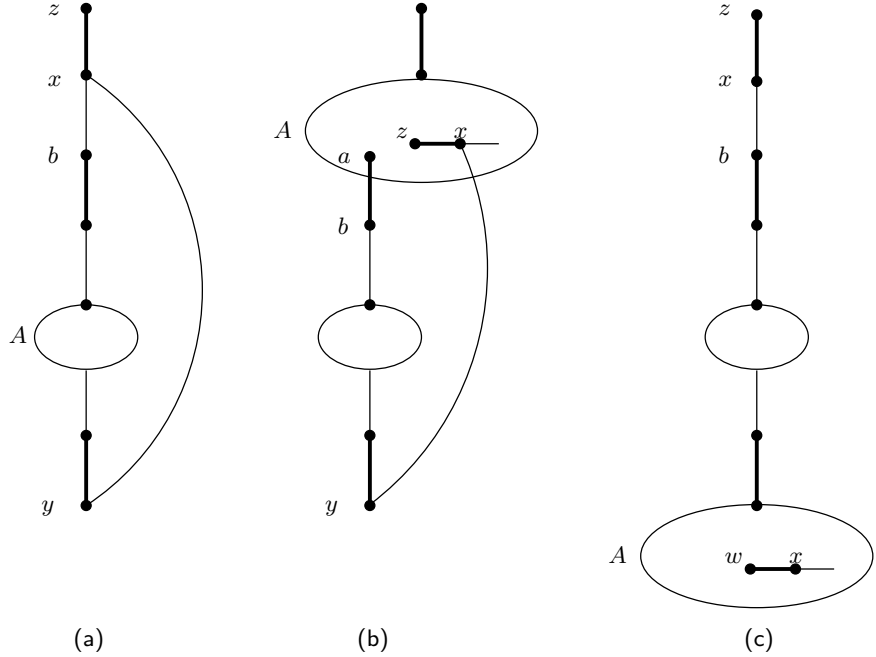

\centering
\input bStepCases.pstex_t
\caption{Three types of blossom steps.
  (a) Initial formation of a blossom.
  The blossom $A$ on $P$ need not exist when the blossom is discovered,
  i.e., when $yx$ is added to $BL(x)$.
Special cases: The entire path 
$P$ can be one edge, the loop 
$xx$. $P$ can have only 1 vertex, $P=x,y,x$.
  (b) Blossom $A$ is enlarged with a blossom trail.
Special case: 
$P$ can be the single edge
$ax$, a noop. 
    (c) Skew blossom.
Special case: $P$ can be a single edge, blossom $A$'s base edge.
}
\label{bStepCasesFig}
\end{figure}

%\paragraph*{Elaboration of the algorithm}
Let us comment on various statements in the algorithm.
An invocation $d(\vtx \varepsilon \alpha)$
made in the main routine is called a {\em search (for an augmenting trail)}.
This invocation is terminated iff an augmenting trail is discovered.
So $d(\vtx \varepsilon \alpha)$ returns if no augmenting trail is discovered.
The discussion following Lemma \ref{PendantProp} 
will show $\alpha$ remains free for the rest of the algorithm.

We make some conventions to justify the terminology for \T..
As usual we call \T. a search tree even though it 
is a forest (each search starts at a different root).
Also we allow \T. to contain loops $xx$ (added in grow steps).
Similarly we allow the path $P$ in a blossom step 
to contain loops.

We continue discussing 
the blossom step.
Fig.\ref{bStepCasesFig} is a high level illustration of the structure of
the three types of blossoms in $d$.
In the blossom base test,
it is unclear why the pop routine returns the correct edge
(i.e., it may have the wrong M-type).
Lemma \ref{oMufirstRemoveLemma}
will show correctness.
In the blossom enlarge test note
\vtx zx is in a blossom is easily checked.
%since it means $B_{\vtx zx}\ne \{\vtx zx\}$.
%NO, BECAUSE OF SINGLETON BLOSSOMS
%but these gives nops anyway
Also let us informally explain this test (the
explanation is proved rigorously below).
The possibility that \vtx zx is in a blossom
covers ordinary blossoms (the possibility holds
when either \vtx zx is already the base vertex of a
blossom, or $zx$ or its reverse occurs on the blossom path).
The possibility that \vtx zx is not in a blossom
but \vtx wx is holds for skew blossoms.
%\vtx zx or some descendant  \vtx wx is in a blossom
%recall an internal node is non-leaf \cite{CLRS}.
Next note that in the blossom step
we identify each arc of $P$ by its nodes in  \T..
%The {\sf for} loops traverse the arcs of $P$ in path order.
\iffalse
For example in the {\sf while} loop, \vtx yx and \vtx zx are vertices of
$P$. Similarly in the loop that merges $B$-values, $v$ and $w$ are vertices of
$G$.
\fi
The blossom step may have $P$  empty, 
specifically $B_{\vtx yx}=B_{\vtx zx}$.
We call such a blossom step a {\em noop} since nothing changes
(there are no $B_{\vtx uv}$ merges or $d(\vtx vu)$ invocations).
An invocation $d(\vtx xy)$ in a grow step adds $xy$ to \T., but
$d(\vtx vu)$ in the blossom step does not cause a similar addition.
So the call chain of $d$ can be a proper subset of
the current search path.
(More precisely,
an invocation $d(\vtx zx)$ in the current call chain has
the corresponding arc $xz$ or $zx$
in the current search path. Within blossoms,
the call chain may omit search path edges.)

%\paragraph*{Examples of the algorithm}

For more  motivation
let us explain the restriction on $uv$ in the blossom-invocation loop.
(The explanation gets rigorously proved in our analysis below.)
Let $ab$ be the first arc of $P$, i.e.,
$d(\vtx ab)$ is not called in the blossom-invocation loop.
($x=a$ in Fig.\ref{bStepCasesFig}(a) and (c).)
Let $B$ denote the blossom being formed.
   Consider two cases.

Suppose  the first node of $P$ is not in a blossom.
This means 
$B$ is either the first blossom formed with base edge $zx$,
or $B$ is a skew blossom (Fig.\ref{bStepCasesFig}(a) or (c)).
There
is no need to invoke $d(\vtx ab)$ because the edges of $GL(x)$
have all been scanned before the blossom step.
(In Fig.\ref{bStepCasesFig}(a) 
the scanning occurs because of edges $zx$ and $yx$.
In Fig.\ref{bStepCasesFig}(c) 
the scanning occurs in blossom $A$ or some subblossom.)

Suppose the first end of $P$ is in a blossom $A$
(Fig.\ref{bStepCasesFig}(b)).
(We remark that this implies the last end of $P$ is also in $A$.)
There is no need for $d(\vtx ab)$ since
\vtx ab belongs to $A$. So just like the skew blossom case,
$A$ gives an invocation
$d(\vtx vu)$ with 
$\vtx vu =\vtx ab$ and $\mu(vu)=\mu(ab)$.

We turn to the issue of blossom completeness.
A blossom $B$ becomes {\em complete} when $\eta(B)$ returns.
(Recall there are two possibilities if $\beta(B)$
is a  free vertex, say $b$.
In a search rooted at $b$, 
i.e., $b=\alpha$,
  $\eta(B) $ is the artificial arc
$\varepsilon \alpha$. 
Alternatively $b$ may occur in a search where it is not
the root. In that case
$\eta(B)$ can be a matched $G$-edge. 
In both cases $B$ can be completed when $d(\eta(B))$ returns.)

\section{Valid search structure}
\label{WellDefinedSec}
The goal of this section is to show \fbs\ constructs a valid search structure, i.e.,
all blossoms are valid and all search paths are alternating \cite{G18}.
We adopt this as an invariant maintained by the algorithm:

\bigskip

{\parindent=0pt
  
  (I1) \hskip 20pt Every step of the algorithm maintains a valid
  search structure, i.e., every blossom is valid and every
  path of the search structure is alternating.

}

\bigskip

\noindent The goal is achieved in Lemma \ref{DefinitionOfBlossomSatisfiedLemma}.

All the proofs and other arguments in this section make the implicit
assumption that all previous steps satisfy (I1).  We
explicitly prove (I1) for every step that modifies the search
structure.  As part of this analysis we will verify that the algorithm
is well-defined. Specifically, in the blossom step it is unclear why
the path $P$ exists for arbitrary $yx\in BL(x)$. The other statements
of the algorithm present no problems in terms of their meaning.

\bigskip

We start with a fundamental concept of the algorithm.
Consider a vertex $x$.  Let $d(\vtx zx)$ be the first invocation of
$d$ for $x$ that returns.  Define $e_1(x)$ to be the edge \ed zx.
This definition requires that $d(\vtx
zx)$ is not terminated.
For example
an invocation $d(\vtx \varepsilon \alpha)$ that gives an unsuccessful
search makes $e_1(\alpha)=\varepsilon \alpha$. If every search for
a vertex $x$ is successful then $e_1(x)$
is an edge of $G$ reached in a search
rooted at some free vertex $\ne x$. $e_1(x)$ need not exist
in this case.

$e_1(x)$ may be either arc $zx$ or
$xz$. In fact we will show the former always holds (Lemma
\ref{e0Lemma}) but this is not required at the moment.  We usually
abbreviate $e_1(x)$ to $e_1$ when context establishes the identity of
the node $x$.

We begin the analysis by showing correctness of the pop routine.

\begin{lemma}
\label{oMufirstRemoveLemma}
When an invocation $d(\vtx zx)$ satisfies the blossom base test,
$\mu(zx)=\o\mu(e_1(x))$. 
\end{lemma}

\remark
    {By definition $e_1(x)$ is the first edge added to $BL(x)$.
      So in $d(\vtx zx)$  pop returns $e_1(x)$, The lemma shows $e_1(x)$
      has the required M-type $\o\mu(zx)$.
      So pop operates correctly in the blossom base test. Correctness of pop
      is not
      an issue in the blossom enlarge test since any edge can be removed
      from $BL(x)$.}

\begin{proof}
  First observe that no operation pop$(BL(x))$ has been performed previously
  in the blossom enlarge test,
  since that would mean a blossom already contains an occurrence of $x$.

If the blossom base test is satisfied then $BL(x)$ is nonempty so $e_1(x)$ exists.
  Suppose for contradiction that $\mu(zx)=\mu(e_1)$.
  Let $xy$ be the edge that satisfies the blossom test, i.e.,
  $xy\in BL(x)\cap \o\mu$.
  So $\mu(xy)=\o\mu(zx)=\o\mu(e_1)$. $d(\vtx yx)$ has returned
  (since $xy\in BL(x)$). $e_1$ was added to $BL(x)$ before $d(\vtx yx)$
  returned (by definition of $e_1$. Also $e_1\ne yx$ since the edges have
  opposite M-types.) So $d(\vtx yx)$ satisfied the blossom test before
  $d(\vtx zx)$. This triggered the operation pop$(BL(x))$, contradiction.
   \end{proof}

A blossom has a simple structure in \T.:

\begin{proposition}
  \label{BlossomSubtreeProp}
The nodes in a blossom $B$ form a subtree of \T..  $\beta(B)$ is the
subtree's root and $\eta(B)$ is the parent arc of $\beta(B)$.
 \end{proposition}

\begin{proof}
  We induct on the size of the blossom.  Let $P$ be the path in $\o {\T.}$
  forming blossom
$B$.  Before the blossom step merges $B_\cdot$ values, the nodes on
path $P$ form a subtree of $\o {\T.}$. The inductive hypothesis,
applied to each blossom on $P$, implies this is a subtree of
\T.. After the merge this subtree corresponds to the set of nodes of
$B$.

Let $d(\vtx zx)$ be the invocation that forms blossom $B$.  Suppose
the blossom base test is satisfied. $P$ is the path from \vtx zx to
$B_{\vtx yx}$. So \vtx zx is $\beta(B)$ and $zx$ is $\eta(B)$, as
claimed in the lemma. Suppose the blossom enlarge test is satisfied.  By
induction \vtx zx is $\beta(B_{\vtx zx})$ before the merge of
$B_\cdot$ values. As in the blossom base test, \vtx zx remains
$\beta(B)$.
\end{proof}

Now we continue to investigate $e_1(x)$.  In a slight abuse of
notation we write $d(e_1(x))$ to denote $d(\vtx zx)$ for the
invocation defining $e_1(x)$.  We next show this invocation is for a
\T.-arc $zx$.
More generally, every
invocation for an occurrence of $x$ in the call chain to $d(e_1(x))$
corresponds to a \T.-arc.
Fig.\ref{dExamplesFig}(f) illlustrates this
with arcs $z_ix$, $i=1,\ldots, 4$, $z_4x=e_1(x)$,
as well as $b'z_4=e_1(z_4)$.
\begin{lemma}
  \label{e0Lemma}
  Every invocation $d(\vtx zx)$ made before $d(e_1(x))$ returns has
  $zx$ an arc of \T..
\end{lemma}

\remark{The lemma even includes invocations made in searches before
  $d(e_1)$ is invoked.}

\begin{proof}
Suppose $d(\vtx zx)$ is executed for an
arc $xz$.  We show some invocation $d(\vtx wx)$ returns before
$d(\vtx zx)$ is invoked. Clearly this implies $d(e_1(x))$ returns before 
$d(\vtx zx)$ is invoked. The lemma follows.

Let $ax$ be the parent arc of $xz$. ($ax$ may be the artificial arc
$\varepsilon \alpha$.) We analyze two cases
  depending on whether $d(\vtx ax)$ returns before or after $\vtx ax$
  has entered a blossom.

  \case {$\vtx ax$ is not in a blossom when $d(\vtx ax)$ returns}
  Since $\vtx ax = \vtx zx$ this case implies $d(\vtx zx)$ has not
  been invoked when $d(\vtx ax)$ returns. So $ax$ is the desired edge $wx$.

\case{$\vtx ax$ is in a blossom when $d(\vtx ax)$ returns}
Let $B$ be the blossom. We claim $ax=\eta(B)$. To prove the
claim suppose the contrary.
Proposition \ref{BlossomSubtreeProp} implies $ax$ is in
some
blossom path $P$. The code for a blossom step implies $d(\vtx ax) $
returns before the blossom is formed. This contradicts the case
definition.  So the claim holds.

Let $yx\in BL(x)$ be the edge triggering the blossom step that
makes \vtx ax a base.  $yx$ was added to $BL(x)$ when $d(\vtx yx)$
returned. So $yx$ returns before \vtx ax becomes a base
vertex. This is before blossom $B$ is formed. So it is before
$\vtx ax=\vtx zx$ is in a blossom. Thus it is before $d(\vtx zx)$ is invoked.
So $yx$ is the desired edge $wx$.
\end{proof}

Consider the special case of the lemma for occurrences of $x$
in searches before $d(e_1(x))$ is invoked.
Every such  occurrence of $x$ 
is on a trail of \A.. In proof, suppose $zx$ is the arc leading to
the occurrence of $x$. 
The corresponding invocation $d(\vtx zx)$ does not return
before $d(e_1)$ returns (definition of $e_1$). So it is terminated, i.e.,
$zx$ is on the augmenting path.
Furthermore $zx$ is not
in a blossom (that would also imply $d(\vtx zx)$ has returned).
So $zx\in E(\A.)$ as claimed.

\iffalse
ORIGINAL PROOF, CORRECT BUT LONGER, AND ONLY WORKS FOR e1
\begin{proof}
We argue by contradiction.  Let $e_1=xz$ be an arc directed from $x$
to $z$.  Let $B$ be the first blossom formed that contains an
occurrence of $x$.  Clearly $B$ is formed before $d(\vtx zx)$ is
invoked.  We will show an invocation $d(\vtx wx)$ returns before $B$
is formed, contradicting $xz=e_1$.  Let $st$ be the base edge of $B$
with $t$ the base vertex.  ($st$ may be the artificial arc
$\varepsilon \alpha$ directed to the free vertex of the search.)

\case {$t=x$}
Let $yx\in BL(x)$ be the edge triggering the blossom step for $B$.
$yx$ was added to $BL(x)$ when $d(\vtx yx)$ returned. So  $yx$
is the desired edge $wx$.

\case {$t\ne x$} $xz$ is in the blossom path $P$ for $B$, and it is
not the first arc of $P$ (as specified in the blossom step code).
This implies $P$ contains an arc $ax$ entering \vtx zx.  (\vtx zx is
not in a blossom on the path $P$, since we have assumed no occurrence
of $x$ is in a blossom before $B$.) We claim $ax$ is the desired edge
$wx$.

To prove the claim, let $rt$ be the edge popped from $BL(t)$
triggering $B$. $d(\vtx ax)$ was invoked before $d(\vtx rt)$, and it
returns before $d(\vtx st)$ pops $rt$. So $d(\vtx ax)$ returns before
$B$ is formed, as desired.
\end{proof}
\fi

The next result  establishes the significance of $e_1$.
We extend our notational convention to abbreviate
$\mu(e_1(x))$ to $\mu$ when $x$ is clear.
Let \A. be defined when $d(e_1)$ is invoked, i.e., \A. contains
the augmenting
trails in the searches before $e_1$ is reached.

\def\ia.{($i.a$) }\def\ib.{($i.b$) }
\def\xia.{($i.a$)}\def\xib.{($i.b$)}

\begin{lemma}
  \label{e1Lemma}
Fix a vertex $x\in V(G)$.
  
\i  Let $zx$ be an arc in \T., $zx\ne e_1$. There are two possibilities for $zx$:

\ia. $zx$ enters \T. before $d(e_1)$ is invoked:  Either $zx\in E(\A.)$ or
$zx$ is an ancestor of $e_1$ in  \T., as well as $\o {\T.}$, the contraction of \T.
 when $e_1$ enters \T..

\ib. $zx$  enters \T. after $d(e_1)$ returns:
$zx$ is a pendant edge of \T. throughout the execution of \fbs.
Furthermore $zx$ has M-type $\mu$.

\ii Consider an  invocation $d(\vtx zx)$ where
$\mu(zx)=\o\mu$.
If $d(\vtx zx)$ pops an edge from $BL(x)$
then $GL(x)$ is empty. So
no edge of $\delta(x)\cup \gamma(x)$ is added to \T.
after the pop
(this includes both the current search and future searches).
\end{lemma}

\remarks{
\ia. applies even if $e_1$ does not exist. If $e_1$ exists, it
  satisfies \ia. ($e_1$ is an ancestor
  of itself). It may also satisfy \ib. ($e_1$ has M-type $\mu$
  and it may be pendant).

 Lemma \ref{lScanningLemma}\iii extends \ia. to describe the ancestors
 of $e_1$ after $e_1$ is popped.
 In Fig.\ref{dExamplesFig}(d.2) the leaf occurrence of
 $v$ illustrates \xib..
 %{PathScanFig} $x_3$ and $x_4$ illustrate \xib..
 
% In Fig.\ref{PathScanFig} $x_3$ and $x_4$ illustrate \xib..

  We later prove that
  \ii holds without the requirement on $\mu(zx)$.}

\begin{proof}
  \i First observe that \i  covers all possibilities for $zx$:
  In proof the only case not covered is that
$d(\vtx zx)$ is invoked after $d(e_1)$ is invoked and before $d(e_1)$ returns. 
Such an invocation  returns before $d(e_1)$ returns. This contradicts the definition
of $e_1$.
 
\ia.
Assume $d(\vtx zx)$ is invoked in the same search as 
$d(e_1(x))$. (If not  $d(\vtx zx)$ is invoked in a previous search,
and as shown above $zx\in E(\A.)$.)
$d(\vtx zx)$ is invoked before 
$d(e_1)$ is invoked (assumption) and does not
return before $d(e_1)$ returns
(definition of $e_1$). So $d(e_1)$ returns during the execution of
$d(\vtx zx)$. (Notice this accounts for $d(\vtx zx)$ being terminated
before it returns.)
We have shown
$e_1$ descends from $zx$ in \T.. $zx$ is not
in a contracted blossom $B$ of $\o{\T.}$: The contrary makes $zx$ an arc
in the blossom path of $B$. That implies 
$d(\vtx zx)$ returns before $B$ is formed, contradiction.
(\vtx zx may be the base of a contracted blossom.)

  \ib.
  $d(e_1)$ returns with $GL(x)\cap \o\mu$ empty.
  $zx$ is added to \T. in a grow step so this implies it has
M-type $\mu$.
Furthermore it implies
$d(\vtx zx)$ does not execute any grow steps, i.e., $zx$ is a pendant
edge of tree \T..  Thus \ib. holds.

\ii $zx\ne e_1$ since the two edges have opposite M-types.
So $d(\vtx zx)$ returns after $d(e_1)$.
\iffalse
Part \i with $zx$'s M-type
shows $zx$ is a proper ancestor of
$e_1$ in \T..
\fi
$d(e_1)$ returns with $GL(x)\cap \o\mu$ empty.
$d(\vtx zx)$ removes every edge of $GL(x) \cap \mu$
before it pops any edge from $BL(x)$. 
So $GL(x)$ becomes empty before 
$d(\vtx zx)$ pops 
an edge from $BL(x)$.
\end{proof}

As an example suppose $x$ becomes a free vertex in an unsuccessful search.
In \ia. observe that there may be a path of multiple occurrences of
$x$. \ib. shows that after the unsuccessful search, every occurrence of $x$ is a leaf. So $x$ will not be on an augmenting trail after the unsuccessful search.

The lemma contains the seeds of the entire algorithm, which we
now sketch.  (The sketch along with other structural details is proved
in Lemmas \ref{Beforee1Lemma}--\ref{Stage3Lemma}.)
Consider a fixed vertex $x$.  Each time
an edge of $BL(x)$ is popped it triggers a blossom step.  The blossom
step may be a noop (i.e., the current blossom containing $x$ does not
get enlarged).  Ignoring these noops there are three successive
``stages'' for vertex $x$: Stage 1 accounts for the time up until
$e_1(x)$ is popped and its blossom is formed.
After that $BL(x)$-pops  trigger blossoms for leaf occurrences of $x$
(corresponds to Lemma \ref{e1Lemma}\xib.).  We call this Stage 2.
After that $BL(x)$-pops form skew
blossoms (corresponds to Lemma \ref{e1Lemma}\xia.). This is Stage 3.
The search may halt, because of an augmenting
trail, at any point during this progression (e.g., before $e_1$ is defined,
before it is popped, etc.).
Also at any point in the progression
arbitrarily many blossoms may be formed by pops from lists
$BL(x')$ for various vertices $x'\ne x$, and such blossoms may contain $x$.
For example in Stage 1 blossoms may absorb
occurrences of vertex $x$ before $e_1$ is popped.

Vertex $x$ can occur in a
blossom in only one search for an augmenting trail, the search where
$d(e_1)$ returns (Lemma \ref{e1Lemma}\xib. and its proof).  So the
pendant edge and skew blossom stages can only occur in the search
where $d(e_1)$ returns.  (Regarding noops, the blossom step for $e_1$
may be a noop, e.g.,
consecutive arcs $ax,xz$ in a blossom path
with $ax=e_1(x)$ -- $d(\vtx zx)$ pops $ax$.
Pendant edges and skew blossoms
do not give noops.  Other noops at arbitrary points may occur.)

\bigskip

As previously mentioned we must show the blossom step is well-defined.
It is convenient to record that  as the following property:

\bigskip

\noindent
$(*)$ \hskip 20pt  \parbox {5.8 in} {At the moment
an invocation  $d(\vtx zx)$ pops an edge $yx$ from $BL(x)$,
$\vtx yx$ descends from  $\vtx zx$ in
$\o {\T.}$.}

\bigskip

\noindent Notice the contraction $\o {\T.}$ need not be 
the same as when $yx$ was added to $BL(x)$.

\begin{figure}[h]
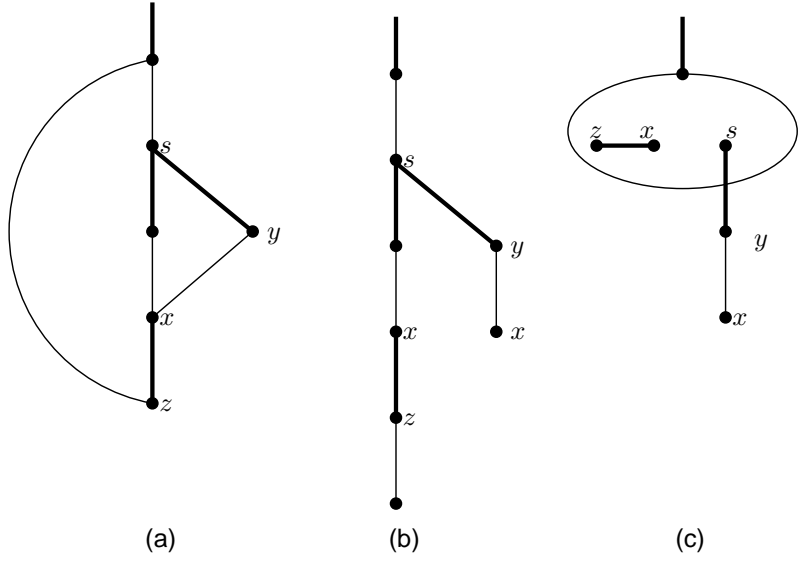

\centering
\input PDescend.pstex_t
\caption{Blossom path $P=sy,yx$ does not descend from
\vtx zx in \T. (b), but does so in $\o {\T.}$ (c).
\iffalse
Note also in the blossom-invocation loop,
$u$ cannot be \vtx yx yet it can be
\vtx xy: The former holds because the
test of the loop requires $u\in V(P-\vtx yx-\vtx zx)$.
For the latter note that $yx$ is the arc leaving $y$ in $P$.
If $y$ has not been blossom scanned then the algorithm sets
$u$ to $y$, $uv$ to arc $yx$, and it invokes
$d(\vtx xy)$.
\fi
}
\label{PDsecendFig}
\end{figure}

It is helpful to give some illustrations for $(*)$.
Fig. \ref{PDsecendFig} shows
$(*)$ needn't hold if we change  $\o {\T.}$ to \T..

\begin{figure}[h]
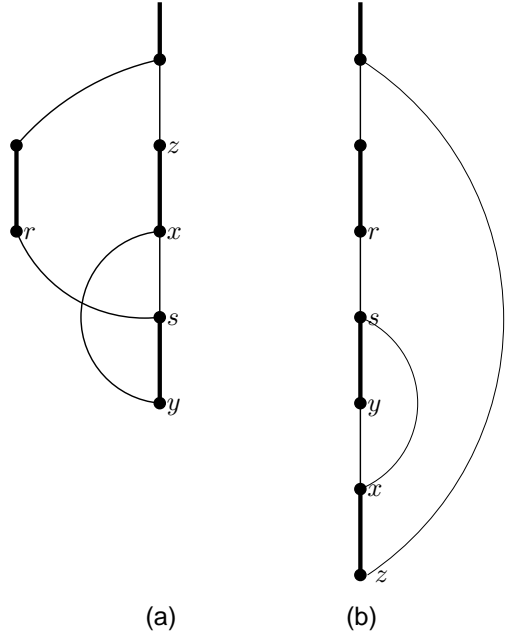

\centering
\input badBlossom.pstex_t
\caption{(a) A flawed counterexample: Blossom step for edge $rs\in BL(s)$ violates invariant (I1). However the configuration is impossible --
  the true configuration
  is (b) (assuming $r$ is visited before $z$). Note the blossom step for $sx\in BL(x)$ is a noop.}
\label{badBlossomFig}
\end{figure}

Next consider Fig.\ref{badBlossomFig}. It might seem to
violate
(I1) because $d$ executes as follows:
The invocation $d(\vtx zx)$ executes
a blossom step for edge $yx\in BL(x)$. The blossom step invokes
$d(\vtx ys)$. It executes a blossom step for edge $rs\in BL(s)$. But 
$B_{\vtx rs}$
is not a descendant of $B_\vtx ys=B_{\vtx zx}$.
This purported counterexample is incorrect because the picture in (a)
is impossible, the true picture is (b).

In our analysis of a fixed vertex $x$ we assume $(*)$ for all previous blossoms.
This implies an enhanced version of the assumption: Suppose we are considering a blossom
formed when an invocation
$d(\vtx zx)$ pops an edge $yx$.
Let $d(\vtx st)$ be in the call chain from $d(\vtx zx)$. $(*)$ implies
that when  $d(\vtx st)$ returns, $\beta(B_{\vtx st})$ descends from \vtx zx.

$(*)$ implies
any blossom formed after $d(\vtx zx)$ is invoked but before the pop
has its base vertex descending from  \vtx zx.

To set the stage for the ensuing discussion we extend Section
\ref{BlossomSec}'s definition of blossom.  That definition specifies
the properties of a blossom $B$ in the given graph $G$.  We now
include the corresponding properties as they occur in the search tree
\T..

\iffalse
HAL, IN THIS DISCUSSION I DISTINGUISH BETWEEN BLOSSOM TRAILS (WHICH OCCUR IN g)
AND BLOSSOM PATHS (WHICH OCCUR IN t)
\fi

\subsection*{Definition of Blossom, in {\boldmath $G$} and  {\boldmath \T.}}

\noindent$\bullet$ {\em Starter:} The starter can be a node $x$ of \T.
where $x$ does not occur in the 
\T.-subtree of any  blossom.
$x$ becomes
the blossom base $\beta$.
The new blossom $B$ is formed by adding a
blossom path $P$ that begins at node $\beta$ and
ends at another \T.-occurrence of $\beta$.  (In the code for $d$,
$P$ ends at $B_{\vtx yx}$, and ordinary blossoms have $B_{\vtx
  yx}=\{\vtx yx\}$.) Blossom $B$'s occurrence in $G$ is formed by identifying the two node
occurrences of $\beta$.  The first and last edges of $P$ must have the
same M-type $\mu$. The parent arc of $\beta$ is the base edge $\eta$, which
must have opposite M-type $\o \mu$. (A special case is where the blossom 
path $P$ is the loop $\beta\beta$. $B$'s occurrence in $G$ is also that loop.)

A blossom $A$ can be enlarged by using $A$ as the starter.
Blossom $B$ is formed by
adding a blossom path $P$ that begins at some node $a$ of $A$, and
ends at an occurrence of some $G$-vertex in subtree $A$.  (In the code for $d$ the
starter is the blossom $A=B_{\vtx zx}$.  As above $P$ ends at
$B_{\vtx yx}$, and ordinary blossoms have $B_{\vtx yx}=\{\vtx yx\}$.) 
Blossom $B$'s occurrence in $G$ is formed by identifying $\vtx yx$ with node
\vtx zx in $A$. (Note from the code of $d$ that $P$ starts with some
node $a$ in $A$, not necessarily $\vtx zx$.)  The M-types of the extreme edges of $P$ are
arbitrary.
The base vertex and edge of $B$ are inductively
defined as those of $A$.

For skew blossoms the starter is a node not in any blossom subtree,
which becomes the base vertex
$\beta$.
%$\beta=\vtx zx$,  $B_{\vtx zx}= \{\vtx zx\}$.
$P$ ends in a contracted blossom containing an occurrence
of $\beta$.  (In the code for $d$ this blossom is
$B_{\vtx yx}\ne  \{\vtx yx\}$.)
  All other requirements for skew blossoms are the same as
  ordinary blossoms.

\nb {\em Blossom trail:} In the code for a blossom step, $P$ is a
path in $\o {\T.}$, the contraction of \T. when 
blossom $B$ is triggered. Assuming $(*)$, the \T.-path $P$ is guaranteed to have
all required properties
of the blossom trail. Specifically,
$P$ is guaranteed to be
alternating, since
the arcs of $P$ are in \T..
If $A$ is a contracted blossom on $P$,
$P$ is guaranteed to contain $A$'s base edge, which is the parent arc of
$A$. This also guarantees that $A$ occurs only once in the trail $C$ of the blossom.
This property extends to a starter blossom $A$: Its
base edge is the parent of the contracted $A$, so $A$ does not reoccur in the blossom trail.
A $G$-vertex may occur
arbitrarily many times in $P$.
\hfill\qed

\bigskip

We start with some simple properties of blossoms.
Consider the moment in time when a blossom $B$ becomes complete,
i.e., $d(\eta(B))$ returns. Let $x$
be a vertex that occurs in $B$.

\begin{proposition}
\label{CompleteBlossomProp}
When $B$ becomes complete, an
  invocation $d(\vtx wx)$ has returned for two edges $wx$ of opposite
  M-type.
Moreover $e_1(x)$ has been popped.
\end{proposition}

\begin{proof}
Wlog assume
$B$ is the first blossom to contain an occurrence of vertex $x$.

\case{$x$ occurs as   $\beta(B)$}
  Let $yx$ be the first edge of $BL(x)$ to trigger a blossom
(not necessarily $B$).
  The invocations $d(\vtx yx)$ and $d(\eta(B))$ are as desired.
  Obviously $e_1(x)=yx$ was popped.

\case{$x$ does not occur as    $\beta(B)$}
%(Similar to Lemma \ref{e0Lemma}.)
$x$ is not the first or last vertex of the blossom path $P$,
since $x$ is not the base vertex and $x$ does not occur in a
blossom when $B$ is formed. So $P$ contains
an arc $ax$. $ax$ is followed by an arc $xb$ in $P$,
since \vtx ax is not in a blossom on $P$
(again by the choice of $B$ as first blossom).
%So $ax$ is followed by an edge $xb$ in $P$.
The invocations $d(\vtx ax)$ and $d(\vtx bx)$ are as
desired.

Consider the moment when $d(\vtx bx)$ executes the blossom enlarge test.
$GL(x)$ has been emptied. In particular $d(e_1(x))$ has been invoked
and has returned, adding $e_1(x)$ to $BL(x)$. So
$d(\vtx bx)$ pops $e_1(x)$ unless some previous invocation popped it.
\end{proof}

The next two results give the properties of the two blossom tests.

\begin{proposition}
  \label{SecBlTstProp}
Suppose $d(\vtx zx)$ executes the blossom enlarge test.
  
\i Suppose \vtx zx is not in a blossom. $d(\vtx zx)$ satisfies
the blossom enlarge test \xiff the pop triggers a skew blossom step.

\ii \vtx zx is in a blossom \xiff some current blossom $B$
has $zx=\eta(B)$ or arc $xz$ in the blossom path of $B$.

\iffalse
\ii Suppose $zx$ is the base edge of a blossom.
$d(\vtx zx)$ satisfies
the blossom enlarge test \xiff  %$e_1(x)$ has been popped and
$BL(x)\ne \emptyset$.

\iii 
In the second blossom test, the possibility 
\vtx zx is in a blossom holds \iff
there is some currently existing blossom $B$ where
$\vtx zx =\eta(B)$ or arc $zx$ or
$xz$ is on the blossom path $P$ of $B$.
\fi
\end{proposition}

\begin{proof} \i is clear. For \ii consider two cases:
  \iffalse
\xi, \xii, and the
\xrimp direction of \iii are obvious.
For the
\ximp direction
consider two cases:
\fi

$zx$ a \T.-arc: 
During the execution of $d(\vtx zx)$, the current  blossom containing
\vtx zx,  $B_{\vtx zx}$, has base vertex \vtx zx (Proposition \ref{BlossomSubtreeProp}). So the first
alternative ($\vtx zx =\eta(B)$) holds.

$xz$ a \T.-arc: Since the test
is executed by the invocation $d(\vtx zx)$, $xz$ is in the blossom path of a
current blossom. So the second alternative holds.
\end{proof}

\begin{lemma}
\label{PendantProp}
\i $d(\vtx zx)$ with $zx\ne e_1(x)$ pendant is a noop, i.e., it returns without
executing a grow or blossom step.

\ii After a search that invokes $d(e_1)$, $x$ only occurs as a leaf of \T..
Such occurrences do not enter  blossoms.
\end{lemma}

\begin{proof}
\i $d(\vtx zx)$ goes directly to the blossom tests.  Clearly \vtx zx is
  not in a blossom at this point.  So the blossom enlarge test fails.
  Suppose the blossom base test is satisfied.  Then $BL(x)$ contains
  an edge $yx$ with
\begin{equation}
\label{yxMuTypeEqn}
\mu(yx)=
\o \mu(zx)=
\o \mu(e_1(x))
\end{equation}
\noindent
(Lemma \ref{e1Lemma}\xib. gives the second equality).  $d(\vtx yx)$ has returned.  $d(e_1(x))$
returns before $d(\vtx yx)$ returns ($yx\ne e_1(x)$
by \eqref{yxMuTypeEqn}).  So $e_1(x)$ is in $BL(x)$ during
the execution of $d(\vtx yx)$. Thus $d(\vtx yx)$ satisfies the blossom
base test (using \eqref{yxMuTypeEqn}).  It performs a blossom step.
The occurrence of \vtx yx in a blossom means $d(\vtx zx)$ does not
execute the blossom base test.  Contradiction.

\ii Lemma \ref{e1Lemma}\ib. shows any occurrence of $x$ in a search after
the search invoking $d(e_1)$ is a leaf of \T..
A pendant edge $yx$ can enter a blossom only as the last edge of the blossom path, i.e.,
when it is popped from $BL(x)$. But
part \i shows the corresponding
execution of $d$ is a noop, i.e., it does not trigger a blossom step.
\end{proof}

For intuition note this extension of part \xii: The occurrences of $x$ in part \ii
never enter augmenting trails. In proof 
no grow step is executed from the $x$ occurrence, by part \i and the
nonoccurrence of $x$ in a blossom. So the \T.-path to $x$ is not extended
and does not lead to a free vertex.
(This property is
not required in the logic of our algorithm analysis.)
In particular this extension shows that if a free vertex $\alpha$
has a corresponding invocation $d(\vtx z \alpha)$ that returns (as in the code for $d$)
$\alpha$ will never enter an augmenting trail in the rest of the execution of \fbs.

We now begin to track the status of a fixed vertex $x\in V(G)$
during the execution of \fbs.
The following result applies to searches at the start of \fbs,
specifically, before $e_1(x)$ is discovered.

%It may also apply to a search after the one invoking $d(e_1)$.

\begin{proposition}
  \label{Noe1ExistsProp}
  Consider a search 
where no invocation $d(\vtx zx)$ returns.  %If $e_1(x)$ does not exist then
Any \T.-arc $zx$ of the search is on an augmenting trail.
Furthermore $x$ does not occur in any blossom of the search.
\end{proposition}

\remarks{The search may contain arcs $xz$ directed from $x$.

  The proposition may or may not apply to a search after $e_1(x)$
is defined.}

\begin{proof}
  Consider a \T.-arc $zx$. (Such an arc always exists;
for a free vertex  $\alpha$ it is the artificial arc
  $\varepsilon \alpha$.)
  The corresponding
invocation $d(\vtx zx)$ %does not return do %(definition of $e_1$). So it
was terminated, i.e., $zx$ is on an augmenting trail.

$zx$ is not an arc in a blossom path, again since $d(\vtx zx)$
did not return. $zx$ is not the base edge of a blossom, since the
first blossom containing $\vtx zx$ contains $e_1(x)$.  So as
claimed, $x$ does not occur in a blossom.
\end{proof}

The following property of our edge-contracted blossoms is  needed 
for both Sections \ref{WellDefinedSec}
and \ref{BlockingSec}:

\bigskip

\noindent
$(\dagger)$ \hskip 20pt  \parbox {5.8 in} {At any moment in \fbs\
  a given vertex $x$ occurs in at most one blossom.}

\bigskip

\noindent
Observe that $x$ can enter a blossom only in the search that invokes $d(e_1)$:
Proposition \ref{Noe1ExistsProp} shows this for searches before
the invocation and Lemma \ref{PendantProp}\ii shows it for after.
To establish $(\dagger)$  we must show it holds throughout the
search invoking $d(e_1)$. We have broken this search up into three
time periods,
Stages 1--3 for $x$. The next several lemmas prove $(\dagger)$ in
each of these stages. Specifically
Lemma \ref{Beforee1Lemma} proves $(\dagger)$
at any moment in Stage 1.
Lemma \ref{lScanningLemma}\i does this for Stage 2, and
Lemma \ref{Stage3Lemma} for Stage 3.

\bigskip

\begin{figure}[h]
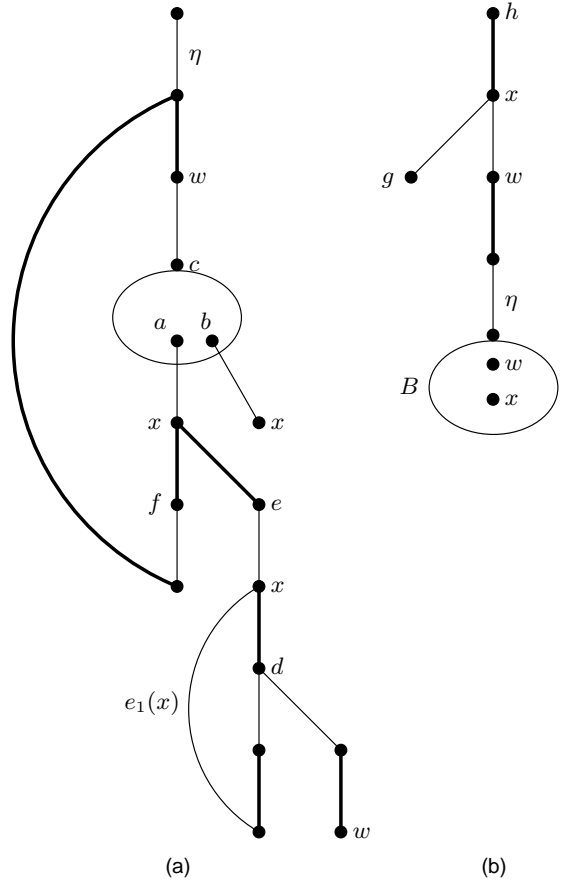

\centering
\input Beforee1.pstex_t
\caption{(a) Stage 1 search:
    $\vtx ax$ is the first occurrence  of $x$ to enter a blossom.
  $\vtx bx$ is a leaf added after $d(\vtx ax)$ returns. 
    $d(\vtx cw)$ forms a blossom containing $\vtx dx$.
  $d(\vtx dx)$ pops $e_1(x)$. Stage 1 ends when that blossom step is complete.
  Only the pendant edge $bx$ remains, it will be processed in Stage 2.
  (b) Stage 3: $B$ is a complete blossom when $d(\eta)$ returns. $B$ is enlarged
  by
  skew blossoms in $d(\vtx xw)$ and $d(\vtx hx)$.}
\iffalse
After that blossom step,
  $d(\vtx dx)$ pops $bx$.  $d(\eta)$ returns.
  The blossom contains every occurrence of $x$ descending from $\eta$.}
\fi
    \label{Beforee1Fig}
\end{figure}

Now assume some invocation $d(\vtx zx)$ in \fbs\ returns, i.e.,
$e_1(x)$ exists.  We account for the time up
to and including the formation of the blossom for the pop of $e_1(x)$.
Recall this time period is called {\em Stage 1} for $x$.
It is possible that the pop of $e_1(x)$ is a noop, but it is still included Stage 1.
\iffalse
until
all blossoms formed up to and including the blossom formed
when 
$e_1$ is popped.

and the resultant blossom is formed.

a blossom containing $e_1$ is formed (if such a blossom
exists). This is Stage 1.
\fi
Fig. \ref{Beforee1Fig}(a) gives an example.
It is also possible that  $e_1(x)$ does not get popped. In that case Stage 1
ends after the last blossom containing $x$ is formed. Again this may be a noop.

\begin{lemma}
  \label{Beforee1Lemma}
$x$ occurs in at most one blossom at any moment in Stage 1.
\end{lemma}

\begin{proof}
  Consider a point in Stage 1 where there is a unique  blossom $B$ containing
  one or more occurrences of $x$. Let $C$ be the next blossom formed in Stage 1.
  If $x$ occurs in $C$, we will show $\eta(C)=\eta(B)$. Clearly this
  implies the lemma holds throughout Stage 1. Note that
  \iffalse
  Let $B$ be the first blossom containing $x$.
  Let $C$ be any Stage 1 blossom formed after $B$ but during Stage 1.
  \fi
  Stage 1 ends when $C$ is formed  by a pop of $e_1$, unless
  $e_1$ is never popped 
  because of an
  augmenting trail.
  
  Observe that $\eta(C)$ descends from $\eta(B)$.
In proof Proposition \ref{CompleteBlossomProp}
shows
that when
$d(\eta(B))$ returns,
$e_1$ has been popped and so Stage 1 has ended.
Thus
$C$ is formed before $d(\eta(B))$ returns. This implies
$\eta(C)$ descends from $\eta(B)$.

\iffalse
  In proof suppose not, i.e.,
  \iffalse
  blossom $C$ formed after $B$ whose base edge
\fi
  $\eta(C)$ is a proper ancestor of
$\eta(B)$.  This implies $C$ is formed after $d(\eta(B))$
returns.  Proposition \ref{CompleteBlossomProp}
implies this is after $e_1$ was popped. So no such $C$ occurs in Stage 1.
\fi

\iffalse
some
invocation $d(\vtx
wx)$ returns before $d(\eta(B))$, after popping
$e_1(x)$.  So $C$ is formed after Stage 1.
\fi

$C$ is irrelevant unless it contains an occurrence of $x$.
If $C$ does not contain a new occurrence of $x$ then $\eta(C)=\eta(B)$ as desired.
Suppose $C$ has a new occurrence
  of $x$, say on edge \ed yx.
  
\ed yx is not a pendant edge $yx$.
In proof, \ed yx is in the blossom path of $C$.
$yx$ pendant must be the last edge of this blossom path.
\iffalse   
  That occurrence $x$ is 
%  Observe that an occurrence of $x$ that enters an early blossom is
  an
ancestor of $e_1$. In proof this holds by Lemma \ref{e1Lemma}\i if $x$
is not a leaf of \T..  (Note $x$ may be the end of arc $e_1$.)  If $x$
is a leaf of \T. then $x$ is the end of the last arc of the blossom
path of $C$. Call that arc $yx$.
\fi
So the blossom is triggered by the pop of $yx$. $yx$ is popped in an invocation $d(\vtx
zx)$. This pop was preceded by the pop of $e_1(x)$. But that ended Stage 1.

We conclude \ed yx is 
%  Observe that an occurrence of $x$ that enters an early blossom is
  an
  ancestor of $e_1$, by Lemma \ref{e1Lemma}\xi.
  %Lemma \ref{e0Lemma}  shows we can assume \ed yx is the arc $yx$.
  Let $P_1$ be the \T.-path from $B$ to $e_1$.
We can assume \ed yx is the arc $yx$ on $P_1$.
%Let $C$ be an early blossom formed after $B$.
$\eta(C)$ is an ancestor of $e_1$ (since $C$ contains $x$, an ancestor of $e_1$).
%by the above observation).
$\eta(C)$ is not an edge of $P_1$, since
  every invocation for an arc on $P_1$ returns before $B$ is formed.
  Thus $\eta(C)=\eta(B)$, as desired.
\end{proof}

Note that if $e_1$ exists but does not get popped, any search after
$d(e_1)$ is invoked can add leaf occurrences of $x$ but no blossom
containing $x$ (Lemma \ref{PendantProp}\xii).

Next we analyze the pop of $e_1$. In particular
part \ii of the next lemma shows that the pop of $e_1$ satisfies
$(*)$. Part \ii 
will also be used to show the Stage 2 blossoms satisfy $(*)$.

\begin{lemma}
  \label{e1PopLemma}
Suppose  $d(\vtx zx)$ pops $e_1(x)$.

\i $\mu(zx)=\o\mu(e_1)$.
So $GL(x)=\emptyset$ 
when $e_1$ is popped, and
no edge of $\delta(x)\cup \gamma(x)$ will be  added to \T.
after that.

\ii 
Let $\eta$ be the base edge of the first blossom containing an occurrence of $x$.
  At every moment until $d(\eta)$ returns, every edge of $BL(x)$ descends from
  $\eta$.
So every edge popped from $BL(x)$ during the execution of $d(\vtx zx)$ descends from $\eta$.
\end{lemma}

\remark{Fig.\ref{Beforee1Fig}(a) gives an example of \xii. For instance
$BL(x)$ is the list $(e_1, ex, ax, bx)$
when the first blossom is
formed. In general $BL(x)$ may contain arcs directed to or from
$x$, e.g., after $d(\vtx dx)$ returns $dx$ enters $BL(x)$.}

\begin{proof}
  \i The second assertion follows immediately from the first using
  Lemma \ref{e1Lemma}\xii. We turn to the first assertion.

  Suppose $e_1$ is popped in the blossom base test.
(In particular this implies $zx$ is a \T.-arc.) The blossom step
makes $zx$ the base edge of a blossom,
$zx=\eta(B)$.
The test implies $\mu(e_1)=\o\mu(zx)$, as desired.

Now suppose  $zx$ satisfies the blossom enlarge test.
(Note that $d(\vtx zx)$ may be invoked strictly after
moment $e_1$ enters a blossom. In that case the pop of $e_1$ is a noop.)

\bclaim {$x$ does not occur in a complete blossom.}

\noindent
    {\bf Proof of Claim:}
Assume it does occur.
Proposition \ref{CompleteBlossomProp}
implies $d(\vtx wx)$ has been invoked for edges $wx$ with opposite M-types.
Choose an invocation $d(\vtx wx)$ that has $\mu(wx)=\o \mu(e_1)$.
$ e_1$ is popped before $d(\vtx wx)$ returns, contradicting completeness.
    \hfill$\spadesuit$

    \bigskip

    \vtx zx must be in a blossom $B$.
If not, Proposition \ref{SecBlTstProp}\i implies
\vtx zx is the base vertex of a skew blossom.
But then the Claim gives a contradiction.

  If $zx=\eta(B)$ the blossom base test has been executed and as shown above
  \i holds. So
  Proposition \ref{SecBlTstProp}\ii shows
$xz$ is an arc in the blossom path  $P$.
  The node $\vtx zx$ is not in a blossom
  on $P$, again by the Claim.
  The algorithm statement shows $xz$ is not the first
  arc of $P$. So $P$ contains an arc $ax$ preceding $xz$.
$ax$ is not a base edge so $\mu(ax)=\mu(e_1)$.
The alternation at $x$ implies $\mu(zx)=\o\mu(ax)=\o\mu(e_1)$.

\bigskip

%\ii Let $B$ be the blossom containing the first occurrence of $x$.
%Note that Lemma \ref{Beforee1Lemma} shows $\eta$ is the base of the blossom containing
%\vtx zx when $d(\vtx zx)$ is invoked.}

\ii We first show $e_1$ descends from $\eta$, i.e., $(*)$ holds for the pop
of $e_1$.
Consider two cases.
If $zx =\eta$ then Lemma \ref{e1Lemma}\i shows $e_1(x)$ descends from $zx$.
Suppose $zx \ne \eta$. So $x$ occurs on the path of a blossom $B$ being
formed, say on consecutive arcs $ax,xb$, and $\eta=\eta(B)$.
(Here we use $(*)$ for $B$.) The invocation $d(ax)$ occurs before $B$ is formed
and pops $e_1$, (So $ax=zx$.)  Lemma \ref{e1Lemma}\i shows $ax$ is an
ancestor of $e_1$, and $\eta$ is an ancestor of $ax$. Thus $\eta$ is an ancestor
of $e_1$ and $(*)$ holds.

We continue with \xii.
Since $e_1$ descends from $\eta$, $BL(x)$ is empty when $d(\eta)$ is invoked.
From that moment on until $d(\eta)$ returns, every edge added to $BL(x)$ descends from
$\eta$.

The second assertion of \ii follows siimilarly:
Lemma \ref{Beforee1Lemma} shows that at the moment
$d(\vtx zx)$ forms
the blossom for $e_1$,
$\eta$ is the base of the blossom containing $\vtx zx$.
\end{proof}

 Note from part \i that if $e_1$ is popped, no search after
this pop contains an occurrence of $x$.

Before proceeding it is worthwhile to give an incorrect proof
for the first part of \xii, i.e.,
that $(*)$ holds for the pop of $e_1$.

\bigskip
\noindent
    {\bf Incorrect Proof:}
Lemma \ref{Beforee1Lemma} shows that at the moment the blossom for $e_1$ is formed,
$\eta$ is the base of the unique blossom containing occurrences of $x$.
This makes $\eta$ an ancestor of $e_1$
since a blossom is a subtree of \T. (Proposition \ref{BlossomSubtreeProp}).

\bigskip

The argument fails since the use of Proposition \ref{BlossomSubtreeProp}
assumes the blossom has already been properly formed, i.e., $(*)$ held
for the blossom popping $e_1$.

\bigskip

We now analyze the time from when $d(\vtx zx)$ pops $e_1(x)$ until it
returns. This is Stage 2.

\begin{figure}[h]
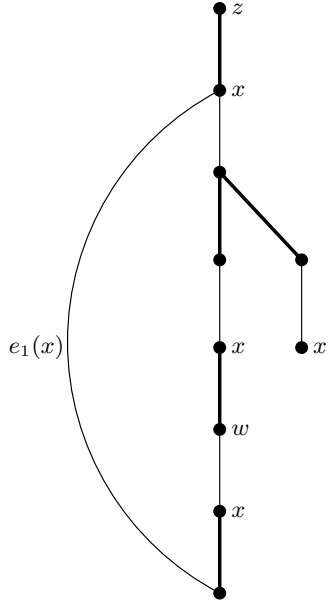

\centering
\input oMuScan.pstex_t
\caption{Preemptive invocation:
  $d(\vtx zx)$ pops $e_1(x)$ but a different invocation,
  $d(\vtx wx)$ for the matched arc $xw$, executes the blossom step for the  leaf occurrence
  of $x$.} 
\label{oMuScanFig}
\end{figure}

\iffalse
OLD STUFF

We first show $(*)$ for $yx=e_1$, specifically we show $\vtx zx$ is a
${\T.}$-ancestor of $e_1$. First suppose $zx$ is a \T.-arc. $zx$ is a \T.-ancestor of
$e_1$ (Lemma \ref{e1Lemma}\i with Proposition \ref{BlossomSubtreeProp}). Also note
this case includes the trivial possibility that \vtx zx is the free vertex $\alpha$,
where $zx$ is the artificial edge $\varepsilon \alpha$.) Next suppose $xz$ is a
\T.-arc. (See Fig.\ref{PathScanFig} for an example of the various possibilities.) Let
$ax$ be the parent arc of \vtx zx. (Note $ax$ need not be on the current blossom path
$P$, e.g., Fig.\ref{uuBlossomFig}.) As in the previous case $ax$ is a \T.-ancestor of
$e_1$. So $\vtx zx$ is a \T.-ancestor of $e_1$.
\fi

\begin{lemma}
\label{lScanningLemma}
Suppose  $d(\vtx zx)$ pops $e_1(x)$.
In parts \i and \ii $yx$ is an arbitrary 
edge  popped from $BL(x)$
during the execution of $d(\vtx zx)$ (i.e., from the moment
$d(\vtx zx)$ is invoked and until it returns).

\i The pop of $yx$ makes it an edge of $B_{\vtx zx}$.
So  $B_{\vtx zx}$ remains the only blossom containing an occurrence of $x$.

\ii The pop of $yx$ satisfies $(*)$.

\iii Suppose $d(\vtx zx)$ returns.  Every pendant edge $yx$ ever
created in the execution of \fbs\ was popped from $BL(x)$ during the
execution of $d(\vtx zx)$.  So at the moment $d(\vtx zx)$ returns,
every occurrence of $x$ in \T. %$\o {\T.}$
%is a proper ancestor of $B_{\vtx zx}$.
is either in $B_{\vtx zx}$ or is a proper ancestor of $\beta(B_{\vtx zx})$
not in any blossom.
\end{lemma}

\remarks{The arc corresponding to $d(\vtx zx)$ can be $zx$ (blossom base test)
  or $xz$ (blossom enlarge test).

  Stage 2 can have blossom steps for pendant edges as well as noops.
  For example in Fig.\ref{Beforee1Fig}(a), where we have noted 
  $BL(x)$ as the list $(e_1, ex, ax, bx)$, $ex$ and $ax$ trigger
  noops and $bx $ is pendant. In general noops are for ancestors of $e_1(x)$.

  The edge $zx$ need
 not be in the first blossom containing an occurrence of $x$. For instance
in Fig.\ref{Beforee1Fig}(a)  $d(\vtx dx)$ pops $e_1(x)$ but
\vtx ax is first blossom occurrence of $x$.

  In \xi, Fig.\ref{oMuScanFig} shows
 a leaf $x$ can be
 popped in an invocation other than $d(\vtx zx)$.}

\begin{proof}
\i Lemma \ref{Beforee1Lemma}
 shows \i holds for $yx=e_1$, the first pop of $BL(x)$.

 Inductively assume \i has always held up to  some point in the execution of $d(\vtx zx)$.
As previously observed (proof of Lemma \ref{PendantProp}\xii)
a pendant edge $yx$ can enter a blossom
only
when it is popped from $BL(x)$.
Let $yx$ be 
the next edge $yx$ to be popped from $BL(x)$.
$yx$ may be popped by an invocation $d(\vtx wx)$, $wx \ne zx$.
(The arc for $d(\vtx wx)$ may be $wx$ or $xw$, Fig.\ref{oMuScanFig}(a)
shows the latter.) $wx$ is not
pendant (Lemma \ref{PendantProp}\xi).  So $\vtx wx$ is a \T.-ancestor of $e_1$
(Lemma \ref{e1Lemma}\xi; $wx$ may be $e_1$).
$d(\vtx wx)$ is on the call chain starting from $d(\vtx zx)$. Hence
$wx$ was in $B_{\vtx zx}$ after the pop of $e_1$ (and possibly before that).
So the pop of $yx$ preserves \xi.
%Thus $B_{\vtx wx}=B_{\vtx zx}$.  So \i holds.
This completes the induction. 

\bigskip

\ii Although blossom steps during the execution of $d(\vtx zx)$ may
enlarge $B_{\vtx zx}$ without adding new occurrences of $x$,
$\eta(B_{\vtx zx})$ remains $\eta$.  So Lemma \ref{e1PopLemma}\ii
shows every edge popped during the execution of $d(\vtx zx)$
descends   from $\eta$.  This gives \xii.

\bigskip

\iffalse
Since Lemma \ref{Beforee1Lemma} applies to $B$, $B$
has the base edge $\eta$ of Lemma \ref{e1PopLemma}\xii.
 Lemma \ref{e1PopLemma}\ii
shows
since $yx$ descends from $\eta$. This gives \xii.

Lemma \ref{PendantProp}\i shows \vtx zx is not a leaf. So Lemma
\ref{e1Lemma}\i shows \vtx zx is a \T.-ancestor of $e_1$. 
This gives $(*)$ for the  pop of $e_1$.
\fi

\iffalse
$yx$ was
added to \T. after $e_1$, so it was added after $d(\vtx ab)$
was invoked.
$d(\vtx yx)$ returned before $d(\vtx zx)$ returned, so it returned  before $d(\vtx ab)$ returned.
Thus $yx$  descends from $ab$, so $(*)$ holds.
===============================================
Also this step ensures that $B_{\vtx zx}$ contains every
nonleaf occurrence of $x$ that descends from its base. In proof, Lemma
\ref{e1Lemma}\i shows that every nonleaf occurrence in the current search
is on the \T.-path to $e_1$.
===============================================================
\iv 
Observe that no edge of $\delta(x) \cup \gamma(x)$ is added to
\T.  after $d(\vtx zx)$ pops $e_1$.  This follows immediately from \iii
and Lemma \ref{e1Lemma}\xii. \iv follows.
\fi

\iii Lemma \ref{e1PopLemma}\i shows
every pendant edge $yx$ is created before $e_1$ is popped.
$yx$ is added to $BL(x)$ when $d(\vtx yx)$ returns.
So $yx$ is in $BL(x)$ before $e_1$ is popped.

$d(\vtx zx)$ returns with $BL(x)$ empty. Thus every pendant
edge gets popped from  $BL(x)$ during the execution of
$d(\vtx zx)$.
With \i this implies
$d(\vtx zx)$ returns with $B_{\vtx zx}$ containing every leaf occurrence
of $x$ in \T.. \i also shows $B_{\vtx zx}$ is the only blossom
in which $x$ occurs.
\end{proof}

%GOOD STUFF

Continuing we will  track the execution of \fbs\ for vertex $x$ after
$d(\vtx zx)$ of Lemma \ref{lScanningLemma} returns.  The return initiates  Stage 3.
First consider Stage 3 for the example of 
Fig.\ref{Beforee1Fig}: $d(\vtx dx)$ (which popped $e_1$)  returns with
$BL(x)=(dx)$. $d(\vtx fx)$ is on the search path to $d(\vtx dx)$,
so eventually it pops $dx$, executes a noop blossom for $dx$, and
returns with $BL(x)=(fx)$.
In Fig.\ref{Beforee1Fig}(b) edge $cw$ triggers the skew blossom
for $w$ and $fx$ triggers the skew blossom for $x$.

For the general case we continue with previous notation:
$d(\vtx zx)$ pops $e_1(x)$. %$B$ is the blossom formed by the pop.
$\eta$ is the base edge of the blossom formed by that pop.

\begin{lemma}
  \label{Stage3Lemma}
  After $d(\vtx zx)$ returns,
  the blossom steps triggered by pops of $BL(x)$ are as follows:

  \pa Until %$d(\eta(B))$
  $d(\eta)$  returns each blossom is a noop.

\pb If $\eta$ is directed to $x$ the blossom is also a noop.
  
  \pc After %$d(\eta(B))$
  $d(\eta)$ returns each blossom is a skew blossom.
  $(*)$ holds for this blossom.

  \noindent At all times $x$ occurs in a unique blossom.
\end{lemma}

\remark{A blossom of \pc may be incomplete, as in
  Fig.\ref{dExamplesFig}(f) if the parent of $b'$ is changed to
  $z_3$.}

%  Fig.\ref{PathScanFig} when $d(\vtx rs)$ pops $e_1(s)$. ?????????}

\begin{proof}
  We will prove the following invariant holds after $d(\vtx zx)$ returns
  and until the current search ends.
  Let $B$ be the current blossom $B_{\vtx zx}$.
  Recall $B$ is a subtree of \T. rooted at $\beta(B)$.
  
  \bigskip
  
  {\parindent=0pt

    (I2) \hskip 20pt
Any occurrence of $x$ either
  belongs to $B$ or is a proper \T.-ancestor
  of $\beta(B)$, not in any blossom.
 $BL(x)$ is a singleton whose edge is in $B\cup \eta(B)$.

}

\bigskip
\noindent
Note that (I2) implies the last assertion of the lemma, property $(\dagger)$.

The invariant holds at the moment $d(\vtx zx)$ returns. In proof,
Lemma \ref{lScanningLemma}\iii gives the condition on occurrences of $x$.
Also
$BL(x)=(zx)$, since the return adds $zx$ to the previously emptied
$BL(x)$, and $zx\in  B\cup \eta(B)$.
%Let $\eta$ be the base edge of $B_{\vtx zx}$ at this moment.

From this moment until $d(\eta)$ returns, noops may change
the entry in $BL(x)$ to some other occurrence of $x$ in $B\cup
\eta(B)$. But (I2) is maintained.
Now consider the return of $d(\eta)$.
If $\eta$ is an arc directed to  $x$, i.e., $x=\beta(B_{\vtx zx})$,
then the blossom is a noop, $BL(x)$ changes to $(\eta)$, and  (I2) is preserved.

Now assume (I2) holds after $d(\eta)$ has returned,
and a blossom step creates the next blossom
$C$. Let $b=\beta(C)$. Let $a$ be the first proper ancestor of
$\beta(B)$ that is an occurrence of $x$, if such exists.

$BL(x)$ cannot change before control returns to
$d(a)$. Suppose $C$ is formed before that.  There are two
possibilities. If $b$ is not an ancestor of $\beta(B)$, then no
occurrence of $x$ enters $C$ (we use (I1) here).  So (I2) is
preserved.  The other possibility is that $b$ properly descends from
$a$.  $C$ is either disjoint from $B$ or contains $B$. Again no new
occurrence of $x$ enters a blossom and (I2) is preserved.

Now assume control returns to $d(a)$.
The blossom enlarge test is satisfied. The entry $yx$ 
in $BL(x)$ is an edge in $B\cup \eta(B)$.
So $C$ is a 
skew blossom with base vertex $a$. After this blossom step
$C$ may get enlarged in $d(a)$ but no occurrence of $x$ is added
(again by (I1)). When $d(a)$ returns $\eta(C)$ (the arc directed to $a$)
is added to $BL(x)$.
So (I2) is preserved.

To show $(*)$ for the skew blossom, observe $a$ is %$\vtx ax$ is
a \T.-ancestor of $\vtx yx$. %So $\vtx ax$ is an ancestor of \vtx bx in the
So $a$ is an ancestor of \vtx yx in the
current contraction $\o {\T.}$.
\end{proof}

Lemma \ref{Stage3Lemma} completes the proof of $(\dagger)$.
As mentioned Lemma \ref{e1PopLemma}\i
shows $x$ does not occur in future searches if Stage 3 has been entered.

Note we have also verified that $(*)$ always holds:
Lemmas \ref{e1PopLemma}\xii,
\ref{lScanningLemma}\xii, and
\ref{Stage3Lemma}(c) establish $(*)$ for Stages 1,2, and 3 respectively.

We can now establish the validity of the search structure \T. and $\o
{\T.}$.  It simply amounts to the validity of the algorithm's
blossoms. 

\begin{lemma}
  \label{DefinitionOfBlossomSatisfiedLemma}
Any blossom formed in the algorithm is
valid, i.e., it satisfies the above
{\em Definition of Blossom, in $G$ and \T.}.
\end{lemma}

\begin{proof}
As noted in the Definition of Blossom in $G$ and \T.,
the requisite properties of the blossom trail hold automatically.
So we need only verify the properties of the Starter.
  The requirements are satisfied trivially but for completeness
  we step through them.
  
The blossom base test forms ordinary blossoms with singleton
starters $\vtx zx =\beta$.  The blossom path correctly begins
with $B_{\vtx zx}=\vtx zx=\beta$ and ends at $B_{\vtx yx}=\vtx
yx=\beta$ (recall $yx=e_1(x)$). $zx$ is the base edge and the
blossom test ensures the first and last edges of the blossom
trail have the same M-type, specificially $\o\mu(zx)$.

Now suppose \vtx zx is in a blossom $B_{\vtx zx}\ne \{\vtx zx\}$.
The blossom step is triggered by the blossom enlarge test.
$B_{\vtx zx}$ is the starter blossom $A$.  As specified in the
code, $P$ starts at a node in $B_{\vtx zx}=A$, called vertex $a$
in the Definition.  $(*)$ implies the blossom path $P$ descends
from $A$.
It ends at $B_{\vtx yx}$.
Consider two possibilities:

\bigskip

\noindent
$B_{\vtx yx}\ne \{\vtx yx\}$: This implies $B_{\vtx
yx}=A$, since $A$ is the only blossom containing an occurrence of
$x$.  So $P$ is empty and the blossom step is a noop.

\noindent
$B_{\vtx yx}=\{\vtx
yx\}$: If $yx$ is the loop $xx$ and $xx$ is already a blossom,
we again have a noop. Otherwise,
identifying \vtx yx with \vtx zx gives a valid blossom.

\bigskip

In the blossom enlarge test, a skew blossom corresponds to the
alternative that \vtx zx is a singleton but some descendant is in a
blossom.
\end{proof}

\section{Blocking set}
\label{BlockingSec}
This section proves \fbs\ returns a valid blocking set. This culminates in the main Theorem
\ref{MainTheorem}, which is followed by a brief discussion of the linear time bound.

\subsection{Search structure}
We first give several additional properties
of the search structure that hold when \fbs\ halts.
They are needed to establish the blocking property.

We start by fleshing out 
Proposition \ref{CompleteBlossomProp}. Again
consider a moment in time when a blossom $B$ becomes complete.
Note that every $x$ occurring in $B$ is in stage 3.
Let $\beta$ and $\eta$ be the base vertex and edge of $B$, respectively.
Let $e=\ed xy$ be any edge incident to $B$ in $G$,
say $x\in V(B) \not\ni y$.

\begin{corollary}
\label{CompleteBlossomCor}
At the moment $B$ becomes complete,
%\i $e$ is in \T. as an arc $zx$ or $xz$, and Either
either

\i $e$ is a proper $\o{\T.}$-ancestor of $\eta$, and $e$ is in the
search path $SP$ of $d(\eta)$,
or

\ii $e=yx=\eta$, or

\iii $e=xy$ and leaves either $B$ or $SP$ in $\o{\T.}$.
\end{corollary}

% NO: EXPAND THIS TO SHOW BLOSSOMS ARE COMPLETE AT THE END OF A SEARCH

\remark{Part \iii is illustrated by arc $xg$ in Fig.\ref{Beforee1Fig}(b).}

\begin{proof}
$e$ is an arc of \T. (in one direction or the other)
by Lemma \ref{e1PopLemma}\xi.
Lemma \ref{lScanningLemma}\iii shows the $x$-end of $e$
is either in $B$ or is a proper ancestor of $\beta$
not in any blossom.

\iffalse
 Proposition \ref{CompleteBlossomProp}
ensures $GL(x)$ is empty, so a grow step has been executed for $e$.

\case{$x$ occurs as   $\beta(B)$}
  Let $yx$ be the edge of $BL(x)$ that triggers blossom $B$.
  The invocations $d(\vtx yx)$ and $d(\eta(B))$ are as desired.

\case{$x$ does not occur as    $\beta(B)$}
%(Similar to Lemma \ref{e0Lemma}.)
The blossom path $P$ for $B$ contains an arc $ax$, since \vtx ax is
not in a blossom before $B$. $ax$ is followed by an arc $xb$ in $P$,
since \vtx ax is not in a blossom before $B$ and $\vtx ax
\ne\beta(B)$.  The invocations $d(\vtx ax)$ and $d(\vtx bx)$ are as
desired.

\bigskip
\fi

\case{$\vtx yx$ is a node of $B$}  The nodes of $B$ form a subtree
of \T.  (Proposition \ref{BlossomSubtreeProp}).  So if $e$ is the arc
$yx$ then $e=\eta$ and \ii holds.  If $e$ is arc $xy$ then it leaves the subtree of $B$ and \iii holds.
%Both possibilities are covered in the lemma.

\case{$\vtx yx$ is a proper \T.-ancestor of $\beta$}
If $e$
is arc $yx$ then the invocation $d(\vtx yx)$ precedes
$d(\eta)$ in the call chain. Thus $e$ is in the current search
path $SP$. $e$ is not in a contracted blossom since
$B$ is the only blossom containing $x$. So \i holds.

Suppose $e=xy$.  Since $\vtx yx$ is not in a blossom $xy$
is in $\o{\T.}$. It
is either in the $\o{\T.}$-path to $\eta$ or it is incident to that path.  In
the first case $e$ is in $SP$ as before.  In the second case $e$ is
incident to $SP$, giving \xiii.
\end{proof}

Now assume $x$ occurs in blossoms that are both complete and incomplete.
Let $B_0$ denote the maximal
complete blossom containing  an occurrence of $x$.
Let the sequence of incomplete blossoms containing $x$ be
$B_1\pcon B_2 \pcon \ldots \pcon B_k$, $B_k$, $k>0$.
$\eta(B_0) $ and $\eta(B_1)$ may be identical or different.
However the base edges $\eta(B_i) $, $i\ge 1$ are identical.  
This follows since when $d(\eta(B_1))$ is terminated every
invocation of $d$ in the call chain is terminated, i.e., no new blossoms
are formed.

\begin{figure}[h]
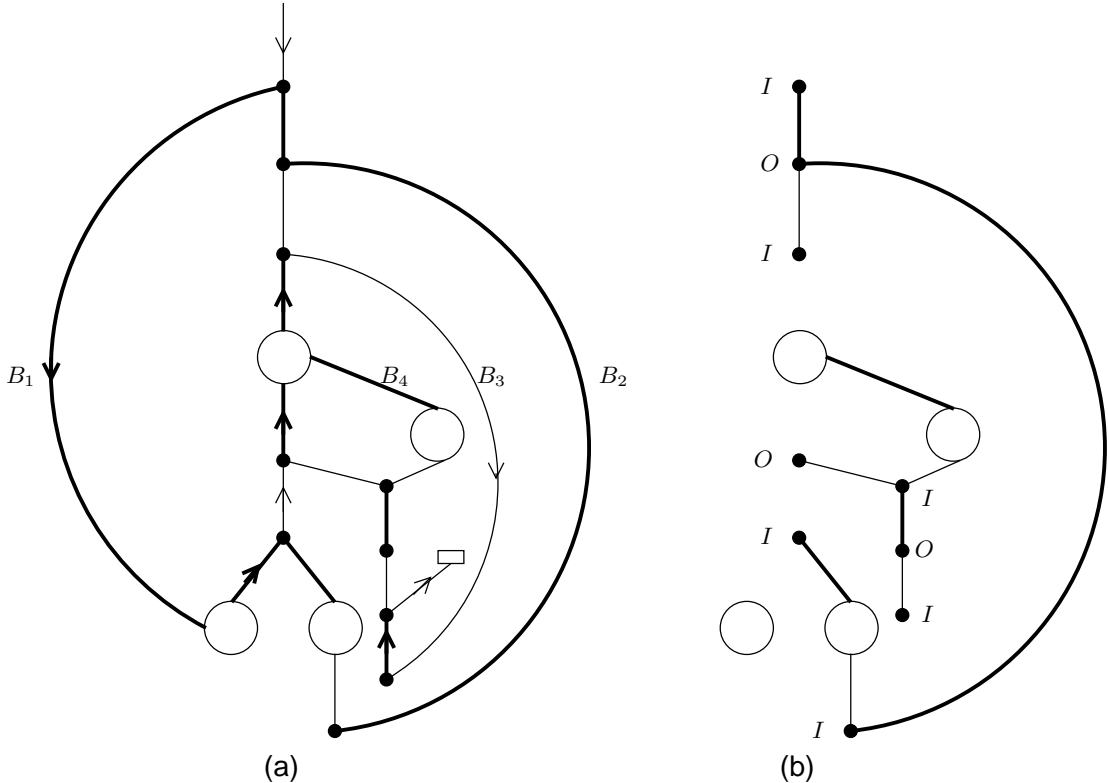

\centering
\input BigIncomplete.pstex_t
\caption{(a) Nested  incomplete blossoms $B_1$--$B_4$.
  The augmenting trail is disjoint from the blossom paths of
  $B_2$ and $B_4$. (b) Residual edges of $B_4$, with vertex labels
  $I,O$ from Section \ref{BlockingSec}.}
\label{BigIncompleteFig}
\end{figure}

The above sequence has some
simple special cases, for an arbitrary vertex $x$:
$B_0$ may exist with $k=0$. If $B_0$ does not exist
define $B_0=\{x\}$. 
$k$ may or may not be 0.

Let \C. be the set of maximal complete blossoms when \fbs\ halts.

\begin{lemma}
  \label{TBarResidualLemma}
  A \T.-arc $zx$ not in any  trail of \A. or any
  set $E(B)\cup \eta(B)$, $B\in \C.$ has M-type $\mu(e_1(x))$.
% An arc $xy$ of $G-\A.-E(C\cup \eta(C))$ has M-type $\mu(e_1(y))$.
\end{lemma}

% HAL: DO WE NEED TO CONSIDER ARCS W PARENT X?

\begin{proof}
  First observe that $e_1(x)$ exists, since invocation $d(\vtx zx)$ returns
  (i.e., it is not terminated).
  Suppose for the sake of contradiction that $zx$ has M-type  $\o \mu(e_1(x))$.

  \bigskip

  \noindent
  {\bf Claim 1} {\it $zx$ is not the base edge of any blossom formed in \fbs.}

  \bigskip
  
  \noindent {\bf Proof of Claim 1:}
  Suppose on the contrary that 
  $zx$ is the base edge of a blossom $B$.

  Suppose  $B$ is incomplete.
By definition the search's augmenting trail
includes $\eta(B)$.
  But this contradicts the lemma's hypothesis on \A..

  Suppose  $B$ is complete. Let $C$ be the maximal complete blossom
  containing $B$. Either $zx=\eta(C)$ or $zx$ is on a blossom
  path contained in $C$ or a subblossom of $C$. Both alternatives
  contradict the lemma's hypothesis on \C..
  \hfill$\spadesuit$

\iffalse
Assume the contrary. Let $B$ be  the maximal blossom with base edge $zx$.

Suppose  $B$ is incomplete.
By definition the search's augmenting trail
includes $\eta(B)$.
  But this contradicts the lemma's hypothesis on \A..
So $B$ is complete.

$B$ is not maximal complete (lemma's hypothesis).
  So $B$ is contained in a larger blossom $C$. Choose $C$ as the minimal
  such blossom.

  $C$ is not incomplete, since that makes $B$ maximal complete. So $C$ is complete.
  
$zx\ne \eta(C)$, by definition of $B$. So $zx \in E(C)$.
  Thus $zx\in E(B)$ for some $B\in \C.$, again contradicting the lemma's  hypothesis.
  \hfill$\spadesuit$
  \fi

\bigskip

\noindent
{\bf Claim 2} {\it $x$ occurs in some blossom when $d(\vtx zx)$
  executes the blossom base test.}

  \bigskip
  
  \noindent
      {\bf Proof of Claim 2:}
$zx\ne e_1(x)$ since the two edges have opposite M-type.
  Lemma \ref{e1Lemma}\i shows $zx$ is a proper ancestor of $e_1$.
  Thus
$e_1$ enters $BL(x)$ during the execution of
$d(\vtx zx)$. In fact it enters before
$d(\vtx zx)$ executes the blossom test.
(It enters during the execution of a grow step in $d(\vtx zx)$.)

  Suppose Claim 2 fails. Then $x$ satisfies the first condition of
  the blossom base test.
  \iffalse
  does not occur in a blossom when 
  $d(\vtx zx)$ executes  
 At that moment 
\fi
 $e_1$ is still in $BL(x)$. (If not it was popped, placing
   $e_1$ in a blossom. This contradicts the test's first condition.)
 The second condition of the base blossom test is satisfied (since
  $\mu(zx)=\o\mu(e_1)$). So $e_1$ is popped and $zx$ becomes the base edge
  of
  a new blossom. This contradicts  Claim 1.
  \hfill $\spadesuit$
  
  \bigskip

  Using Claim 2, let $wx$ be an arc (directed to $x$)
  with \vtx wx in a blossom  when $d(\vtx zx)$ executes the blossom base
  test. 
  $wx$ and $zx$ are both ancestors of $e_1$, so one is an ancestor of the other.
For the moment assume $zx\ne wx$.
Let $B$ be the first blossom containing vertex \vtx wx.
(It is possible that $B$ is incomplete.)  There are four possibilities for $zx$, but none can hold:

\bigskip

{\parindent=0pt
  
\case{$zx$ an ancestor of $\eta(B)$}
The execution of $d(\vtx zx)$ makes $zx$ the base of a skew blossom,
  contradicting Claim 1.

\case{$zx=\eta(B)$} Contradicts Claim 1 again.

\case{$zx$ is on the blossom path of $B$}
  $d(\vtx zx)$ executes the blossom base test before
$B$ is formed.  This contradicts the definition of $wx$.
The same contradiction occurs if $zx=wx$.
(Note $wx$ is either $\eta(B)$ or it is on the blossom path of $B$.
The latter must hold if $wx=zx$.)

\case{$zx$ descends from $B$} This implies $zx$ descends from $wx$.
Again
$d(\vtx zx)$ executes the blossom base test before $B$ is formed,
contradiction.
}
\end{proof}

\iffalse

  We can now assume $x$ occurs in a blossom and the blossom enlarge test
  is executed. Suppose $\vtx zx$ is in a blossom $B$.
Proposition \ref{SecBlTstProp}\ii shows this means makes $zx=\eta(B)$.
  (The second possibility of Proposition \ref{SecBlTstProp}\ii does not
  apply to arcs $zx$, rather arcs $xz$). So Claim 1 holds.

  Next suppose a descendant \vtx wx is in a blossom.
  Then a skew blossom step is executed.
  $zx$ is the base edge, so again Claim 1 holds.
  
The remaining possibility is that
$x$ occurs in a blossom
but that occurrence, say \vtx wx, does not descend from \vtx zx.
We can assume the occurence is a node \vtx wx, for arc $wx$ directed
to $x$. %So \vtx wx is an ancestor of \vtx zx.
In particular $e_1$ has not been popped.
So \vtx wx is not pendant and must be an ancestor of $e_1$.
Thus \vtx wx is an ancestor of \vtx zx.
If $wx$ is in the blossom path it has returned before $B$ is
formed.  So \vtx zx has returned before $B$  is formed, contradiction.
If \vtx wx is the blossom base then $B$  is formed after
$e_1$ and $zx$ return, contradiction.

%Wlog choose $xy$ so $d(\vtx xy)$ returns as early as possible.
$e_1(x)$ remains in $BL(x)$ until it is popped
by some invocation $d(\vtx wx)$. 

================================
\fi

\iffalse
HAL
IT CANNOT BE POPPED BY A PENDANT EDGE. MAKE SURE
THIS IS PROVED EVEN IN THE FIRST LEMMA!!!!!!
\fi

\begin{lemma}
  \label{FreeBlossomLemma}
  A vertex $x$ that is free at end of a search wherein it occurs in a
  complete blossom $B$ has
  \begin{equation}
    \label{FreeVtxEqn}
  \text{$def(x)=1$ and $x=\alpha$}.
  \end{equation}
  Furthermore choosing $B$
  to be maximal makes    $\eta(B)=\varepsilon \alpha$.
\end{lemma}

\remarks{$x$ will be free when \fbs\ halts.

  The hypothesis of maximality is necessary:
  In 
Fig.\ref{dExamplesFig}\ii assume $b=\alpha$ and $def(\alpha)=1$.
The blossom of the figure is complete but
does not have base vertex $b$. 
But the maximal complete blossom, formed as a skew blossom at $\alpha$,
has base vertex $\alpha$.}

\begin{proof}
Let $B$ be the first blossom formed with an occurrence of $x$.
$x$ occurs on an edge of the blossom path $P$. 

\case {$x$ does not occur as an interior vertex of $P$}
Since $x$ is not in a prior blossom, $x$ is
the base vertex of $B$ and the base edge, say arc $zx$, is unmatched.
$d(\vtx zx)$ 
starts by executing the test for an augment
step. The test fails (since $B$ has not been formed yet).
This implies \eqref{FreeVtxEqn}.

\case{$x$ occurs as an interior vertex of $P$}
$P$ alternates at $x$ so $x$ is on an unmatched edge \ed zx of $P$.
The invocation $d(\vtx zx)$ is made before $B$ becomes complete.
(The invocation may be for arc $zx$ or $xz$.)
As before it  starts by executing the test
for an augment, which fails ($B$ has not been completed).
So again  \eqref{FreeVtxEqn} holds.

If $\eta(B)\ne \varepsilon \alpha$, eventually control returns to the
invocation $d(\varepsilon \alpha)$. It executes a skew blossom step.
This guarantees the second condition of the lemma.
\end{proof}

%%%%%%%%%%%%%%%%%%%%%%%%%%%%%%%%%%%%%%%%

%% file: code.tex
\def\myif #1{{\sf {if}} (#1)}
\def\myelif #1{{\sf else if} (#1)}
\def\myifel #1 #2 #3 {{\sf if} (#1) {{#2}} {\sf else} {{#3}}}
\def\myelse {{\sf else}}

\def\myfor #1{{\sf for} (#1)} 
\def\mywhile #1{{\sf while} (#1)}
\def\myloop {{\sf loop}}

\def\mycom #1{{\tt /* {\hskip-3pt#1\hskip-3pt} */}}

\def\myand {{\sf and }}
\def\myor {{\sf or }}
\def\myto {{\sf to }}
\def\myreturn{{\sf return}}
\def\mybreak{{\sf break}}

\def\sm.{\mathy{\cal S^-}}

    \setlength{\Efigwidth}{\linewidth}
    %\addtolength{\Efigwidth}{-.1in}
\setlength{\vmargin}{15pt}

\begin{figure}[p]

%\begin{center}
  %\fbox{

%USE THE framed ENVIRONMENT FROM THE FRAMED PACKAGE
%IT IS MORE ROBUST THAN frame OR framebox, WHICH DOESNT WORK HERE
\begin{framed} 
   \begin{minipage}[t]{\Efigwidth}

  %\setlength{\parindent}{.02in} %{.2in}
%\narrower{
%\setlength{\parindent}{0pt}
\vspace{\vmargin}
\setlength{\parindent}{50pt}

{\sf procedure} {\fbs}

\bigskip

initialize \T. to an empty forest and \A. to an empty set
%,and all $b$ entries to $\emptyset$

\myfor {$x \in V(G)$}

{\hi

initialize $def(x)$ to the deficiency of $x$ in the current
matching

initialize lists $GL(x)$ to $\delta(x)\cup \gamma(x)$
and $BL(x)$ to $\emptyset$

}

\bigskip

\myfor {$\alpha \in V(G)$}

{\hi

\myif {$def(\alpha)>0$ and no invocation $d(\vtx z\alpha)$ has returned}
%$\alpha$ was not the root of a previous  unsuccessful search}
$d(\vtx \varepsilon \alpha)$ 

\mycom{$\varepsilon$ is an artificial vertex,
$\epsilon \alpha$ a  matched artificial arc} % directed to $\alpha$}

}

%rematch each augmenting trail of \A.

{\sf return} \A.

\bigskip

{\sf procedure} {\em d}(\vtx zx)    %($e$)
%\mycom{$e$ is an edge of $G$ directed to vertex $x$}

%$x \gets$ the head of edge $e$, 

%$e\gets$ arc $zx$, $\mu \gets \mu(e)$
%$\mu \gets \mu(zx)$

\bigskip

\myif {$def(x)>0$ \myand $ \mu(zx)= \o M$ \myand 
$(x\ne \alpha$ \myor  $def(\alpha)\ge 2)$} 

{\hi 

\mycom{augment step}

add the $\o {\T.}$-path  from $\alpha$ to $x$ to \A.

decrement $def(\alpha)$ and $def(x)$ 

terminate every currently executing invocation of  $d$, including this one
}

\bigskip

\mywhile {$\exists \text{ edge } xy\in GL(x) \cap \o \mu(zx) $}

{\hi

remove $xy$ from $GL(x)$ and $GL(y)$

\mycom{grow step} 

add an arc $xy$, from node $\vtx zx$ to a new node $y$, 
to \T. 

$B_{\vtx xy}\gets \{\vtx xy\}$

$d(\vtx xy)$

}
\bigskip

\myloop

{\hi

\mycom{blossom base test}

\myif {no occurrence of $x$ is in a blossom} 

{\hi

\myifel 
{$\exists \text{ edge } xy\in BL(x) \cap \o \mu(zx)$} %
{{$xy \gets {\rm pop}(BL(x))$}} {{\mybreak}}
}

\mycom{blossom enlarge test}

\myelif{\vtx zx or some descendant  \vtx wx is in a blossom} 
%{$\vtx zx$ is an internal node of \T.}
%$\vtx zx$ is in a blossom, i.e., $B_{\vtx zx}\ne \{\vtx zx\}$}

{\hi

\myifel{$\exists \text{ edge } xy\in BL(x)$} % 
{$xy \gets {\rm pop}(BL(x))$} {\mybreak}

}

\mycom{blossom step}

let $P$ be the path in $\o {\T.}$ from 
$B_\vtx z x$ to $B_{\vtx yx}$

\iffalse
\myfor {every node $v\in V(\T.)$ with $B_v\in P$} $b(v)\gets b(x,e)$
add $B_{\vtx v u}$ to $B_\vtx z x$

for every node v on an edge of P

\fi

\myfor {every arc $uv$  of $P$}
merge $B_{\vtx uv}$ into  $B_{\vtx zx}$

\myfor {every arc $uv$  of $P$, as ordered in $P$ but skipping the first arc}
\iffalse
\myfor {every arc $uv$  of $P$, as ordered in $P$, $u\ne \vtx zx$}
\mycom{blossom-invocation loop}
\fi

{\hi

\mycom{blossom-invocation loop}

$d(\vtx v u)$

}

}

\iffalse
\mywhile {$\exists$ a vertex of 
$P- \vtx yx -\vtx zx $ 
that has not been blossom scanned}

{\hi
$u \gets$ the first such vertex in 
$P- \vtx yx -\vtx zx $ 

%\mycom{by definition $x$ has been blossom scanned}

$uv\gets $  the arc leaving $u$ in $P$

$d(\vtx v u)$

\mycom{$u\in V(G)$ is blossom-scanned if $d(\vtx v u)$ returns}

%\vtx vu 

\mycom{(but not if $d(\vtx v u)$  is terminated)}

}

}
\fi

\bigskip

add $zx$ to $BL(x)$

\myreturn
%}

\vspace{\vmargin}

\end{minipage}
\end{framed}

  \caption{Blocking algorithm for $f$-factors.}
\label{FBSFig}

\end{figure}

%% file: dExamples.pstex_t
\begin{picture}(0,0)%
\includegraphics{dExamples.pstex}%
\end{picture}%
\setlength{\unitlength}{1934sp}%
\begingroup\makeatletter\ifx\SetFigFont\undefined%
\gdef\SetFigFont#1#2#3#4#5{%
  \reset@font\fontsize{#1}{#2pt}%
  \fontfamily{#3}\fontseries{#4}\fontshape{#5}%
  \selectfont}%
\fi\endgroup%
\begin{picture}(16844,12113)(279,-14552)
\put(5026,-4261){\makebox(0,0)[lb]{\smash{{\SetFigFont{8}{9.6}{\rmdefault}{\mddefault}{\updefault}{\color[rgb]{0,0,0}$a$}%
}}}}
\put(1051,-8011){\makebox(0,0)[lb]{\smash{{\SetFigFont{8}{9.6}{\sfdefault}{\mddefault}{\updefault}{\color[rgb]{0,0,0}(a)}%
}}}}
\put(4651,-8011){\makebox(0,0)[lb]{\smash{{\SetFigFont{8}{9.6}{\sfdefault}{\mddefault}{\updefault}{\color[rgb]{0,0,0}(b)}%
}}}}
\put(13576,-11836){\makebox(0,0)[lb]{\smash{{\SetFigFont{8}{9.6}{\rmdefault}{\mddefault}{\updefault}{\color[rgb]{0,0,0}$b$}%
}}}}
\put(12151,-11611){\makebox(0,0)[lb]{\smash{{\SetFigFont{8}{9.6}{\rmdefault}{\mddefault}{\updefault}{\color[rgb]{0,0,0}$b'$}%
}}}}
\put(9676,-14461){\makebox(0,0)[lb]{\smash{{\SetFigFont{8}{9.6}{\sfdefault}{\mddefault}{\updefault}{\color[rgb]{0,0,0}(f)}%
}}}}
\put(11101,-7636){\makebox(0,0)[lb]{\smash{{\SetFigFont{8}{9.6}{\sfdefault}{\mddefault}{\updefault}{\color[rgb]{0,0,0}(d.1)}%
}}}}
\put(13276,-8011){\makebox(0,0)[lb]{\smash{{\SetFigFont{8}{9.6}{\sfdefault}{\mddefault}{\updefault}{\color[rgb]{0,0,0}(d)}%
}}}}
\put(16126,-6286){\makebox(0,0)[lb]{\smash{{\SetFigFont{8}{9.6}{\rmdefault}{\mddefault}{\updefault}{\color[rgb]{0,0,0}$v$}%
}}}}
\put(13801,-6286){\makebox(0,0)[lb]{\smash{{\SetFigFont{8}{9.6}{\rmdefault}{\mddefault}{\updefault}{\color[rgb]{0,0,0}$v$}%
}}}}
\put(14851,-7636){\makebox(0,0)[lb]{\smash{{\SetFigFont{8}{9.6}{\sfdefault}{\mddefault}{\updefault}{\color[rgb]{0,0,0}(d.2)}%
}}}}
\put(15001,-6811){\makebox(0,0)[lb]{\smash{{\SetFigFont{8}{9.6}{\rmdefault}{\mddefault}{\updefault}{\color[rgb]{0,0,0}$\cal T$}%
}}}}
\put(14701,-3586){\makebox(0,0)[lb]{\smash{{\SetFigFont{8}{9.6}{\rmdefault}{\mddefault}{\updefault}{\color[rgb]{0,0,0}$v$}%
}}}}
\put(8401,-8011){\makebox(0,0)[lb]{\smash{{\SetFigFont{8}{9.6}{\sfdefault}{\mddefault}{\updefault}{\color[rgb]{0,0,0}(c)}%
}}}}
\put(9076,-4411){\makebox(0,0)[lb]{\smash{{\SetFigFont{8}{9.6}{\rmdefault}{\mddefault}{\updefault}{\color[rgb]{0,0,0}$b'$}%
}}}}
\put(6901,-4411){\makebox(0,0)[lb]{\smash{{\SetFigFont{8}{9.6}{\rmdefault}{\mddefault}{\updefault}{\color[rgb]{0,0,0}$b$}%
}}}}
\put(10801,-3586){\makebox(0,0)[lb]{\smash{{\SetFigFont{8}{9.6}{\rmdefault}{\mddefault}{\updefault}{\color[rgb]{0,0,0}$v$}%
}}}}
\put(12151,-4486){\makebox(0,0)[lb]{\smash{{\SetFigFont{8}{9.6}{\rmdefault}{\mddefault}{\updefault}{\color[rgb]{0,0,0}$v'$}%
}}}}
\put(16051,-4486){\makebox(0,0)[lb]{\smash{{\SetFigFont{8}{9.6}{\rmdefault}{\mddefault}{\updefault}{\color[rgb]{0,0,0}$v'$}%
}}}}
\put(9676,-9061){\makebox(0,0)[lb]{\smash{{\SetFigFont{8}{9.6}{\rmdefault}{\mddefault}{\updefault}{\color[rgb]{0,0,0}$\alpha$}%
}}}}
\put(9601,-10036){\makebox(0,0)[lb]{\smash{{\SetFigFont{8}{9.6}{\rmdefault}{\mddefault}{\updefault}{\color[rgb]{0,0,0}$z_1$}%
}}}}
\put(9676,-10936){\makebox(0,0)[lb]{\smash{{\SetFigFont{8}{9.6}{\rmdefault}{\mddefault}{\updefault}{\color[rgb]{0,0,0}$x$}%
}}}}
\put(7276,-12886){\makebox(0,0)[lb]{\smash{{\SetFigFont{8}{9.6}{\rmdefault}{\mddefault}{\updefault}{\color[rgb]{0,0,0}$z_2$}%
}}}}
\put(9601,-12886){\makebox(0,0)[lb]{\smash{{\SetFigFont{8}{9.6}{\rmdefault}{\mddefault}{\updefault}{\color[rgb]{0,0,0}$z_3$}%
}}}}
\put(11776,-12886){\makebox(0,0)[lb]{\smash{{\SetFigFont{8}{9.6}{\rmdefault}{\mddefault}{\updefault}{\color[rgb]{0,0,0}$z_4$}%
}}}}
\put(4126,-14461){\makebox(0,0)[lb]{\smash{{\SetFigFont{8}{9.6}{\sfdefault}{\mddefault}{\updefault}{\color[rgb]{0,0,0}(e)}%
}}}}
\put(5401,-13636){\makebox(0,0)[lb]{\smash{{\SetFigFont{8}{9.6}{\rmdefault}{\mddefault}{\updefault}{\color[rgb]{0,0,0}$z_3$}%
}}}}
\put(5401,-12736){\makebox(0,0)[lb]{\smash{{\SetFigFont{8}{9.6}{\rmdefault}{\mddefault}{\updefault}{\color[rgb]{0,0,0}$x_3$}%
}}}}
\put(3751,-9061){\makebox(0,0)[lb]{\smash{{\SetFigFont{8}{9.6}{\rmdefault}{\mddefault}{\updefault}{\color[rgb]{0,0,0}$z_1$}%
}}}}
\put(3826,-10036){\makebox(0,0)[lb]{\smash{{\SetFigFont{8}{9.6}{\rmdefault}{\mddefault}{\updefault}{\color[rgb]{0,0,0}$x_1$}%
}}}}
\put(3826,-12736){\makebox(0,0)[lb]{\smash{{\SetFigFont{8}{9.6}{\rmdefault}{\mddefault}{\updefault}{\color[rgb]{0,0,0}$z_2$}%
}}}}
\put(3826,-13636){\makebox(0,0)[lb]{\smash{{\SetFigFont{8}{9.6}{\rmdefault}{\mddefault}{\updefault}{\color[rgb]{0,0,0}$x_2$}%
}}}}
\put(5026,-5911){\makebox(0,0)[lb]{\smash{{\SetFigFont{8}{9.6}{\rmdefault}{\mddefault}{\updefault}{\color[rgb]{0,0,0}$b$}%
}}}}
\put(5026,-2761){\makebox(0,0)[lb]{\smash{{\SetFigFont{8}{9.6}{\rmdefault}{\mddefault}{\updefault}{\color[rgb]{0,0,0}$\alpha$}%
}}}}
\end{picture}%

%% file: bStepCases.pstex_t
\begin{picture}(0,0)%
\includegraphics{bStepCases.pstex}%
\end{picture}%
\setlength{\unitlength}{2131sp}%
\begingroup\makeatletter\ifx\SetFigFont\undefined%
\gdef\SetFigFont#1#2#3#4#5{%
  \reset@font\fontsize{#1}{#2pt}%
  \fontfamily{#3}\fontseries{#4}\fontshape{#5}%
  \selectfont}%
\fi\endgroup%
\begin{picture}(10080,7636)(8836,-8552)
\put(11926,-2536){\makebox(0,0)[lb]{\smash{{\SetFigFont{9}{10.8}{\rmdefault}{\mddefault}{\updefault}{\color[rgb]{0,0,0}$A$}%
}}}}
\put(9301,-1111){\makebox(0,0)[lb]{\smash{{\SetFigFont{9}{10.8}{\rmdefault}{\mddefault}{\updefault}{\color[rgb]{0,0,0}$z$}%
}}}}
\put(17101,-1936){\makebox(0,0)[lb]{\smash{{\SetFigFont{9}{10.8}{\rmdefault}{\mddefault}{\updefault}{\color[rgb]{0,0,0}$x$}%
}}}}
\put(17101,-1111){\makebox(0,0)[lb]{\smash{{\SetFigFont{9}{10.8}{\rmdefault}{\mddefault}{\updefault}{\color[rgb]{0,0,0}$z$}%
}}}}
\put(9226,-6886){\makebox(0,0)[lb]{\smash{{\SetFigFont{9}{10.8}{\rmdefault}{\mddefault}{\updefault}{\color[rgb]{0,0,0}$y$}%
}}}}
\put(13351,-2536){\makebox(0,0)[lb]{\smash{{\SetFigFont{9}{10.8}{\rmdefault}{\mddefault}{\updefault}{\color[rgb]{0,0,0}$z$}%
}}}}
\put(14026,-2536){\makebox(0,0)[lb]{\smash{{\SetFigFont{9}{10.8}{\rmdefault}{\mddefault}{\updefault}{\color[rgb]{0,0,0}$x$}%
}}}}
\put(9601,-8461){\makebox(0,0)[lb]{\smash{{\SetFigFont{9}{10.8}{\sfdefault}{\mddefault}{\updefault}{\color[rgb]{0,0,0}(a)}%
}}}}
\put(12901,-8461){\makebox(0,0)[lb]{\smash{{\SetFigFont{9}{10.8}{\sfdefault}{\mddefault}{\updefault}{\color[rgb]{0,0,0}(b)}%
}}}}
\put(17401,-8461){\makebox(0,0)[lb]{\smash{{\SetFigFont{9}{10.8}{\sfdefault}{\mddefault}{\updefault}{\color[rgb]{0,0,0}(c)}%
}}}}
\put(12526,-6886){\makebox(0,0)[lb]{\smash{{\SetFigFont{9}{10.8}{\rmdefault}{\mddefault}{\updefault}{\color[rgb]{0,0,0}$y$}%
}}}}
\put(12676,-2836){\makebox(0,0)[lb]{\smash{{\SetFigFont{9}{10.8}{\rmdefault}{\mddefault}{\updefault}{\color[rgb]{0,0,0}$a$}%
}}}}
\put(17926,-7486){\makebox(0,0)[lb]{\smash{{\SetFigFont{9}{10.8}{\rmdefault}{\mddefault}{\updefault}{\color[rgb]{0,0,0}$x$}%
}}}}
\put(17176,-7486){\makebox(0,0)[lb]{\smash{{\SetFigFont{9}{10.8}{\rmdefault}{\mddefault}{\updefault}{\color[rgb]{0,0,0}$w$}%
}}}}
\put(15826,-7486){\makebox(0,0)[lb]{\smash{{\SetFigFont{9}{10.8}{\rmdefault}{\mddefault}{\updefault}{\color[rgb]{0,0,0}$A$}%
}}}}
\put(9301,-2836){\makebox(0,0)[lb]{\smash{{\SetFigFont{9}{10.8}{\rmdefault}{\mddefault}{\updefault}{\color[rgb]{0,0,0}$b$}%
}}}}
\put(17101,-2836){\makebox(0,0)[lb]{\smash{{\SetFigFont{9}{10.8}{\rmdefault}{\mddefault}{\updefault}{\color[rgb]{0,0,0}$b$}%
}}}}
\put(12676,-3661){\makebox(0,0)[lb]{\smash{{\SetFigFont{9}{10.8}{\rmdefault}{\mddefault}{\updefault}{\color[rgb]{0,0,0}$b$}%
}}}}
\put(8851,-4936){\makebox(0,0)[lb]{\smash{{\SetFigFont{9}{10.8}{\rmdefault}{\mddefault}{\updefault}{\color[rgb]{0,0,0}$A$}%
}}}}
\put(9301,-1936){\makebox(0,0)[lb]{\smash{{\SetFigFont{9}{10.8}{\rmdefault}{\mddefault}{\updefault}{\color[rgb]{0,0,0}$x$}%
}}}}
\end{picture}%

%% file: PDescend.pstex_t
\begin{picture}(0,0)%
\includegraphics{PDescend.pstex}%
\end{picture}%
\setlength{\unitlength}{2368sp}%
\begingroup\makeatletter\ifx\SetFigFont\undefined%
\gdef\SetFigFont#1#2#3#4#5{%
  \reset@font\fontsize{#1}{#2pt}%
  \fontfamily{#3}\fontseries{#4}\fontshape{#5}%
  \selectfont}%
\fi\endgroup%
\begin{picture}(8281,5835)(2835,-15152)
\put(10426,-12736){\makebox(0,0)[lb]{\smash{{\SetFigFont{10}{12.0}{\familydefault}{\mddefault}{\updefault}{\color[rgb]{0,0,0}$x$}%
}}}}
\put(10651,-11911){\makebox(0,0)[lb]{\smash{{\SetFigFont{10}{12.0}{\familydefault}{\mddefault}{\updefault}{\color[rgb]{0,0,0}$y$}%
}}}}
\put(10351,-10786){\makebox(0,0)[lb]{\smash{{\SetFigFont{10}{12.0}{\familydefault}{\mddefault}{\updefault}{\color[rgb]{0,0,0}$s$}%
}}}}
\put(8926,-10786){\makebox(0,0)[lb]{\smash{{\SetFigFont{10}{12.0}{\familydefault}{\mddefault}{\updefault}{\color[rgb]{0,0,0}$z$}%
}}}}
\put(4426,-13636){\makebox(0,0)[lb]{\smash{{\SetFigFont{10}{12.0}{\familydefault}{\mddefault}{\updefault}{\color[rgb]{0,0,0}$z$}%
}}}}
\put(4426,-10936){\makebox(0,0)[lb]{\smash{{\SetFigFont{10}{12.0}{\familydefault}{\mddefault}{\updefault}{\color[rgb]{0,0,0}$s$}%
}}}}
\put(4426,-12736){\makebox(0,0)[lb]{\smash{{\SetFigFont{10}{12.0}{\familydefault}{\mddefault}{\updefault}{\color[rgb]{0,0,0}$x$}%
}}}}
\put(5551,-11836){\makebox(0,0)[lb]{\smash{{\SetFigFont{10}{12.0}{\familydefault}{\mddefault}{\updefault}{\color[rgb]{0,0,0}$y$}%
}}}}
\put(6976,-13786){\makebox(0,0)[lb]{\smash{{\SetFigFont{10}{12.0}{\familydefault}{\mddefault}{\updefault}{\color[rgb]{0,0,0}$z$}%
}}}}
\put(6976,-11086){\makebox(0,0)[lb]{\smash{{\SetFigFont{10}{12.0}{\familydefault}{\mddefault}{\updefault}{\color[rgb]{0,0,0}$s$}%
}}}}
\put(6976,-12886){\makebox(0,0)[lb]{\smash{{\SetFigFont{10}{12.0}{\familydefault}{\mddefault}{\updefault}{\color[rgb]{0,0,0}$x$}%
}}}}
\put(8101,-11986){\makebox(0,0)[lb]{\smash{{\SetFigFont{10}{12.0}{\familydefault}{\mddefault}{\updefault}{\color[rgb]{0,0,0}$y$}%
}}}}
\put(8101,-12886){\makebox(0,0)[lb]{\smash{{\SetFigFont{10}{12.0}{\familydefault}{\mddefault}{\updefault}{\color[rgb]{0,0,0}$x$}%
}}}}
\put(9451,-10786){\makebox(0,0)[lb]{\smash{{\SetFigFont{10}{12.0}{\familydefault}{\mddefault}{\updefault}{\color[rgb]{0,0,0}$x$}%
}}}}
\end{picture}%

%% file: badBlossom.pstex_t
\begin{picture}(0,0)%
\includegraphics{badBlossom.pstex}%
\end{picture}%
\setlength{\unitlength}{2368sp}%
\begingroup\makeatletter\ifx\SetFigFont\undefined%
\gdef\SetFigFont#1#2#3#4#5{%
  \reset@font\fontsize{#1}{#2pt}%
  \fontfamily{#3}\fontseries{#4}\fontshape{#5}%
  \selectfont}%
\fi\endgroup%
\begin{picture}(5182,6660)(2778,-15977)
\put(4426,-10936){\makebox(0,0)[lb]{\smash{{\SetFigFont{10}{12.0}{\familydefault}{\mddefault}{\updefault}{\color[rgb]{0,0,0}$z$}%
}}}}
\put(4426,-11836){\makebox(0,0)[lb]{\smash{{\SetFigFont{10}{12.0}{\familydefault}{\mddefault}{\updefault}{\color[rgb]{0,0,0}$x$}%
}}}}
\put(4426,-12736){\makebox(0,0)[lb]{\smash{{\SetFigFont{10}{12.0}{\familydefault}{\mddefault}{\updefault}{\color[rgb]{0,0,0}$s$}%
}}}}
\put(4426,-13636){\makebox(0,0)[lb]{\smash{{\SetFigFont{10}{12.0}{\familydefault}{\mddefault}{\updefault}{\color[rgb]{0,0,0}$y$}%
}}}}
\put(2926,-11836){\makebox(0,0)[lb]{\smash{{\SetFigFont{10}{12.0}{\familydefault}{\mddefault}{\updefault}{\color[rgb]{0,0,0}$r$}%
}}}}
\put(6526,-12736){\makebox(0,0)[lb]{\smash{{\SetFigFont{10}{12.0}{\familydefault}{\mddefault}{\updefault}{\color[rgb]{0,0,0}$s$}%
}}}}
\put(6526,-13636){\makebox(0,0)[lb]{\smash{{\SetFigFont{10}{12.0}{\familydefault}{\mddefault}{\updefault}{\color[rgb]{0,0,0}$y$}%
}}}}
\put(6526,-11836){\makebox(0,0)[lb]{\smash{{\SetFigFont{10}{12.0}{\familydefault}{\mddefault}{\updefault}{\color[rgb]{0,0,0}$r$}%
}}}}
\put(6526,-14536){\makebox(0,0)[lb]{\smash{{\SetFigFont{10}{12.0}{\familydefault}{\mddefault}{\updefault}{\color[rgb]{0,0,0}$x$}%
}}}}
\put(6601,-15436){\makebox(0,0)[lb]{\smash{{\SetFigFont{10}{12.0}{\familydefault}{\mddefault}{\updefault}{\color[rgb]{0,0,0}$z$}%
}}}}
\end{picture}%

%% file: Beforee1.pstex_t
\begin{picture}(0,0)%
\includegraphics{Beforee1.pstex}%
\end{picture}%
\setlength{\unitlength}{1934sp}%
\begingroup\makeatletter\ifx\SetFigFont\undefined%
\gdef\SetFigFont#1#2#3#4#5{%
  \reset@font\fontsize{#1}{#2pt}%
  \fontfamily{#3}\fontseries{#4}\fontshape{#5}%
  \selectfont}%
\fi\endgroup%
\begin{picture}(7027,11260)(4620,-11777)
\put(9601,-5536){\makebox(0,0)[lb]{\smash{{\SetFigFont{9}{10.8}{\familydefault}{\mddefault}{\updefault}{\color[rgb]{0,0,0}$B$}%
}}}}
\put(7951,-5986){\makebox(0,0)[lb]{\smash{{\SetFigFont{9}{10.8}{\familydefault}{\mddefault}{\updefault}{\color[rgb]{0,0,0}$x$}%
}}}}
\put(6901,-2836){\makebox(0,0)[lb]{\smash{{\SetFigFont{9}{10.8}{\familydefault}{\mddefault}{\updefault}{\color[rgb]{0,0,0}$w$}%
}}}}
\put(6376,-5986){\makebox(0,0)[lb]{\smash{{\SetFigFont{9}{10.8}{\familydefault}{\mddefault}{\updefault}{\color[rgb]{0,0,0}$x$}%
}}}}
\put(6901,-3961){\makebox(0,0)[lb]{\smash{{\SetFigFont{9}{10.8}{\familydefault}{\mddefault}{\updefault}{\color[rgb]{0,0,0}$c$}%
}}}}
\put(7051,-4711){\makebox(0,0)[lb]{\smash{{\SetFigFont{9}{10.8}{\familydefault}{\mddefault}{\updefault}{\color[rgb]{0,0,0}$b$}%
}}}}
\put(6451,-4711){\makebox(0,0)[lb]{\smash{{\SetFigFont{9}{10.8}{\familydefault}{\mddefault}{\updefault}{\color[rgb]{0,0,0}$a$}%
}}}}
\put(7951,-8086){\makebox(0,0)[lb]{\smash{{\SetFigFont{9}{10.8}{\familydefault}{\mddefault}{\updefault}{\color[rgb]{0,0,0}$x$}%
}}}}
\put(7951,-9136){\makebox(0,0)[lb]{\smash{{\SetFigFont{9}{10.8}{\familydefault}{\mddefault}{\updefault}{\color[rgb]{0,0,0}$d$}%
}}}}
\put(6076,-9586){\makebox(0,0)[lb]{\smash{{\SetFigFont{9}{10.8}{\familydefault}{\mddefault}{\updefault}{\color[rgb]{0,0,0}$e_1(x)$}%
}}}}
\put(9001,-11236){\makebox(0,0)[lb]{\smash{{\SetFigFont{9}{10.8}{\familydefault}{\mddefault}{\updefault}{\color[rgb]{0,0,0}$w$}%
}}}}
\put(7951,-7036){\makebox(0,0)[lb]{\smash{{\SetFigFont{9}{10.8}{\familydefault}{\mddefault}{\updefault}{\color[rgb]{0,0,0}$e$}%
}}}}
\put(6901,-1261){\makebox(0,0)[lb]{\smash{{\SetFigFont{9}{10.8}{\familydefault}{\mddefault}{\updefault}{\color[rgb]{0,0,0}$\eta$}%
}}}}
\put(6376,-7036){\makebox(0,0)[lb]{\smash{{\SetFigFont{9}{10.8}{\familydefault}{\mddefault}{\updefault}{\color[rgb]{0,0,0}$f$}%
}}}}
\put(10951,-4411){\makebox(0,0)[lb]{\smash{{\SetFigFont{9}{10.8}{\familydefault}{\mddefault}{\updefault}{\color[rgb]{0,0,0}$\eta$}%
}}}}
\put(10951,-2836){\makebox(0,0)[lb]{\smash{{\SetFigFont{9}{10.8}{\familydefault}{\mddefault}{\updefault}{\color[rgb]{0,0,0}$w$}%
}}}}
\put(10951,-1786){\makebox(0,0)[lb]{\smash{{\SetFigFont{9}{10.8}{\familydefault}{\mddefault}{\updefault}{\color[rgb]{0,0,0}$x$}%
}}}}
\put(10951,-5686){\makebox(0,0)[lb]{\smash{{\SetFigFont{9}{10.8}{\familydefault}{\mddefault}{\updefault}{\color[rgb]{0,0,0}$x$}%
}}}}
\put(10951,-5236){\makebox(0,0)[lb]{\smash{{\SetFigFont{9}{10.8}{\familydefault}{\mddefault}{\updefault}{\color[rgb]{0,0,0}$w$}%
}}}}
\put(10951,-736){\makebox(0,0)[lb]{\smash{{\SetFigFont{9}{10.8}{\familydefault}{\mddefault}{\updefault}{\color[rgb]{0,0,0}$h$}%
}}}}
\put(9376,-2836){\makebox(0,0)[lb]{\smash{{\SetFigFont{9}{10.8}{\familydefault}{\mddefault}{\updefault}{\color[rgb]{0,0,0}$g$}%
}}}}
\end{picture}%

%% file: oMuScan.pstex_t
\begin{picture}(0,0)%
\includegraphics{oMuScan.pstex}%
\end{picture}%
\setlength{\unitlength}{1934sp}%
\begingroup\makeatletter\ifx\SetFigFont\undefined%
\gdef\SetFigFont#1#2#3#4#5{%
  \reset@font\fontsize{#1}{#2pt}%
  \fontfamily{#3}\fontseries{#4}\fontshape{#5}%
  \selectfont}%
\fi\endgroup%
\begin{picture}(3930,7734)(4036,-9301)
\put(4051,-6136){\makebox(0,0)[lb]{\smash{{\SetFigFont{9}{10.8}{\familydefault}{\mddefault}{\updefault}{\color[rgb]{0,0,0}$e_1(x)$}%
}}}}
\put(6901,-1786){\makebox(0,0)[lb]{\smash{{\SetFigFont{9}{10.8}{\familydefault}{\mddefault}{\updefault}{\color[rgb]{0,0,0}$z$}%
}}}}
\put(6901,-2836){\makebox(0,0)[lb]{\smash{{\SetFigFont{9}{10.8}{\familydefault}{\mddefault}{\updefault}{\color[rgb]{0,0,0}$x$}%
}}}}
\put(6901,-7186){\makebox(0,0)[lb]{\smash{{\SetFigFont{9}{10.8}{\familydefault}{\mddefault}{\updefault}{\color[rgb]{0,0,0}$w$}%
}}}}
\put(7951,-6136){\makebox(0,0)[lb]{\smash{{\SetFigFont{9}{10.8}{\familydefault}{\mddefault}{\updefault}{\color[rgb]{0,0,0}$x$}%
}}}}
\put(6901,-6136){\makebox(0,0)[lb]{\smash{{\SetFigFont{9}{10.8}{\familydefault}{\mddefault}{\updefault}{\color[rgb]{0,0,0}$x$}%
}}}}
\put(6901,-8236){\makebox(0,0)[lb]{\smash{{\SetFigFont{9}{10.8}{\familydefault}{\mddefault}{\updefault}{\color[rgb]{0,0,0}$x$}%
}}}}
\end{picture}%

%% file: BigIncomplete.pstex_t
\begin{picture}(0,0)%
\includegraphics{BigIncomplete.pstex}%
\end{picture}%
\setlength{\unitlength}{2131sp}%
\begingroup\makeatletter\ifx\SetFigFont\undefined%
\gdef\SetFigFont#1#2#3#4#5{%
  \reset@font\fontsize{#1}{#2pt}%
  \fontfamily{#3}\fontseries{#4}\fontshape{#5}%
  \selectfont}%
\fi\endgroup%
\begin{picture}(12827,9131)(12661,-8345)
\put(19576,-3661){\makebox(0,0)[lb]{\smash{{\SetFigFont{9}{10.8}{\rmdefault}{\mddefault}{\updefault}{\color[rgb]{0,0,0}$B_2$}%
}}}}
\put(17026,-3661){\makebox(0,0)[lb]{\smash{{\SetFigFont{9}{10.8}{\rmdefault}{\mddefault}{\updefault}{\color[rgb]{0,0,0}$B_4$}%
}}}}
\put(21451,-286){\makebox(0,0)[lb]{\smash{{\SetFigFont{9}{10.8}{\rmdefault}{\mddefault}{\updefault}{\color[rgb]{0,0,0}$I$}%
}}}}
\put(21451,-1186){\makebox(0,0)[lb]{\smash{{\SetFigFont{9}{10.8}{\rmdefault}{\mddefault}{\updefault}{\color[rgb]{0,0,0}$O$}%
}}}}
\put(21451,-2236){\makebox(0,0)[lb]{\smash{{\SetFigFont{9}{10.8}{\rmdefault}{\mddefault}{\updefault}{\color[rgb]{0,0,0}$I$}%
}}}}
\put(22051,-7786){\makebox(0,0)[lb]{\smash{{\SetFigFont{9}{10.8}{\rmdefault}{\mddefault}{\updefault}{\color[rgb]{0,0,0}$I$}%
}}}}
\put(21451,-5536){\makebox(0,0)[lb]{\smash{{\SetFigFont{9}{10.8}{\rmdefault}{\mddefault}{\updefault}{\color[rgb]{0,0,0}$I$}%
}}}}
\put(21376,-4636){\makebox(0,0)[lb]{\smash{{\SetFigFont{9}{10.8}{\rmdefault}{\mddefault}{\updefault}{\color[rgb]{0,0,0}$O$}%
}}}}
\put(12676,-3661){\makebox(0,0)[lb]{\smash{{\SetFigFont{9}{10.8}{\rmdefault}{\mddefault}{\updefault}{\color[rgb]{0,0,0}$B_1$}%
}}}}
\put(23326,-5086){\makebox(0,0)[lb]{\smash{{\SetFigFont{9}{10.8}{\rmdefault}{\mddefault}{\updefault}{\color[rgb]{0,0,0}$I$}%
}}}}
\put(23251,-5686){\makebox(0,0)[lb]{\smash{{\SetFigFont{9}{10.8}{\rmdefault}{\mddefault}{\updefault}{\color[rgb]{0,0,0}$O$}%
}}}}
\put(23326,-6436){\makebox(0,0)[lb]{\smash{{\SetFigFont{9}{10.8}{\rmdefault}{\mddefault}{\updefault}{\color[rgb]{0,0,0}$I$}%
}}}}
\put(18151,-3661){\makebox(0,0)[lb]{\smash{{\SetFigFont{9}{10.8}{\rmdefault}{\mddefault}{\updefault}{\color[rgb]{0,0,0}$B_3$}%
}}}}
\end{picture}%

%% file: blocking.tex
\subsection{Min-max  relation}
\label{MinmaxSec}
\iffalse
line 551 begins discards
\fi

\def\CO{\mathy{\widehat C O}}

For  disjoint sets of vertices $S,T \con V(G)$ let
%$E[U]$ denote the set of all edges with at least one end in $U$, and
$E[S,T]$ be the set of all edges with  
one end in $S$ and the other in $T$.
%(A loop $(v,v)$ is in  $E[U,U']$ exactly when $v\in U\cap U'$.)
The maximum size an $f$-matching (i.e., a subgraph where every
vertex $x$ has degree $\le f(x)$) is the minimum value of the expression
\begin{equation}
\label{fFactorSizeEqn}
f(I)+ |\gamma(O)| + \sum _C \bigg\lfloor \frac{f(C)+|E[C,O]|}{2} \bigg\rfloor  
\end{equation}
\noindent where 
$I$ and $O$  
range over all pairs of disjoint vertex sets,
and in the summation
$C$ ranges over all
connected components of $G-I-O$.
\iffalse
\min\ \bigg\{f(I)+ |\gamma(O)| + \sum _C \Big\lfloor \frac{f(C)+|E[C,O]|}{2} \Big\rfloor   \bigg\}

%\f{\frac{f(C)+|E[C,O]|}{2}}\bigg\}
\noindent where 
the set is formed by letting $I$ and $O$  
range over all pairs of disjoint vertex sets,
and in the summation
$C$ ranges over all
connected components of $G-I-O$. 

Intuitively $I$ is the set of inner atoms, $O$ the set of
outer atoms, and $C$ the set of all vertices in blossoms.
\fi
This 
min-max relation
is proved in \cite[Theorem 32.1]{S} by reduction;
\cite[Sec. 5.3]{G18} gives an algorithmic proof.
This section gives another self-contained algorithmic proof, in the process
of establishing the blocking property of our algorithm.

We first give some intuition for the relation.
The set names are mnemonics for the tight case of the bound:
We will show that equality holds for our matching by taking
$I$ as the set of inner atoms and $O$ as the set of
outer atoms. The connected components $C$ are formed
by the remaining vertices, specifically vertices in blossoms
and vertices not reached in any search. 
In the special case of $f\equiv 1$ our min-max relation
is Edmonds' odd set cover formula for a maximum cardinality matching \cite{Ed, L, LP}.
We view our relation as a ``generalized odd set formula''.
Just like Edmonds' formula the nontrivial part of the bound is
due to rounding odd sets down by $1/2$. For 1-matching the 
$C$ components are the blossoms and one more, the unreached vertices.

Let us show the expression \eqref{fFactorSizeEqn} always upper bounds
the size of an $f$-matching.  Classify the edges of G as type I (one
or both ends in $I$), $OO$ (both ends in $O$), and \CO \ (joins a C
vertex to a C or O vertex). Fix an $f$-matching $M$. $M$ trivially
contains at most $f(I)$ $I$-edges and $|\gamma(O)|$ OO edges.
Consider a connected component $C$ of $G-I-O$.  The number of edges of
$M$ in \CO is $\frac{1}{2} (\sum _{x\in C\cup O} deg (x, M\cap \CO))$.
The terms for $x\in O$ trivially sum to at most $|E[C,O]|$.  The terms
for $x\in C$ sum to at most $f(C)$.  So $|M\cap \CO|$ is at most
$\frac{1}{2} (f(C)+ |E[C,O]|)$.  Since $|M\cap \CO|$ is integral we
can take the floor, thus giving the last term of
\eqref{fFactorSizeEqn}.

We will show our algorithm finds a blocking set even if we make
augmenting paths through complete blossoms residual. To be precise
consider the graph when \fbs\  halts.
Let \A. be the set of all augmenting trails at that point.
As before 
\C. is the set of maximal complete blossoms.
\iffalse
HAL

MORE GENERAL-CAN BE BEFORE FBS HALTS????
\fi
Define the {\em residual graph} 
to be $RG=G-\A.\cup \C.$.
Let $M$ be the matching on $RG$ when \fbs\ halts.  Notice $M$
contains the matched edges of augmenting trails
when they occur in blossoms of \C.. For $x\in V(G)$ let $def(x)$ be the
deficiency
computed by \fbs, i.e.,
the deficiency of $x$ in the matching on $G$
after augmenting the trails of \A..
The degree constraint function on $RG$ is
\begin{equation*}
  f'(x)= deg(x,M) + def(x).
\end{equation*}
Throughout this section we will write $f'$ and $def'$
as $f$ and $def$. This causes no problem. For instance
the definition of $f'$ shows deficiencies are identical in $G$ and $RG$.
(We will use $def$ values when we apply
\eqref{FreeVtxEqn} of Lemma \ref{FreeBlossomLemma}.)
We will show \fbs\ finds a maximum cardinality matching of $RG$.

An edge of $RG$ is a {\em residual edge}. Note that a blossom $B\in
\C.$ need not have a base edge in $RG$. This occurs when a trail of
\A.  passes through $B$ (or it simply contains an edge incident of
$\delta(\beta(B), \delta(B))$).

\begin{figure}[ht]
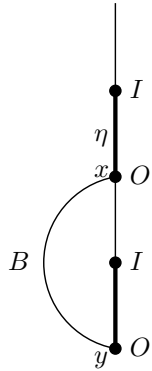

\centering
\input EraseLabels.pstex_t
\caption{Erasing labels. Vertices are  labelled before blossom $B$ is
  formed. The bound of \eqref{fFactorSizeEqn} is not tight, since
  the unmatched edge $xy$ contributes to  $|\gamma(O)|$.
  If an augmenting trail is discovered before $d(\eta)$ returns
  $xy$ is on the trail, removing the contribution.
  Otherwise blossom $B$ is formed, again removing the contribution.}
\label{EraseLabelsFig}
\end{figure}

%Let $M$ be the matching on $RG$ when \fbs\ halts.
We will  use \eqref{fFactorSizeEqn} to
show $M$ is a maximum cardinality matching of $RG$.
To define sets $I$ and $O$ consider a vertex
$x\in V(G)$
where the invocation $d(e_1(x))$  returned 
but $x$ does not occur in any complete blossom.
\begin{equation}
  \label{LabelDefEqn}
x\in
  \begin{cases}
  I&\mu(x)=\o M\\
  O&\mu(x)=M.  
  \end{cases} 
\end{equation}
As an example a free vertex with deficiency greater than 1 is in $O$
(Lemma \ref{FreeBlossomLemma}),
so its
$f$ value is not used in \eqref{fFactorSizeEqn}.
We often refer to I and O as vertex labels, and call the vertices
of $G-I-O$ {\em unlabelled}.
Mnemonically
note that $x$ is labelled according to arc $e_1$: $x$ is I (O) if
$e_1$ makes $x$ an inner (outer) vertex.
Intuition for the labelling scheme is given in Fig.\ref{EraseLabelsFig}.

%Say $x$ has been {\em grow-scanned} in this case.)

There are two types of unlabelled vertices:
If $e_1(x)$ exists then an unlabelled $x$ is in a complete blossom
of $G$. The second possibility for unlabelled $x$ is that
$e_1(x)$ does not exist, i.e., no invocation $d(\vtx zx)$
 ever returned. Equivalently,
every
occurrence of $x$ in \T. was in an augmenting trail $A$.
Call such a vertex $x$
an {\em orphan}, i.e., $x$ does not occur in \T. or
$x$ occurs in $\T.\cap RG$ but has no parent, e.g., vertex $a$
in Fig.\ref{dExamplesFig}(b).
(Note that a free vertex such a parent.)
\T. may contain residual arcs directed from $x$.
There may also be
edges incident to $x$ that are not in \T., i.e., no corresponding grow
step was executed.

\iffalse
I THINK INCOMPLETE BLOSSOMS ARE HANDLED AUTOMATICALLY. CHECK THIS. AND REMARK
ABOUT IT IN THE PROOF.
\fi

We proceed to analyze the number of matched residual edges for each of the three types
I, OO, \CO.
An $I$ vertex is not free (a free vertex is O or unlabelled).
So exactly $f(I)$ type I edges are matched. Thus $M$ achieves equality in
\eqref{fFactorSizeEqn} for type I edges.

Next consider an OO edge. Say it occurs as the \T.-arc $zx$.
It satisfies Lemma \ref{TBarResidualLemma}. 
(In particular $zx$ is not the base edge of a \C.-blossom $B$, 
since $\beta(B)$ is unlabelled.) 
Thus $zx$ has M-type $\mu(e_1(x))=M$. So every OO edge
is matched and $M$ achieves equality in \eqref{fFactorSizeEqn} for OO edges.

\iffalse

We claim every type OO edge is matched. In proof suppose $xy$ is an unmatched OO edge.
Since $x$ is type O, the unmatched edge $xy$ was removed from $GL(x)$. Wlog it was
removed in an invocation of the form $d(\vtx zx)$ and added to \T. as arc $xy$.
$xy$ is not edge $e_1(y)$ ($y$ an O vertex means $\mu(e_1(y))=M$).
At some point after $d(e_1(y))$ returns, control returns to $d(\vtx xy)$ and
the blossom base test pops $e_1(y)$, forming a blossom $B$
with base edge $xy$. Since $xy$ is residual, $B$ is complete. Thus $y$ is a C vertex, NO!!!!!!!!!!!!!!! NEW PROOF is has an arrow to it 
contradiction. So the claim holds and $M$ achieves equality for OO edges.
\fi

It remains to consider the \CO\ edges.
Let $C$ be a 
connected component of unlabelled vertices.
There are two similar cases, depending on whether or not $C$ contains an orphan.
We analyze them in turn.

\case{{\bf 1.} $C$ consists of \C.-blossoms}
$C$ is a subtree of \T. (Proposition \ref{BlossomSubtreeProp}, with augmenting
trails included).
$C$
consists of one or more 
\C.-blossoms joined by their base edges.
The root of this subtree is the base vertex of the ``root'' blossom $B_0$.
So a blossom $B$ in $C$ has its base edge in $C$ iff
$B\ne B_0$.

Next we analyze the edges of $G$ incident to a blossom $B\in \C.$.
Consider a residual edge $e=\ed xy$, $x\in B\notin y$.
%$e$ occurs as a \T.-arc since $GL(x)$ is empty when $\eta(B)$ returns.

\bigskip
\noindent
{\bf Claim} {\em Either $e=\eta(B)$ ($e$ being arc $yx$) or $e$ is arc $xy$ with
  \begin{equation}
    \label{AdjacentYEqn}
\text{$y$ labelled 
and
$\mu(e)=\mu(e_1(y))$ or $y$ unlabelled and $y=\beta(A)$ for a
%base vertex of a
blossom $A\in \C.$.}
\end{equation}
}

\noindent
    {\bf Proof of Claim:}
We will apply
Corollary \ref{CompleteBlossomCor} to  blossom $B$, adjusting for
events that occur after $B$ becomes complete.
Consider the corollary's three alternatives \xi--\iii for $e$:
   
\i $e$ is one of two oppositely directed arcs, say $ax$ and $xb$.
$B$ is maximal complete so
the skew blossom $S$ for $x$ was not completed.
$S$ was either never triggered or was triggered and is
incomplete. The first case has both arcs on the augmenting
trail, so neither arc is residual. In the second case $ax$ is on the augmenting
trail. $xb$ is not, since it
is not on the search path of
any $d$ invocation made in the blossom step.
$xb$ is residual, and in particular $b$ is not an orphan.
So if $b$ is unlabelled then $xb$ is the base edge of a blossom $A \in \C.$.
The Claim holds.
If $b$ is labelled then 
Lemma \ref{TBarResidualLemma} applies (with $zx$ taken as $xb$). It shows 
$\mu(xb)=\mu(e_1(b))$. 
Again the Claim holds.

\ii $e=\eta(B)$ as in the Claim. 

\iii $e$ is arc $xy$. The analysis of $xb$ in \i applies, i.e., the same two
possibilities of the Claim hold. Note that $e$ may originate from
any number of  occurrences of $x$ on $SP$.
\hfill$\spadesuit$

\bigskip

\iffalse
The skew blossom for $x$, 
p corresponds to
two possibilities
If $e$ is on the search path to $d(\eta)$ then
the skew blossom for $e$'s $x$-end was not executed.
So $e$ is on the search's
augmenting trail and is not residual.
\fi

\iffalse
$y$ is not an orphan,
so if $y$ is unlabelled then $xy$ is the base edge of a blossom.
If $y$ is labelled then 
Lemma \ref{TBarResidualLemma} shows 
$\mu(e)=\mu(e_1(y))$. Again the claim holds.
\fi

\bigskip

We now show the component $C$ achieves equality in
\eqref{fFactorSizeEqn}. The only fact we will use
about edges incident
to $C$ is the Claim. We will reuse the Claim when we analyze
components $C$ that contain an orphan.

Let $M$ be the set of matched residual edges with one or both ends in $C$.
Define $M_I$ to be the set of $M$-edges of type $I$,
and similarly for $M_O$ and $M_\CO$.
We will show the number of matched edges of type
\CO\ for our component $C$ 
equals its corresponding term in the summation of \eqref{fFactorSizeEqn}.
To do this we show the number of ends of matched  \CO\ edges is exactly
\begin{equation}
  \label{MatchedEndsEqn}
  2|M_\CO|=f(C) +|E[C,O]| - \epsilon
\end{equation}
where $\epsilon \in \{0,1\}$.
(Recall $f$ is the residual degree constraint function.)
Since the left-hand side is even,
$\epsilon=1$ implies $f(C) +|E[C,O]|$ is odd.
So the desired equality always holds.

\begin{figure}[ht]
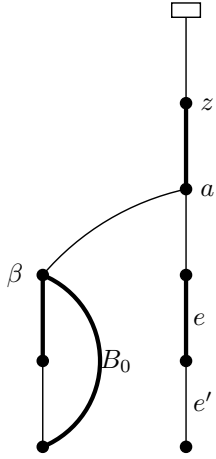

\centering
\input CIEdge.pstex_t
\caption{Possibilities for the base edge of root blossom $B_0$:
If $e=e_1(a)$ then $a$ is labelled O when
$d(\vtx za)$ returns and $\epsilon=\epsilon_3=1$. If
$e'=e_1(a)$  and an augment occurs after $d(e_1(a))$ returns
but before the blossom for $e_1(a)$ is formed then $a$ is labelled I
and $\epsilon=0$. If the blossom for $e_1(a)$ is formed then it does not
get completed, since $B_0$ is the root. Again $a$ remains labelled I.
A final possibility with $\epsilon=0$ is when
arc $a\beta$ is not residual. This can occur
if a blossom step is triggered by an 
arc from $B_0$ to $z$, and an augmenting trail passes through $B_0$.}
\label{CIEdgeFig}
\end{figure}

Let
$\beta$ be the base vertex of the root blossom $B_0$,
and let $\eta$ be its base edge, also denoted as arc $a\beta$.
The above quantity $\epsilon$ is the sum of
three quantities $\epsilon_i\in \{0,1\}, i=1,2,3$,
defined by
\[
\epsilon_i=1 \xiff \eta \text{ is }
\begin{cases}
  \text{the artifical edge $\varepsilon \alpha$}&i=1\\
  \text{matched with $a$ labelled I}&i=2\\
    \text{unmatched with $a$ labelled O}&i=3.
\end{cases}
\]
Clearly at most one of these quantities is 1.
Fig. \ref{CIEdgeFig} illustrates the possibilities for $\epsilon$.

\iffalse
$a$ is either
\begin{array*}[l]
  \text{the artifical vertex \varepsilon, or}
  \text{a matched edge in $I$, or}
    \text{an unmatched edge in $O$'}
\end{eqnarray*}
Otherwise $\epsilon=0$.
\fi

The number of ends of matched edges of type \CO\ is exactly
\begin{equation}
  \label{ObviousMatchedEndsEqn}
  2|M_\CO|=\sum_{x\in C} deg(x,M) - deg(x,M_I) + |M_\CO|.
\end{equation}
To analyze this let $e=\ed xy$ be a
residual edge incident to $C$, with $x\in C$.
The Claim  and \eqref{AdjacentYEqn}
show either $e=\eta(B_0)$ for the root blossom $B_0$
or $e$ occurs as arc $xy$ with $y$ labelled and
$e$ is either
\[\text{
matched with $y$ labelled O or
unmatched with $y$ labelled I.}\]
This implies
\begin{equation}
  \label{Epsilon23Eqn}
  \sum_{x\in C} deg(x,M_I) = \epsilon_2 \text{ and }
  |E [C,O]|=  |M_\CO| + \epsilon_3.
\end{equation}
Recall (Lemma \ref{FreeBlossomLemma})
$C$ contains at most one free vertex,
which must be $\beta$, with $\beta=\alpha$ and $f(\beta)=deg(\beta,M) +1$.
Thus
\begin{equation}
  \label{Epsilon1Eqn}
\sum_{x\in C} deg(x,M)= f(C) - \epsilon_1.
\end{equation}
Combining \eqref{ObviousMatchedEndsEqn}
with \eqref{Epsilon23Eqn} and \eqref{Epsilon1Eqn}
gives \eqref{MatchedEndsEqn}.

\case{{\bf 2.} $C$ contains  an orphan}
We start by analyzing
the adjacencies
of an orphan.

\begin{lemma}
\label{OrphanLemma}
Consider a residual edge $e=\ed xy$ incident to an orphan $x$.
Either $y$ is unlabelled and an orphan or $y$ satisfies
\eqref{AdjacentYEqn}.
\end{lemma}

\iffalse
If $y$ is labelled then $\mu(xy)=\mu(e_1(y))$.
If $y$ is unlabelled then either $y$ is an orphan or
$xy$ is a \T.-arc and the base edge of a complete blossom.
\fi

\begin{proof}
  Let $e$ denote edge \ed xy.

  \case{$y$ is labelled} We will show $\mu(e)=\mu(e_1(y))$,
  so \eqref{AdjacentYEqn} holds.
  If edge $e$ occurs as a \T.-arc it must be arc $xy$ ($x$ is an orphan).
  Lemma \ref{TBarResidualLemma}
  applies (in particular $e$ is not
  the base edge of a \C.-blossom). It proves
 $\mu(e)=\mu(e_1(y))$ as claimed.
  Suppose $e$ is not a \T.-arc.
  The invocation $d(e_1(y))$ exists and returns (because of $y$'s label). 
  $e$ was
  not removed in $d(e_1(y))$ ($x$ is an orphan).
  So again $ \mu(e)=\mu(e_1(y))$.
  
\case{$y$ is unlabelled}
If $y$ is an orphan the lemma's condition holds.
The other possibility is that $y$ is in a complete blossom $B'$.
We  can assume $B'\in \C.$. We have $y\in B'\not\ni  x$ and
\ed xy is the arc $xy$ ($x$ an orphan). The Claim above shows
$xy=\eta(B')$. So \eqref{AdjacentYEqn} holds for $A$ taken as $B'$.
\end{proof}

\iffalse

Apply
Corollary \ref{CompleteBlossomCor} to $B$. $e$ occurs in \T. in all three
alternatives \xi--\xiii.
$e$ does not occur as arc $yx$, since the orphan $x$ has no residual parent arc.
So $e$ is arc $xy$ (in the notation of the corollary this is arc $yx$).
Consider the three alternatives:

\bigskip

\i This alternative has arc $xy$ a proper ancestor of $\eta(B)$. 
Since arc $xy$ is residual, the invocation $d(\vtx xy)$ returned.
Since $y$ occurs in $B$ the invocation  executes a skew blossom step.
This new blossom does not get completed ($B\in \C.$).
So the augmenting path contains $xy$, contradiction.

\ii $e=\eta(B)$ so \eqref{AdjacentYEqn} holds.

\iii As before this alternative is impossible, the orphan $x$
has no parent arc.
\fi

Define an
{\em orphanage} to be  a
connected component of the graph induced on the residual graph
by the set of orphans. (Its edges are the residual edges joining two orphans.)
Clearly $C$ contains an orphanage, say $R$.
Lemma
\ref{OrphanLemma}
and \eqref{AdjacentYEqn} show
$R$ can be joined to other unlabelled vertices
only by \T.-arcs directed from $R$ to the base vertex of a complete blossom.
Thus $C$ consists of $R$ and 0 or more
subtrees $S$ that consist entirely
of complete blossoms and are joined to $R$ by the root blossom's
base edge $\eta$.
We will show $C$ satisfies the tightness relation
\eqref{MatchedEndsEqn} with $\epsilon=0$.
To do this we 
examine the contribution to the right-hand side made from
the orphanage $R$ and 
from a typical subtree $S$.

First consider $S$. Case 1 applies and shows \eqref{MatchedEndsEqn}.
Furthermore $\eta$ does not leave the current component $C$.
So each $\epsilon_i=0$ and
$\epsilon=0$.

Now consider $R$.
\eqref{ObviousMatchedEndsEqn}
holds by definition.
%THE BELOW IS ON SINCE WE ARE ANALYZING THE contribution to rhs
Since \eqref{AdjacentYEqn} holds for $R$,
\eqref{Epsilon23Eqn} holds with $\epsilon_2=\epsilon_3=0$.
 $R$ does not contain any free vertex,
since a free vertex  occurs in an unsuccessful search.
So \eqref{Epsilon1Eqn} holds with $\epsilon_1=0$.
As before combining these equations gives \eqref{MatchedEndsEqn} 
with $\epsilon =0$.

Now combining the instances of  \eqref{MatchedEndsEqn}
 for $R$ and all subtrees $S$ gives 
 \eqref{MatchedEndsEqn} for $C$ with $\epsilon=0$.
We have now verified the tightness of \eqref{fFactorSizeEqn} for our labelling
and the residual graph  matching of \fbs. So our main result follows:

\begin{theorem}
  \label{MainTheorem}
  \fbs\ finds a blocking trail set \A..
  In fact \fbs\ halts with a maximum cardinality matching of
  the residual graph. \hfill$\Box$
\end{theorem}

\iffalse
\begin{proof}
  We prove the second assertion, a stronger version of the first.
  The edge set of the residual graph $G-\A.\cup \C.$ consists of the set \Tp. plus
all edges not reached in the search.

The theorem follows.
\end{proof}
\fi
\iffalse
Lemma \ref{LabellingLemma}\i shows the final $I,O,C$ labelling satisfies the criterion of
Corollary \ref{ValidLabelCor}.
\fi

\paragraph{Linear-time bound}
It is straightforward to see that \fbs\ operates in linear time $O(m)$.
Here $m$ counts each edge according to its multiplicity.
Note the blossom enlarge test is easy to implement using
Proposition \ref{SecBlTstProp}\xi, since testing for a skew blossom is trivial.

For an algorithm to use our blocking set
the trails of \A. must be rematched.
The inclusion of \C. in the residual graph allows any alternating trail through a complete
blossom to be rematched. Thus the $P_i$ trails of \cite{G18} may be used (even for skew blossoms). So postprocessing the augmenting trails
uses $O(m)$ time.

%%%%%%%%%%%%%%%%%%%%%% DISCARDS 8.11.21

%% file: EraseLabels.pstex_t
\begin{picture}(0,0)%
\includegraphics{EraseLabels.pstex}%
\end{picture}%
\setlength{\unitlength}{2368sp}%
\begingroup\makeatletter\ifx\SetFigFont\undefined%
\gdef\SetFigFont#1#2#3#4#5{%
  \reset@font\fontsize{#1}{#2pt}%
  \fontfamily{#3}\fontseries{#4}\fontshape{#5}%
  \selectfont}%
\fi\endgroup%
\begin{picture}(1305,3882)(3211,-12002)
\put(4126,-9586){\makebox(0,0)[lb]{\smash{{\SetFigFont{10}{12.0}{\familydefault}{\mddefault}{\updefault}{\color[rgb]{0,0,0}$\eta$}%
}}}}
\put(3226,-10936){\makebox(0,0)[lb]{\smash{{\SetFigFont{10}{12.0}{\familydefault}{\mddefault}{\updefault}{\color[rgb]{0,0,0}$B$}%
}}}}
\put(4501,-9136){\makebox(0,0)[lb]{\smash{{\SetFigFont{10}{12.0}{\familydefault}{\mddefault}{\updefault}{\color[rgb]{0,0,0}$I$}%
}}}}
\put(4501,-10036){\makebox(0,0)[lb]{\smash{{\SetFigFont{10}{12.0}{\familydefault}{\mddefault}{\updefault}{\color[rgb]{0,0,0}$O$}%
}}}}
\put(4501,-10936){\makebox(0,0)[lb]{\smash{{\SetFigFont{10}{12.0}{\familydefault}{\mddefault}{\updefault}{\color[rgb]{0,0,0}$I$}%
}}}}
\put(4501,-11836){\makebox(0,0)[lb]{\smash{{\SetFigFont{10}{12.0}{\familydefault}{\mddefault}{\updefault}{\color[rgb]{0,0,0}$O$}%
}}}}
\put(4126,-9961){\makebox(0,0)[lb]{\smash{{\SetFigFont{10}{12.0}{\familydefault}{\mddefault}{\updefault}{\color[rgb]{0,0,0}$x$}%
}}}}
\put(4126,-11911){\makebox(0,0)[lb]{\smash{{\SetFigFont{10}{12.0}{\familydefault}{\mddefault}{\updefault}{\color[rgb]{0,0,0}$y$}%
}}}}
\end{picture}%

%% file: CIEdge.pstex_t
\begin{picture}(0,0)%
\includegraphics{CIEdge.pstex}%
\end{picture}%
\setlength{\unitlength}{2368sp}%
\begingroup\makeatletter\ifx\SetFigFont\undefined%
\gdef\SetFigFont#1#2#3#4#5{%
  \reset@font\fontsize{#1}{#2pt}%
  \fontfamily{#3}\fontseries{#4}\fontshape{#5}%
  \selectfont}%
\fi\endgroup%
\begin{picture}(2062,4744)(2461,-12733)
\put(2476,-10936){\makebox(0,0)[lb]{\smash{{\SetFigFont{10}{12.0}{\familydefault}{\mddefault}{\updefault}{\color[rgb]{0,0,0}$\beta$}%
}}}}
\put(4501,-10036){\makebox(0,0)[lb]{\smash{{\SetFigFont{10}{12.0}{\familydefault}{\mddefault}{\updefault}{\color[rgb]{0,0,0}$a$}%
}}}}
\put(4501,-9136){\makebox(0,0)[lb]{\smash{{\SetFigFont{10}{12.0}{\familydefault}{\mddefault}{\updefault}{\color[rgb]{0,0,0}$z$}%
}}}}
\put(3451,-11836){\makebox(0,0)[lb]{\smash{{\SetFigFont{10}{12.0}{\familydefault}{\mddefault}{\updefault}{\color[rgb]{0,0,0}$B_0$}%
}}}}
\put(4426,-12286){\makebox(0,0)[lb]{\smash{{\SetFigFont{10}{12.0}{\familydefault}{\mddefault}{\updefault}{\color[rgb]{0,0,0}$e'$}%
}}}}
\put(4426,-11386){\makebox(0,0)[lb]{\smash{{\SetFigFont{10}{12.0}{\familydefault}{\mddefault}{\updefault}{\color[rgb]{0,0,0}$e$}%
}}}}
\end{picture}%

%% file: app.tex
\setcounter{section}{0}
\renewcommand{\thesection}{\Alph{section}}
\renewcommand{\thetheorem}{\Alph{section}.\arabic{theorem}}

\setcounter{equation}{0}
\renewcommand{\theequation}{\Alph{section}.\arabic{equation}}

\def\Myfor #1{{\bf for} {\em #1} {\bf do}}
\def\Myif #1{{\bf if} {\em #1} {\bf then}}
\def\Myelseif #1{{\bf else if} {\em #1} {\bf then}}
\def\Myelse #1{{\bf else} {#1}}

\newpage

\section{Blocking 1-matchings}
\label{GTApp}
Fig.\ref{AppFig} is a verbatim statement of the
blocking algorithm for 1-matching of  \cite{GT,
G17}. It is the jumping off point for our $f$-factor algorithm.

\bigskip

\input GTdfs

%%%%%%%%%%%%%%%%%%%%%%%%%%%%%%%%%%%%%%%%%%%%%%%%

%% file: GTdfs.tex
\iffalse
\def\myif #1{{\bf if} {\em {#1}} {\bf then}}
\def\myelif #1{{\bf else if} {\em {#1}} {\bf then}}
\def\myifel #1 #2 #3 {{\bf if} {\em {#1}} {\bf then} {\em {#2}} {\bf else} {\em {#3}}}

\def\myfor #1{{\bf for} {\em {#1}}}
\def\mycom #1{{\tt /* {\hskip-3pt#1\hskip-3pt} */}}

\def\sm.{\mathy{\cal S^-}}
\fi

\begin{figure}
\begin{center}
  \fbox{
    %\newlength{\Efigwidth}
    \setlength{\Efigwidth}{\textwidth}
    \addtolength{\Efigwidth}{-.3in}
\setlength{\vmargin}{15pt}
\begin{minipage}{\Efigwidth}%{\figwidth}
\setlength{\parindent}{.1in}%{.2in}

\narrower{
\setlength{\parindent}{0pt}
\vspace{\vmargin}
\setlength{\parindent}{0 pt}%{20pt}
%\noindent

{\bf procedure} {\em find\_ap\_set}

initialize \S. to an empty graph and \P. to an empty set

\myfor {\em each vertex $v\in V$} $b(v)\gets v$
\mycom{$b(v)$ maintains the base vertex of $B_v$}

\myfor {\em each  free vertex $f$}

{\hi 

{\bf if} {$f\notin V(\P.)$} {\bf then}

{\hi

add $f$ to \S. as the root of a new search tree

%\lnl{FfLine} 
{\em find\_ap}$(f)$

}}

{\bf return} \P.

\bigskip

{\bf procedure} {\em find\_ap}($x$)
\mycom{$x$ is an outer vertex}

%\lnl{FLine}
\myfor {\em each edge $xy\notin M$} $/*$ {scan $xy$ from $x$} $*/$

{\hi

{\myif {$y \notin V(\S.)$}

{\hi

\myif {$y$ is free} $/*$ {$y$ completes an augmenting path} $*/$ 

{\hi

add $xy$ to \S. and add path $y P(x)$ to \P.\;

%\lnl{TermLine}
terminate every currently executing recursive call to {\em find\_ap}
}

{\bf else} $/*$ {grow step} $*/$

{\hi

add $xy, yy'$ to \S., where $yy'\in M$

{\em find\_ap$(y')$}
}
}

{\bf else} \myif {$b(y)$ is an outer, proper descendant of $b(x)$ 
in \sm.} \mycom{blossom step}

{\hi

\mycom{equivalent test:\ $b(y)$ became outer strictly after $b(x)$}

let $u_i$, $i=1,\ldots, k$ be the inner vertices in 
$\os.(B_y,B_x)$, ordered so $u_i$ precedes $u_{i-1}$

%$P(y,b(x))$, ordered so $u_i$ precedes $u_{i-1}$

%\goin \tcc{HAL: NEED TO COMPUTE IN LINEAR TIME} 

\myfor {$i\gets 1\ {\bf to}\ k$} 

%\goin\goin $\ell(u_i) \gets (y,x)$ \tcc{$P(u_i)$ is a prefix of $P(u_{i-1})$} 

{\hi

\myfor {every vertex $v$ with $b(v)\in \{u_i,u'_i\}$, where $u_i u'_i\in M$} $b(v)\gets b(x)$

\mycom{this adds $B_{u_i} \cup B_{u'_i}$ to the new outer blossom}

}

\myfor {$i\gets 1\ {\bf to}\ k$}  {\em find\_ap$(u_i)$} 
\mycom{process $u_i$ in order of increasing depth}

}

{\bf return}

}
}

}%narrower

\vspace{\vmargin}

\end{minipage}
}%fbox
\caption{Depth-first search blocking algorithm for ordinary matching.}
\label{AppFig}
\end{center}
\end{figure}

%% file: maximal.tex
\def\X.{\mathy{\cal X}}
\def\il(#1){\mathy{\iota\text{-list}(#1)}}
\def\ibl(#1){\mathy{\bar\iota\text{-list}(#1)}}
\def\ol(#1){\mathy{o\text{-list}(#1)}}
\def\inl(#1){\mathy{i\text{-list}(#1)}}
\def\ib{\mathy{\overline\iota}}

\def\fas{\em find\_ap\_set}
\def\fa(#1){{\em find\_ap}$(#1)$}

\def\sm.{\mathy{\cal S^-}}
\def\os.{\mathy { \o{\cal S} }}

\section{Maximal augmenting trail set}
\label{MaxlAPSetAppendix}
Recall the M-type of an edge is $M$ (matched) or $U$ (unmatched).
If $\mu\in \{M,U\}$ is an M-type, $\o\mu$ denotes the opposite M-type.

\begin{lemma}
Suppose an augmenting trail joining from free vertices
$a$ and  $b$ exists. Then there exists such a trail
in which every vertex has degree $\le 4$.
\end{lemma}

\remarks{As usual we may have $a=b$. A loop $vv$
in the trail contributes 2 to $v$'s degree.}

\begin{proof}
Let $P$ be a given $ab$-augmenting trail.
We will assume $P$ has no loops. (If $P$ contains
a loop, the reader can supply the obvious modifications.
Alternatively assume we replace each loop $vv$ by its expansion
$vx,xx',x'v$ appropriately matched.)
Consider a vertex $v$ in $P$.
First assume $v$ is not either end $a,b$ of $P$.

Traverse $P$ from end $a$ to $b$.
Suppose $P$ enters $v$ for the first time on an edge
$e\in \delta(v,P)$ of M-type $\mu$. Let $f$ be the next
edge in $P$. It leaves 
$v$ and has  M-type $\o\mu$.
We can assume  $f$ is the last edge in $P$  of
M-type $\o \mu$ that leaves $v$.
(If not omit the subtrail between $e$ and $f$.)
If $f$ is the last edge incident to $v$ in $P$
we are done.
Otherwise let $P$ enter $v$ for the first time after $f$
on edge $e'$, and leave on $f'$.
$f'$ does not have M-type $\o\mu$, so its M-type is $\mu$.
(So $e'$ has M-type $\o\mu$.)
As before, we can assume $f'$ is the last edge leaving $v$ with M-type $\mu$.
$f'$ must be the last edge of $\delta(v,P)$, since 
an edge leaving $v$ after $f'$
must have M-type different from $f$ and $f'$, impossible.

Now suppose $v$ is the end $b$. The argument above applies.
The only modification is that $e$ may be the last edge of
$P$, $f$ is never the last edge of $P$, and $e'$ is the last
edge of $P$ if it exists.

Finally suppose $v=a$. Simply traverse $P$ in the opposite direction and apply the previous argument. (Clearly it does not
affect vertex $b$.)???
\end{proof}

\iffalse
\cite{G83}
H.N. Gabow,
``An efficient reduction technique for degree-constrained subgraph and
bidirected network flow problems'',

\begin{figure}[t]
\centering
\input{substitute.pstex_t}
\caption{Vertex substitute}
 \label{SubstituteFig}
\end{figure}
\fi

The {\em io types} are i (for inner) and o (for outer).
For
a vertex $x$, $x_i$ ($x_o$) refers to $x$
being processed as inner (outer), respectively.
$\iota$ always denotes an io-type,
so $x_\iota$ refers to both possibilities for $x$.
$\ib$ denotes the opposite type from $\iota$.
So $\{\iota,\overline\iota\}=\{i,o\}$.

Each vertex $x$ has two adjacency lists:
\inl(x) contains every matched edge $xy$, i.e.,
the edges that alternate at $x$ for $x_i$.
Similarly \ol(x) contains every unmatched edge $xy$, i.e.,
the edges that alternate at $x$ for $x_o$.
\il(x) refers to both of the adjacency lists.

An edge $xy$ of $G$ appears in two $\iota$-lists. Thus
it can potentially be scanned
twice, once in an invocation \fa(x,\iota) and once in an invocation
\fa(y,\kappa). (An invocation \fa(x,\iota) may lead to a subsequent invocation
\fa(x,\iota), but $xy$ does not get rescanned.) Furthermore
if $xy$ gets added to \S. when it is scanned from $x$,
it is not eligible for scanning from $y$.%
\footnote{This restriction seems unnecessary because
the \S.-edge leading to $y$ does not alternate with $y$.
But a blossom step can change $y$ from io-type $\iota$ to
$\bar \iota$, making the \S.-edge alternate.}
So for example consider the construct

{\hi {\sf while} $\exists\; xy \in \ol(x)$ unscanned
}

\noindent
traverses \ol(x), taking recursive invocations into account,
processing each edge in turn, skipping over
the \S. edge leading to
$x$, and then it halts.

\setlength{\figwidth}{\textwidth}
\addtolength{\figwidth}{-.5in}
%\newlength{\figindent}
\setlength{\figindent}{.5in}
%\newlength{\vmargin}
\setlength{\vmargin}{.1in}

\begin{figure}[t]
\begin{center}
\fbox{
  \begin{minipage}{\figwidth}
\setlength{\parindent}{.2in}
\narrower{
\setlength{\parindent}{0pt}
\vspace{\vmargin}

{\sf Algorithm} {\em find\_ap\_set}

%initialize \X. and \S. to empty graphs and \M. to an empty set

\iffalse
{\sf for} {\em each vertex $v\in V$} {\sf do} $b(v)\gets v$
\tcc{$b(v)$ maintains the base vertex of $B_v$}
\fi

{\sf while} $\exists$ free vertex $a$ s.t. $\ol(a)$ has an unscanned edge

{\hi

initialize a new search structure \S. $/*$ old $b$ values are irrelevant $*/$

add $a$ to \S. as the root of a new search tree; {\em find\_ap}$(a,o)$

}

\bigskip

{\sf Algorithm} {\em find\_ap}($x,\iota$)

{\sf if} $x$ is free, $\iota=i$, and either $x\ne a$ or  $def(a)\ge 2$ 

\hspace{10pt} $/*$ a vertex $a\in \B.$ has the same deficiency as its base 
vertex $*/$

{\hi
add $xy$ to \S.

rematch the augmenting trail $yP(x_\iota)$ 

terminate every currently executing invocation of {\em find\_ap}

}

{\sf while} $\exists\; xy\in \il(x)$ unscanned
 {}
{\addtolength{\parindent}{20pt}

{\sf if} $B_y$ is a proper descendant of $B_x$ in \os. 

{\addtolength{\parindent}{10pt}

$/*$ equivalent test: $b(y)$ became outer strictly after $b(x)$ $*/$  %A

and $xy$ alternates with $B_y$
}

{\hi

$/*$ blossom step $*/$

let $u^j$, $j=1,\ldots, k$ be the atoms in 
$\os.(B_x,B_y)$, ordered so $u^j$ precedes $u^{j+1}$

%{HAL: NEED TO COMPUTE IN LINEAR TIME} 

{\sf for} every vertex $v\in V(G)$ with $B_v\in \os.(B_x,B_y)$
{\sf do} $b(v)\gets b(x)$

$/*$ this adds $B_{v}$ to the new outer blossom $*/$

{\sf for} {$j\gets 1\ {\sf to}\ k$}  {\em find\_ap$(u^j_\ib)$},
where $u^j$ has io-type $\iota$ 

$/*$ process $u^j$ in order of increasing depth $*/$

}%A

{\sf else if} $y\notin V(\S.)$ or $y\in \A.$ and $B_y$ a proper  ancestor of $B_x$  

\hspace{10pt} $/*$ may have $x=y\in \A.$ $*/$

{\hi %A

add $xy$ to \S.

{\sf if} $y\in \A.$ {\sf  then} $\kappa=$ the opposite M-type from $xy$

{\sf else} {\sf if} $xy=\eta(y)$ {\sf then} $\kappa=o$ {\sf else}  $\kappa=i$

 \fa(y,\kappa)

}%A

{\addtolength{\parindent}{10pt}

$/*$ {\bf nops:} MAYBE SOME TEST CONDITIONS CAN BE ELIMINATED$*/$

$/*$ $B_y$ unrelated to $B_x$ in \os.  $*/$

$/*$ $y \in \B.$ an ancestor of $B_x$ $*/$

$/*$ $B_y$ proper descendant of $B_x$ and $xy$ not alternating with $B_y$ $*/$

$/*$ $B_x=B_y$ $*/$

}

\iffalse
{\sf else} 
{\sf if} $b(y)$ is an  ancestor of $b(x)$ 
in \sm. $/*$ possibly $b(y)=b(x)$ $*/$

{\hi $/*$ equivalent test: $b(y)$ entered \S. before $b(x)$ or
$b(y)=b(x)$
$*/$  %A

add $xy$ to \S.; {\em find\_ap}($y_{\ib}$)
} %A

{\sf else}

\fi

}%addtolength parindent

{\sf return}

\vspace{\vmargin}
}%narrower
\end{minipage}
}%fbox
\caption{Path-preserving depth-first search.}
\label{PPdfsAlg}
\end{center}
\end{figure}

\iffalse
{$f\notin V(\P.)$} {\bf then}

{\bf for} {\em each  vertex $v\in V$} {\bf do}

{\hi

{\bf while} {$f\notin V(\P.)$} {\bf then}
????????????
-------------------------------------------

\goin {\bf else} \tcc{grow step}

\goin\goin add $xy, yy'$ to \S., where $yy'\in M$

\goin\goin {\em find\_ap$(y')$}

{\bf return}
\fi

\bigskip

The graph for augmentation \G. has 
vertex set $\A.\cup \B.$, where \A. is the set of atoms of $G$
and \B. is the set of maximal blossoms with positive $z$-value.     
The edge set consists of all edges that are eligible.
From hereon in we use the words "atom" and "blossom" only to refer to atoms and 
blossoms of \G..
Such a \G.-blossom $B$ is necessarily outer, but can be heavy or light.
$B$ can have $\delta(B,M)-\eta(B)\ne\emptyset$. $I(B)$ is irrelevant.

The notation $B_x$ refers to \G..

NO!!: 
We use a slight extension of this
notation wherein $B_x=x$ means
that $x$ is an atom, and $B_x\ne x$ means $x$ is in a blossom.

For intuition observe that a \B.-vertex is like a 1-matching blossom
in that an augmenting path may pass through it on two edges of the same M-type;
on the other hand it is like a 1-matching vertex in that it can be on
only 
one augmenting trail.
(Also a free \B.-vertex may occur as an end of an augmenting trail, on a matched or unmatched edge. It may occur only once, since a free \B.-vertex must
have deficiency 1.)

The opposite holds for a \A.-vertex: any augmenting
path passes through it on edges of opposite M-type, but it can be on
an augmenting path multiple times, as well as on many  augmenting paths.

$\ol(v)$ and $\inl(v)$ obviously differ  for \A. and \B. vertices.
A \B. vertex has $\inl(v)=\eta(v)$ and $\ol(v)=\delta(v)-\eta(v)$.

\bigskip

\ibl(y) has an unscanned edge implies
$y \notin \S.$ or $b(y)$ is related to $b(x)$

\bigskip

When a vertex $v$ is added to \S. ($v=a$ at the start of the search, or
$v=y\notin V(\S.)$ in the {\sf else if} step)
its $b$ value is initialized, $b(v)=v$.

Note $b(v)=v$ 
means 
$find\_ap$ has been invoked for at least one of
$v_i, v_o$, not necessarily both.

\bigskip

The blossom step involves two design decisions.
The first is delaying blossoms until the downward scan, i.e., postponing
the possible blossom step when the blossom edge is first detected in the upward scan.
Simple examples show this will necessitate rescans, so we MUST delay.
See my figure "Design Considerations for the Blossom Step."

The second decision is to process the new outer nodes in order of increasing depth.
This seems to be done for simplicity only. ie I cant construct an example leading to
rescans if we process by decreasing depth. But the algorithm is less easily stated,
since we accumulate pending blossom calls.

%% file: fin.bbl
\begin{thebibliography}{99}
\footnotesize

%\def\pp{}
\def\referencesheading{0}
\ifcase\referencesheading
%CASE 0: NO "REFERENCES" HEADING
\or
%CASE 1: "REFERENCES" IN ORDINARY FONT
\bigskip\penalty-2000%
%\aorbsec{References}
\noindent{\bf References \hfill\bigskip}
\parindent=0pt
\or
%CASE 2: "REFERENCES" IN BIG FONT
%\def\it{\bif }%
\line{\quad}\penalty-2000%
\noindent{\twelvebf References \hfill}
%\line{\quad}
\medskip
\parindent=0pt
\fi
%
\font\ninerm=cmr9
\font\ninebf=cmbx9
\font\nineit=cmti9        
\font\ninei=cmmi9
\font\ninesy=cmsy9
\def\ninepoint{%
   \def\rm{\ninerm}\def\bf{\ninebf}%
   \def\it{\nineit}\def\smc{\ninerm}\baselineskip=11pt\rm%
        \textfont0=\ninerm \scriptfont0=\sevenrm
	\textfont1=\ninei \scriptfont1=\seveni
	\textfont2=\ninesy \scriptfont2=\sevensy
	\textfont3=\tenex \scriptfont3=\tenex
}
%\ninepoint
\def\rsize{39pt}
% normal:50pt; focs91: 35 pt; stoc 91:39pt
\def\r #1]{\medskip%
\hangafter=-10\hangindent=\rsize%
\hskip0pt  \llap{\hbox to \rsize{#1]\hfill}}\ignorespaces}
%
%6.11.96: comma inserted after issue number
%
\def\al #1,{{\it Algorithmica, #1,}}
\def\comb #1,{{\it Combinatorica, #1,}}
\def\cu,{Comp.\ Sci.\ Dept., Univ.\ Colorado, Boulder, CO,}
\def\focs #1,{{\it Proc.\ #1 Annual Symp.\ on Found.\ of Comp.\ Sci.,}}
\def\ipl #1,{{\it Inf.\ Proc.\ Letters, #1,}}
\def\ja #1,{{\it J.\ Algorithms,  #1,}}
\def\jacm #1,{{\it J.\ ACM,  #1,}}
\def\jcss #1,{{\it J.\ Comp.\ and System Sci., #1,}}
\def\mprog #1,{{\it Math.\ Programming,  #1,}}
\def\mprogb #1,{{\it Math.\ Programming B,  #1,}}
\def\net #1,{{\it Networks, #1,}}
\def\phd{Ph.\ D.\ Dissertation}
\def\sicomp #1,{{\it SIAM J.\ Comput.,  #1,}}
\def\siad #1,{{\it SIAM J.\ Alg.\ Disc.\ Meth., #1,}}
\def\sidm #1,{{\it SIAM J.\ Disc.\ Math., #1,}}
\def\soda #1,{{\it Proc.\ #1 Annual ACM-SIAM Symp.\ on Disc.\ Algorithms,}} 
\def\stoc #1,{{\it Proc.\ #1 Annual ACM Symp.\ on Theory of Comp.,}}
\def\tr {Tech. Rept.\ }
%
\def\pp#1-#2.{pp.\ #1--#2.}
\def\spp#1-#2;{pp.\ #1--#2;}
\iffalse
\def\al #1,{{\it Algorithmica, #1}}
\def\cu,{Comp.\ Sci.\ Dept., Univ.\ Colorado, Boulder, CO,}
\def\focs #1,{{\it Proc.\ #1 Annual Symp.\ on Found.\ of Comp.\ Sci.,}}
\def\ipl #1,{{\it Inf.\ Proc.\ Letters, #1}}
\def\ja #1,{{\it J.\ Algorithms,  #1}}
\def\jacm #1,{{\it J.\ ACM,  #1}}
\def\jcss #1,{{\it J.\ Comp.\ and System Sci., #1}}
\def\mprog #1,{{\it Math.\ Programming,  #1}}
\def\mprogb #1,{{\it Math.\ Programming B,  #1}}
\def\net #1,{{\it Networks, #1}}
\def\phd{Ph.\ D.\ Dissertation}
\def\sicomp #1,{{\it SIAM J.\ Comput.,  #1}}
\def\siad #1,{{\it SIAM J.\ Alg.\ Disc.\ Meth., #1}}
\def\sidm #1,{{\it SIAM J.\ Disc.\ Math., #1}}
\def\soda #1,{{\it Proc.\ #1 Annual ACM-SIAM Symp.\ on Disc.\ Algorithms,}} 
\def\stoc #1,{{\it Proc.\ #1 Annual ACM Symp.\ on Theory of Comp.,}}
\def\tr {Tech. Rept.\ }
\fi

%
\def\talg #1,{{\it ACM Trans.~on  Algorithms},  #1,}
%

\bibitem{CCPS}
W.J. Cook, W.H. Cunningham, W.R. Pulleyblank, and A.~Schrijver,
{\it Combinatorial Optimization},
Wiley and Sons, NY, 1998.

\iffalse
\bibitem{CLRS}
T.H.~Cormen, C.E.~Leiserson, R.L.~Rivest and C.~Stein,
{\em Introduction to Algorithms}, 2nd Ed.,
McGraw-Hill, NY, 2001.


\bibitem{CMSV}
M.B. Cohen, A. Madry, P. Sankowski, and A. Vladu,
"Negative-weight shortest paths and unit capacity minimum cost flow
in $\widetilde O (m^{10/7} \log W)$ time",
\soda 28th, 2017, \pp 752-771.
\fi

\bibitem{D}
E.A. Dinic,
"Algorithm for solution of a problem of maximum flow in a network with power
estimation",
{\em Soviet Mathematics Doklady}, 11, 1970, \pp 1277-1280. (In Russian.)


\bibitem{DHZ}
R. Duan, H. He, and T. Zhang,
"A scaling algorithm for weighted f-factors in general graphs",
%arXiv:2003.07589v1, 2020.
{\em Proc. of the  47th International
Colloquium on Automata, Languages, and Programming (ICALP 2020)}, 
Vol. 168 of LIPIcs, \pp 41:1-41:17, 2020.

\bibitem{DPS}
R. Duan, S. Pettie, and H-H. Su,
"Scaling algorithms for weighted matching in general graphs",
{\em ACM Trans. Algorithms} 14, 1, 
2018,
Article 8, 35 pages.

\bibitem{Ed}
J. Edmonds, 
"Paths, trees, and flowers", 
Canad. J. Math. 17, 1965, \pp 449-467.

\bibitem{E}
J. Edmonds, ``Maximum matching and a polyhedron with 0,1-vertices'', 
{\it  J.\ Res.\ Nat.\ Bur.\ Standards 69B}, 1965, \pp 125-130.

\bibitem{ET}
S. Even and R.E. Tarjan,
``Network flow and testing graph connectivity'',
\sicomp 4, 1975, \pp 507-518.


\iffalse
\bibitem{FT}
M.L. Fredman and R.E. Tarjan, ``Fibonacci heaps and their uses in 
improved
network optimization algorithms'',
\jacm 34, 3, 1987, \pp 596-615.
\fi

\bibitem{G76}
H.N. Gabow, "An efficient implementation of Edmonds’ algorithm for 
maximum matching on
graphs", \jacm 23, 2, 1976, \pp 221-234.

\bibitem{G}
H.N. Gabow, 
 ``A scaling algorithm for weighted matching on general graphs,''
\focs 26th, 1985, \pp 90-100. 

\bibitem{G17}
H.N. Gabow,
"The weighted matching approach to maximum cardinality matching,"
{\it Fundamenta Informaticae}
154, 1-4, 2017, \pp 109-130.

\bibitem{G20}
H.N. Gabow,
"A weight-scaling algorithm for $f$-factors of multigraphs",
	arXiv:2010.01102, 2020.  

\bibitem{Gnca}
H.N.~Gabow,
"A data structure for nearest common ancestors with linking",
\talg 13, 4, 2017, Article 45, 28 pages.

\bibitem{G18}
H.N.~Gabow,
"Data structures for weighted matching and extensions to
$b$-matching and $f$-factors,"
\talg 14, 3, 2018, Article 39, 80 pages.

\bibitem{GT85}
H.N. Gabow and R.E. Tarjan,
``A linear-time algorithm for a special case of disjoint set union'',
\jcss 30, 2, 1985, \pp 209-221.

\bibitem{GT89}
H.N. Gabow and R.E. Tarjan,
``Faster scaling algorithms for network problems,''
\sicomp 18, 5, 1989, \pp 1013-1036.

\bibitem{GT}
H.N. Gabow and R.E. Tarjan,
``Faster scaling algorithms for general graph matching problems'',
{\it J.\ ACM} 38, 4, 1991, \pp 815-853.

\bibitem{GMG}
Z. Galil, S. Micali and H.N. Gabow,
``An  $O(EV \log V)$ algorithm
for finding a maximal weighted matching in general graphs'',
\sicomp 15, 1, 1986, \pp 120-130.

\iffalse
\bibitem{GoT}
A.V. Goldberg and R.E. Tarjan,
``Finding minimum-cost circulations by successive approximation'',
{\it Math. of Oper. Res.} 15, 3, 1990, \pp 430-466.
 \fi

\bibitem{HK}
J. Hopcroft and R. Karp, ``An $n^{5 / 2}$ algorithm for maximum
matchings in bipartite graphs'',
\sicomp 2, 4, 1973, \pp 225-231.

\bibitem{HP}
D. Huang and S. Pettie, ``Approximate generalized matching:
$f$-matchings and $f$-edge covers",
arXiv:1706.05761, 2017.

\iffalse
\bibitem{K}
H.W. Kuhn, ``The Hungarian method for the assignment problem",
{\em Naval Research Logistics Quarterly} 2, 1955, \pp 83-97.
\fi

\bibitem{K}
A.V. Karzanov,
"On finding maximum flows in network with special structure and
some applications",
{\em Matematicheskie Voprosy Upravleniya Proizvodstvom}, Vol 5,
1973, \pp 81-94. (In Russian.)

\iffalse
"An exact estimate of an algorithm for finding a maximum flow,
applied to the problem 'of representatives'",
Voprosy Kibernetiki, Trudy Seminara po Kombinatornoi Matematike,
1973, \pp 66-70. (In Russian.) 
\fi

\bibitem{L}
E.L. Lawler,
{\it Combinatorial Optimization: Networks and Matroids},
Holt, Rinehart and Winston, New York, 1976.

\bibitem{LP}
L. Lov\'asz and M.D. Plummer,
{\it Matching Theory}, North-Holland Mathematic Studies 121,
North-Holland, New York, 1986.

\iffalse
\bibitem{LS} DOESNT DO WEIGHTED
Y.T. Lee and A. Sidford, ``Path finding methods for linear programming: Solving
linear programs in $o(vrank)$ 

$\sqrt {rank}$
iterations and faster algorithms for maximum flow",
\focs 55th, 2014, \pp 424-433.
\fi

\bibitem{MV}
S. Micali and V.V. Vazirani, ``An $O(\sqrt {|V|} \cdot |E|)$ algorithm 
for finding
maximum matching in general graphs'',
\focs 21st, 1980, \pp 17-27.

\bibitem{S}
A.~Schrijver,
{\it Combinatorial Optimization: Polyhedra and Efficiency},
Springer, NY, 2003.

\iffalse
\bibitem{T}
R.E. Tarjan, ``Applications of path compression on balanced trees'',
\jacm 26, 4, 1979, \pp 690-715.

\bibitem{Th}
M. Thorup, "Undirected single-source shortest paths with positive
integer weights in linear time, \jacm 46, 3, 1999, \pp 362-394.

\bibitem{V}
V.V. Vazirani, "A proof of the MV matching algorithm",
arXiv:2012.03582v1, 2020.
\fi

\end{thebibliography}
